\documentclass[twocolumn]{aastex62}
\makeatletter
\let\frontmatter@title@above=\relax
\makeatother

\graphicspath{{./}{figures/}}
\usepackage{graphicx}	
\usepackage{amsmath}	
\usepackage{amssymb}	
\usepackage{subfigure}
\usepackage{float}
\usepackage{ae,aecompl,amsmath,amsbsy}
\usepackage{epstopdf}
\usepackage{multirow}

\received{2022 February 19}
\revised{2022 June 7}
\accepted{2022 June 7}
\submitjournal{ApJ}

\shorttitle{X-ray studies of CCBs}
\shortauthors{G. Singh and J. C. Pandey}
\begin{document}

\title{An X-ray study of coronally connected active eclipsing binaries}

\email{gurpreet@aries.res.in, jeewan@aries.res.in}

\author{Gurpreet Singh}

\affil{Aryabhatta Research Institute of Observational Sciences (ARIES)\\
Manora Peak, Nainital, 263001, India}
\affil{Deen Dayal Upadhyaya Gorakhpur University, Gorakhpur 273009, India}
\author{J. C. Pandey$^{1}$}

\begin{abstract}
We present a detailed X-ray analysis and imaging of stellar coronae of five coronally connected eclipsing binaries, namely, 44 Boo, DV Psc, ER Vul, XY UMa, and TX Cnc.  Both components of these binaries are found to be active.  The  X-ray light curves of detached and semidetached type systems  show eclipsed-like features,  whereas no evidence for coronal eclipsing is shown by the contact type systems.
The X-ray light curve of DV Psc shows the O'Connell-like effect where the first maximum is found to be brighter than that of the second.
Results of the coronal imaging using three-dimensional deconvolution of X-ray light curves show the coronae of all these binaries are either in the contact or over-contact configuration, with the primary being 1.7 -- 4 times X-ray brighter than its companion.
 In the current sample, a minimum of 30--50 \% of total UV emission is found to originate from the photosphere and positively correlated with the X-ray emission. 
 X-ray spectra of these systems are well explained by two-temperature plasma models. The temperature corresponding to cool and hot components of plasma are found to be in the ranges of 0.25--0.64 and 0.9--1.1 keV, respectively.
 For the majority of binaries in the sample, the phase-resolved X-ray spectral analysis shows the orbital modulation in X-ray luminosity and emission measure corresponding to the hot component.  A total of seven flaring events are also detected in the four systems with the flare energy in the range of (1.95--27.0)$\times$10$^{33}$ erg and loop length of the order of 10$^{9-11}$ cm.
\end{abstract}
\keywords{\href{http://astrothesaurus.org/uat/444}{Eclipsing binary stars (444)} --- \href{http://astrothesaurus.org/uat/909 }{Late-type stars (909)} --- 	\href{http://astrothesaurus.org/uat/1603} {Stellar flares (1603)} --- \href{http://astrothesaurus.org/uat/1823} {X-ray stars (1823)} --- \href{http://astrothesaurus.org/uat/305}{Stellar coronae (305)}}



\section{Introduction}
The magnetic field generated deep within the convective envelope of cool stars by dynamo when penetrating stellar atmosphere gives rise to several activity phenomena like star spots, plages, activity cycles, flares, etc. These activity phenomena are observable from X-ray to radio waves  \citep[see reviews][]{favata2003stellar,gudel2004x,2009A&ARv..17..251S}.
The X-ray emission in these stars is usually due to a million-degree thermal plasma inhabiting the magnetic-coronal loops. The X-ray study gives insights into the outermost parts of stellar atmospheres and thus is crucial to understanding the coronal heating problem.
Magnetically driven activities  increase with decreasing  rotation period and saturate around  the Rossby number  (R$_0$) of 0.13 where X-ray luminosity follows the relation $L_X/L_{bol} \approx 10^{-3}$ \citep[]{pizzolato2003stellar}. 
Various theories have been proposed for such a saturation regime, e.g. the internal dynamo saturates and produces no further magnetic flux \citep[]{vilhu1987chromospheric}, coverage of active regions on the stellar surface to the maximum extent \citep[]{vilhu1984nature}, the corona is being stripped as a result of increased centrifugal force caused by rotation \citep[]{jardine1999coronal} and concentrated
magnetic flux to poles \citep[]{solanki1997polar}. 

\begin{table*}
    \centering
    \caption{Basic parameters of active binaries in the sample.  }
    \label{tab:basic_parameters_of_sample}
    \begin{tabular}{lccccccccc}
    \hline
    \hline
    System & Sp. Type & Distance$^*$ & $T_0$ (HJD) & Period & $M_1$/& $R_1$/& $T_1$/ & $a$ &i\\
           &          &              &             &        & $M_2$ & $R_2$ & $T_2$  &   & \\
           &          & (pc)         &             &  (d)   & ($M_\odot$) & ($R_\odot$) & (K)& ($R_\odot$) &($^o$)\\
    \hline
        44 Boo$^1$ & G0 Vn     & 12.5$\pm$0.2 & 2439852.49030 & 0.2678159 & 0.9/  & 0.85/     & 5300/ &1.94  &77.5\\
                   &           &              &               &           & 0.47  & 0.58      & 5035  &      &    \\
        DV Psc$^2$ & K4V+M0V   &42.5$\pm$0.1  & 2454026.14320 & 0.30853593& 0.714/& 0.71/     & 4450/ &2.034 &74.2\\
                   &           &              &               &           & 0.477 & 0.52      & 3692  &      &    \\
        ER Vul$^3$ & G1-2V+G3V & 50.8$\pm$0.1 & 2445220.40964 & 0.698095  & 1.108/& $1.04-1.25$/& 5900/ &4.28  & 66.63\\
                   &           &              &               &           & 1.052 & $1.05-1.21$ & 5750  &      &      \\
        XY UMa$^4$ & G2V+K5V   & 68.4$\pm$0.2 & 2435216.39630 & 0.47899824& 1.10/ & 1.16 /    & 5310/ &3.107 & 80.86\\
                   &           &              &               &           & 0.66  & 0.63      & 3806  &      &      \\
        TX Cnc$^5$ & F8V       & 188 $\pm$2   & 2455289.34028 & 0.3828824 & 1.276/& 1.255/    & 6400/ &2.46  &63.5        \\
                   &           &              &               &           & 0.580 & 0.866     & 6058  &      &            \\
      \hline    
    \end{tabular}
    ~~\\
    Here $T_0$ is HJD  corresponds to orbital phase 0, $M_1$ and $M_2$,  $R_1$ and  $R_2$, $T_1$ and $T_2$ are mass,  radius, and the effective temperature of the primary and secondary components, respectively, $a$ is binary separation, and $i$ is orbital inclination angle. \\ 
     $^1$ \cite{2001AJ....121.2148G,1978A&AS...32..361D,2004yCat.5124....0S,1989A&A...211...81H};     $^2$ \cite{2019ApJ...877...75P}; $^3$ \cite{2003A&A...402..745D,2004A&A...415..289H,2019CoSka..49..278O,1990A&A...238..145H}; $^4$ \cite{2001MNRAS.326.1489L,2001A&A...371..997P,2010ASPC..435..351G}; $^5$ \cite{2006AJ....132..769P,2007PASJ...59..607L}.

\end{table*}

The saturation regime consists of cool short-period binaries and young stars, whereas old single stars and long period binaries follow the activity-R$_0$  relationship.   Young single stars slow down with age due to angular momentum loss (AML) caused by winds, CMEs, and other factors. Thus, as they grow into old-aged stars, they are not expected to lie in the saturation regime unless they manage to overcome AML.
On the other hand, binary stars exhibit increased activity due to tidal locking between binary components.  The tidal locking may have a detrimental impact on activity because it impedes the differential rotation and decreases the efficiency of the internal dynamo. This has been observed in the case of contact binaries, where decreased differential rotation results in weak magnetic activities and thus weak emission in X-rays \citep[see, e.g.][]{2006AJ....131..990C}.  Based on the lack of correlation between  X-ray luminosity and the Roche lobe filling fraction,  \citet{dempsey1993rosat,dempsey1997rosat} suggested that the magnetic activities are primarily governed by the interaction of convection and differential rotation within the star and are unaffected by other binary components.   
 
Theoretical models predict that semidetached systems evolve into a contact system via orbital AML. During this evolution, the binary system oscillates between contact and broken-contact phases due to thermal relaxations in which AML controls the time spent in each phase \citep[e.g.][]{vilhu1982detached,rucinski1986contact}. Such oscillations should also be observable in X-rays. 
Therefore, X-ray emitting short-period active binaries are ideal candidates for understanding the lower limits of differential rotation for dynamo operation \citep[e.g. ][]{bailyn1995luminosity}.
Such near-contact binaries are believed to contain inter-binary magnetic components \citep[]{gudel2004x}; however, it is unclear if such inter-binary magnetic fields play any role in flaring frequency and overall binary activity.
The existence of inter-binary magnetic fields is more plausible in short-period active binaries by direct contact of coronae of binary components.  Thus, short-period eclipsing binaries are the best objects for constraining theories about the extent of stellar coronae.

In the above context, we present a systematic analysis of  X-ray and UV observations of five short-period eclipsing binaries, namely, 44 Boo, DV Psc, ER Vul, XY UMa, and TX Cnc. We organize the paper as follows:
Section \ref{sample} explains the sample selection, Section \ref{obs} discusses observations and data reduction, Section \ref{analysis} addresses X-ray spectral and temporal analyses, in Section \ref{discussion}, we discuss our results, and we present our conclusions in Section \ref{conclusion}.

\begin{table*}
    \centering
    \caption{Log of observations of stars in the sample.}
    \begin{tabular}{lcllcccc}
    \hline
    \hline
        Star Name & Obs. ID & Instrument & ~~Filter & Start Date and Time & Exposure & Offset &Radius$^*$ \\
    &                   &               &       &                       &   Total/Useful        & &\\
                    &   &     & &  (yyyy-mm-dd hh:mm:ss)  & (ks) & (') & (")\\
    \hline
44 Boo  & 0100650101 & MOS1 & THICK   & 2001-6-8 10:49:36 & 24.2/17.4& 0.005&12-80$^{\dagger}$\\
       &            & MOS2 & THICK   & 2001-6-8 10:49:35 & 24.2/17.6  &      &12-80$^{\dagger}$\\
DV Psc & 0603980101 & MOS1  & MEDIUM & 2009-6-19 14:49:19 & 55.5/50.0  &0.002 & 50\\
       &            & MOS2  & MEDIUM & 2009-6-19 14:49:17 & 55.5/50.0  &      & 50\\
       &            & PN    & MEDIUM & 2009-6-19 15:11:39 & 53.9/49.4  &      & 50\\
       &            & OM    & UVW1   & 2009-6-19 14:57:43 & 55.8/53.0  &      & \\
ER Vul & 0785140601 & MOS1  & THICK  & 2016-4-22 09:08:12 & 14.4/14.4  & 0.018& 50\\
       &            & PN    & THICK  & 2016-4-22 09:14:02 & 14.2/14.2  &      & 50\\
       &            & OM    & UVW2   & 2016-4-22 09:17:08 & 15.8/13.2  &      & \\
       & 0781500101 & MOS1  & MEDIUM & 2016-4-22 13:51:33 & 75.4/60.6  &0.016 & 50\\
       &            & MOS2  & MEDIUM & 2016-4-22 13:52:16 & 75.3/60.6  &      & 50\\
       &            & PN    & MEDIUM & 2016-4-22 13:57:22 & 75.2/60.6  &      & 50\\
       &            & OM    & UVM2   & 2016-4-22 14:00:29 & 76.7/66.0  &\\
XY UMa & 0200960101 & MOS1  & MEDIUM & 2005-3-28 23:11:06 & 86.4/59.2  &0.014 & 52\\
       &            & MOS2  & MEDIUM & 2005-3-28 23:11:07 & 86.4/59.2  &      & 52\\
       &            & PN    & MEDIUM & 2005-3-28 23:33:26 & 87.1/57.8  &      & 50\\
       &            & OM    & UVM2   & 2005-3-28 23:19:34 &102.6/96.4  &\\
TX Cnc & 0761921201 & MOS1  & MEDIUM & 2015-5-11 06:07:50 & 58.0/58.0  &4.501 & 30\\
       &            & MOS2  & MEDIUM & 2015-5-11 05:24:47 & 60.6/60.6  &      & 30\\
       &            & PN    & MEDIUM & 2015-5-11 05:47:03 & 59.0/59.0  &      & 33\\
        
    \hline
    \end{tabular}\\
    $^*$ Radius of the circular region for the extraction of light curve and spectra for the source.   $^\dagger$  Annulus region was taken for the source extraction with inner and outer radii of 12"  and 80", respectively. For each source, a similar area of source-free  regions was taken for background extraction.
    \label{tab:observation_log}
\end{table*}
\section{The sample}
\label{sample}

For the coronally connected systems, eclipsing binaries (EBs) with both components being late-type stars were picked from the catalogs available in VizieR\footnote{\url{https://vizier.u-strasbg.fr/viz-bin/VizieR}}.  A total of 973 such EBs were available in different catalogs.  For selecting possible coronally connected binaries (CCBs), we have assumed that the corona has to be extended up to 1 $R_\odot$ from the photosphere \citep[][]{gudel2004x} and it should satisfy the relation, $R_1+R_2 \ge a/2$, where $R_1$ and $R_2$ are  radii of primary and secondary stars, and $a$  is the binary separation. Adopting these criteria, a total of 671 EBs are found with a  possible coronal connection. These  671 binaries  were then searched for X-ray data in \textit{XMM-Newton} archives\footnote{\url{http://nxsa.esac.esa.int/nxsa-web/\#search}}.
A total of 34 binaries were observed by the \textit{XMM-Newton} observatory until December 2021.  Only seven binaries are found to show significant X-ray detection for a detailed analysis.  X-ray data of one EB, VW Cep, was heavily contaminated by background soft proton flares.  A detailed study of the flaring events observed in another EB, SV Cam, was carried out by \cite{2006A&A...445..673S} and the quiescent state X-ray light curve of SV Cam covers only half an orbital cycle. Therefore, we omitted these two binaries from the present study. Finally, the sample was reduced to only five potential CCBs for a detailed study.  The fundamental parameters of these seven  EBs are given in  Table \ref{tab:basic_parameters_of_sample} and described below.

\subsection{44 Boo}
The system 44 Boo is a short-period closest eclipsing binary, which forms visual binary with the ADS 9494 system.   The first X-ray observation of this system indicated that 44 Boo is the X-ray source in this system \citep{1984ApJ...277..263C}.  \cite{1998ASPC..154.1093K} studied orbital phase-dependent X-ray emission of 44 Boo using ROSAT observations and found the signatures of the formation of hot X-ray emitting plasma components between the binary companion. These signatures directly confirm the existence of a coronal connection between binary components of this system.  The outside eclipse variations in the optical band are explained by the presence of dark spots on the surface of binary components \citep[][]{1984ChA&A...8..126L}. \cite{1999ApJ...524..295S} reported thate 44 Boo is a highly active star with a possible magnetic activity cycle of a period of 3.3 yr. The XMM-Newton observation in the present work was also analyzed by \cite{2004A&A...426.1035G} using the RGS and PN data. They fitted the X-ray spectra from the PN detector with three temperature plasma models with temperatures 0.25, 0.61, and 1.05 keV. They also found three flaring events near phases 0.3, 0.65, and 0.9. A detailed discussion of these results is given in Section \ref{discussion}.


\subsection{DV Psc}
DV Psc is a highly active short-period ($\approx 0.308$ days)  eclipsing binary. 
 \citet{beers1994emission}
investigated the chromospheric activity from Ca H\&K emission lines in spectra and concluded the active nature of this binary. Later, it was detected and confirmed as a soft X-ray source in the \textit{ROSAT} survey \citep[]{bade1998hamburg}. A year later, \citet{robb1999photometry} discovered asymmetries in photometric light curves, which \citet{salehi2003analysis} later linked to its magnetic activities. \citet{2007MNRAS.382.1133Z} investigated the physical nature of the system, and concluded that the system is a near-contact binary and is in process of evolving to a W UMa type contact binary in the next $1.5 \pm 0.7$ Gyr. 
 \citet{2010NewA...15..362Z} detected a first flare-like event in the system in photometric time series. Based on photometric data, \citet{2012arXiv1209.1862P} calculated a flare rate of approximately one flare per 2.5 days ($\approx0.13$ flare per cycle), showing that DV Psc is a highly active binary.
\citet{2019AcA....69..261G} proposed a third body in an eccentric orbit with a period of $9.79 \pm 0.60$ yr and a magnetic cycle of period $14.74 \pm 0.84$ yr. 

\subsection{ER Vul}
ER Vul was classified as a short-period RS CVn binary system by \cite{1976ASSL...60..287H}. The presence of cool spots has been used to explain optical light-curve variation outside the eclipse by several authors in the past \citep[see, e.g.][]{1994A&A...291..110O,2002AN....323...31E,2013NewA...23...27P}. The active nature of both the components of this binary was confirmed by \cite{2003A&A...400..257C} and revealed that the secondary component is more active than the primary. \cite{2019CoSka..49..278O}  updated the physical parameters for this binary and suggested the presence of cool spots at high latitudes for both binary components. \cite{2002ASPC..277..223B} and \cite{2004IAUS..215..334B} studied the coronal activity of ER Vul using coordinated Chandra and VLA radio observations. Throughout the X-ray observations, they found a low level of variability without the presence of any eclipse like feature in the light curve.  A solar-like flare with a duration of $\sim$ 30 minutes was also observed during the Chandra observation.

 \subsection{XY UMa}
 XY UMa was first detected as a short-period eclipsing binary by \citet{geyer1955} and \cite{1976IAUS...73..313G}. Magnetic activities of both components have been studied by several authors \citep[][etc]{1986MitAG..67..305B,1994Hilditch} by linking photometric variability outside the eclipse with the presence of star spots.  Two flaring events in the optical band have been reported in the literature \citep[i.e. ][]{1983AJ.....88..532Z,2001A&A...366..202O} where both optical flares were observed near the secondary eclipse.  
 X-ray observations from EXOSAT detected a flaring event and orbital modulation in a quiescent state \citep[][]{1990MNRAS.243..557B, 1990MNRAS.246..337J}.  Recently, three X-ray flares were detected by \cite{2016RAA....16..131G} during the primary eclipse using CHANDRA data.
 
 \subsection{TX Cnc}
TX Cnc is a well-known contact binary of particular interest
as being the member of the youngest open clusters Praesepe harboring such binaries \citep[][]{1998AJ....116.2998R}. 
The optical light curves 
 of TX Cnc reveal the O’Connell effect \citep{1951PRCO....2...85O}, which is indicative of magnetic activities in this system \citep[][]{2009AJ....138..680Z,2021MNRAS.501.2897G}.

\section{OBSERVATIONS AND DATA REDUCTION}
\label{obs}
All the target stars were observed by the \textit{XMM-NEWTON} observatory at different epochs with different detector configurations. The \textit{XMM-NEWTON} has six detectors (EPIC; one PN \citep[]{2001A&A...365L..18S} and two MOS \citep[]{2001A&A...365L..27T}), two reflection grating spectrometers \citep[RGS;][]{2001A&A...365L...7D}, and one optical monitor \citep[OM][]{2001A&A...365L..36M}. Aside from the OM which is used for observations in the optical and UV bands, all other detectors are devoted to observations in the X-ray band with energy coverage of $\approx$ 0.15 -- 15 keV. Since all the detectors are co-aligned, it is feasible to conduct simultaneous observations in multiple energy bands. A log of observations for all the target stars is provided in Table \ref{tab:observation_log}.

The data were reduced using standard \textit{XMM-NEWTON} Science Analysis System (\textsc{SAS}) v18.0.0 software. 
We adopted standard procedures to generate calibrated event lists from EPIC data. We constrained our analysis in the energy range of 0.3 - 10.0 keV as at high energies coronal sources have very low and undetected flux, and also the background contribution is significant.  Each data set was searched and corrected for high particle background events. The data of the stars 44 Boo, ER Vul (obs ID: 0781500101), DV Psc, and XY UMa are found to be affected by high background flaring events, and the observations affected by this were excluded. The pile-up effect was also checked using the task \textsc{epatplot}  within \textsc {SAS}, for each star. Excluding the star 44 Boo, observations of all other stars were free from this effect.  X-ray light curves and spectra were generated from on-source counts obtained from circular regions with radii 30-80".  In the case of 44 Boo,  pile-up was unavoidable for PN observations, thus we  constrained our analysis using the observation from MOS. We took the annulus region of inner radius 12" and outer radius of 80" to avoid the pile-up effect. For each detector, the background was extracted from source-free regions in such a way that it accounted for the same amount of area as the source selection region in the same CCD.
The X-ray light curves were corrected for background contribution and other effects by using the task \textsc{epiclccorr}. The X-ray spectra were generated using the task \textsc{especget}
 which also computes both the photon redistribution and ancillary matrix.   We rebinned the X-ray spectra with at least 20 counts per spectral bin for the spectral analysis.
OM fast mode data was reduced using the standard {\sc omfchain} task in \textsc{SAS}. However, fast mode data can have artifacts like jumps in count rate for some exposures due to some events falling outside the window, whereas the image mode data of each exposure do not suffer such artifacts. Therefore, we recalibrated the light curves of fast mode data from the image mode data using the following strategy.  If M(i) is the mean value of the count rate corresponding to the time series  T(i) of the i$^{th}$ exposure and C(i) is the count rate in the CORR\_RATE column of image mode data file '*SWSRLI0000*', then the scaled time series can be calculated as Tscaled(i) = T(i) * C(i) / M(i). 
The corrected time series obtained from each OM exposure were then merged using \textsc{fmerge} task of \textsc{FTOOLS}.

\section{Analyses and Results}
\label{analysis}

\subsection{Light curves}
\label{temporal_analysis}
\begin{figure*}
\centering
\subfigure[44 Boo]{\includegraphics[height=6.5cm,width=0.45\linewidth]{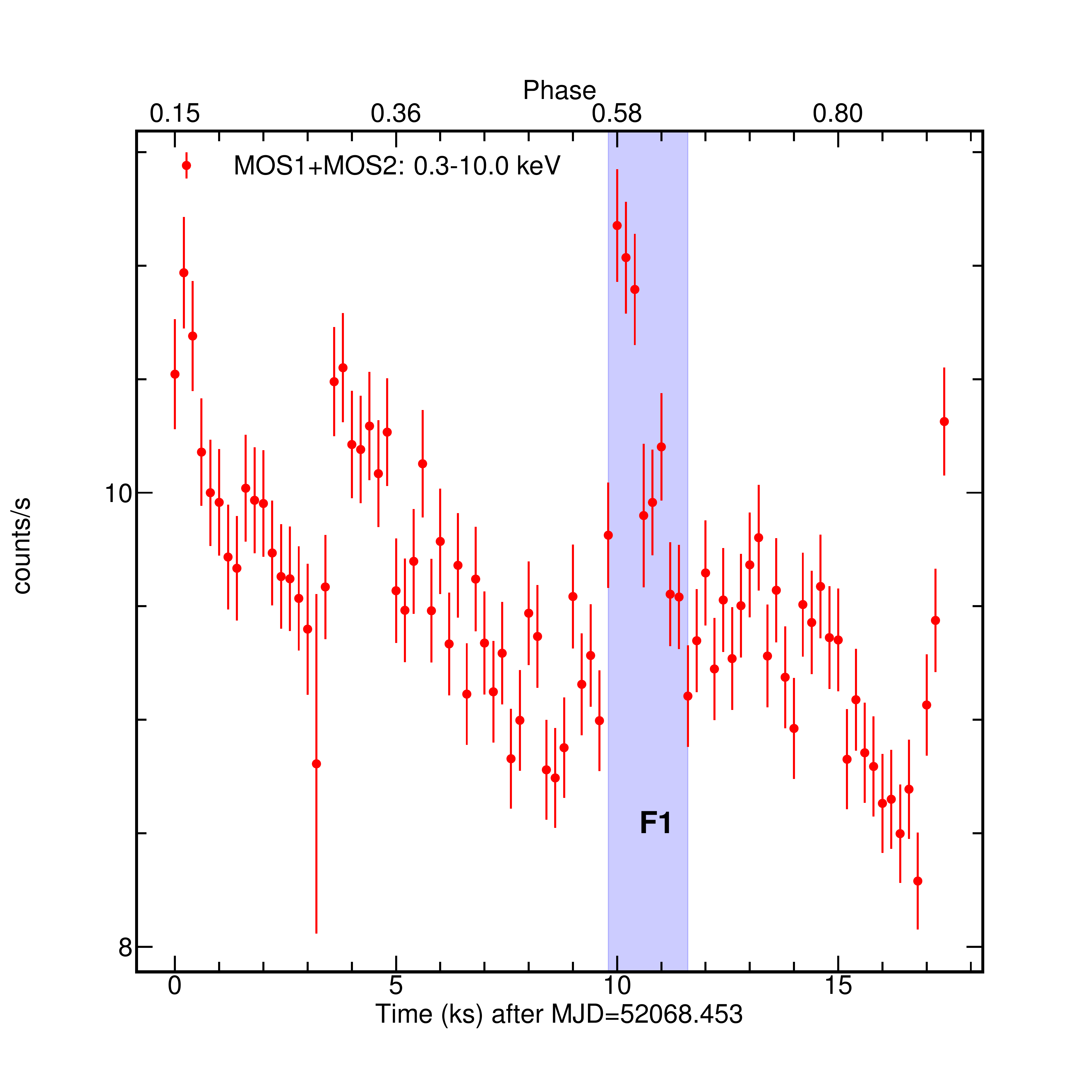}}
\subfigure[DV Psc]{\includegraphics[height=6.5cm,width=0.45\linewidth]{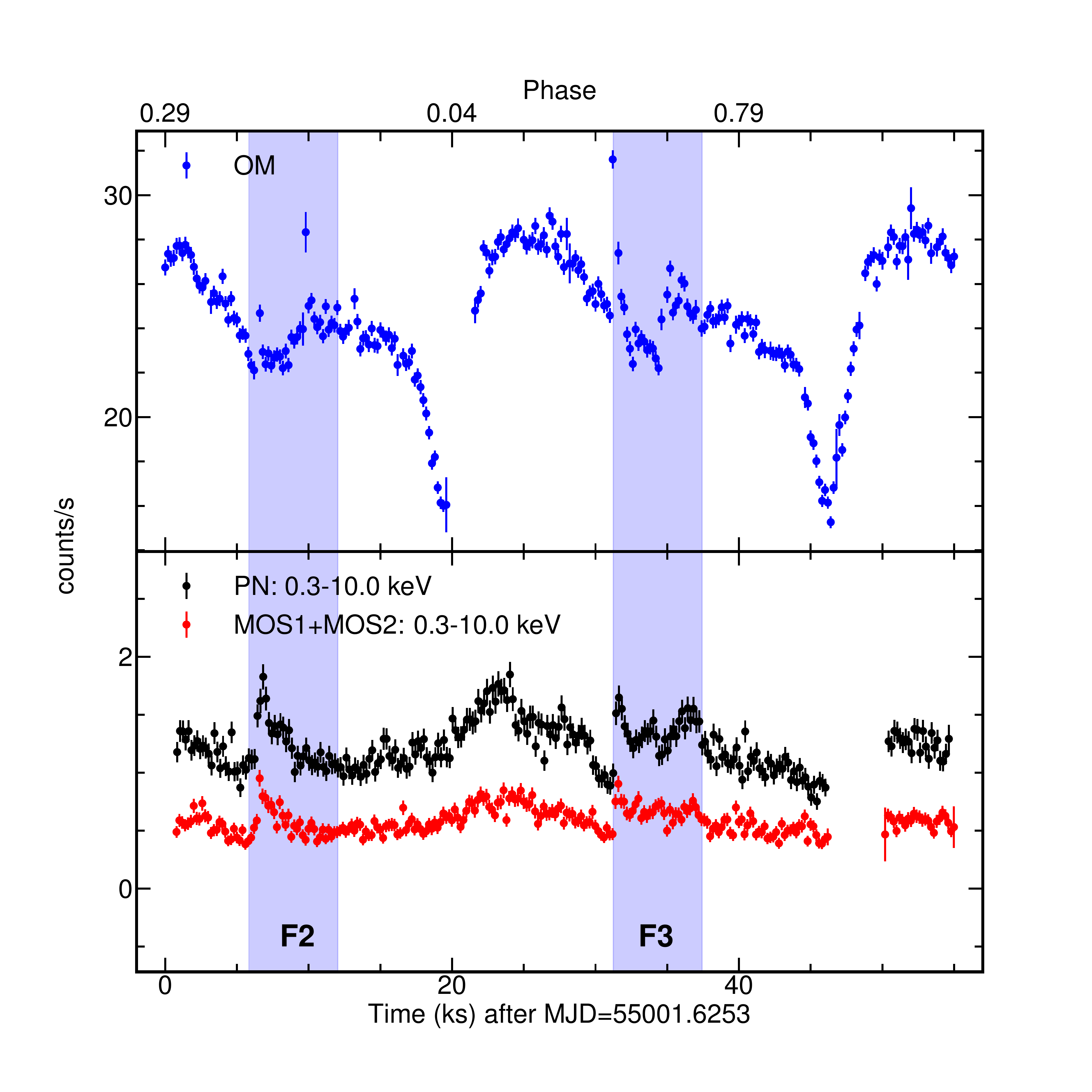}}
\subfigure[ER Vul (obs.ID: 0781500101)]{\includegraphics[height=6.5cm,width=0.45\linewidth]{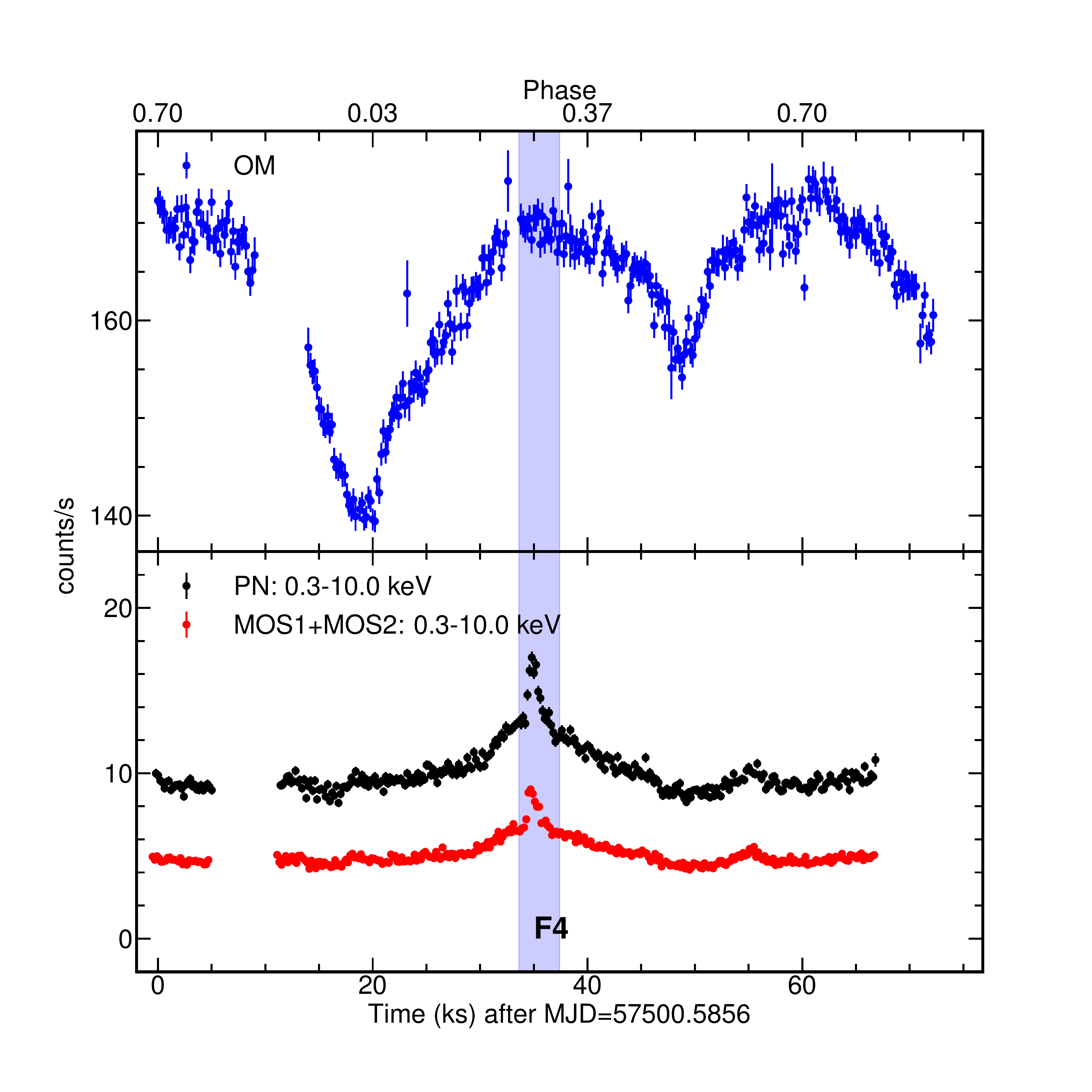}}
\subfigure[ER Vul (obs. ID: 0785140601)]{\includegraphics[height=6.5cm,width=0.45\linewidth]{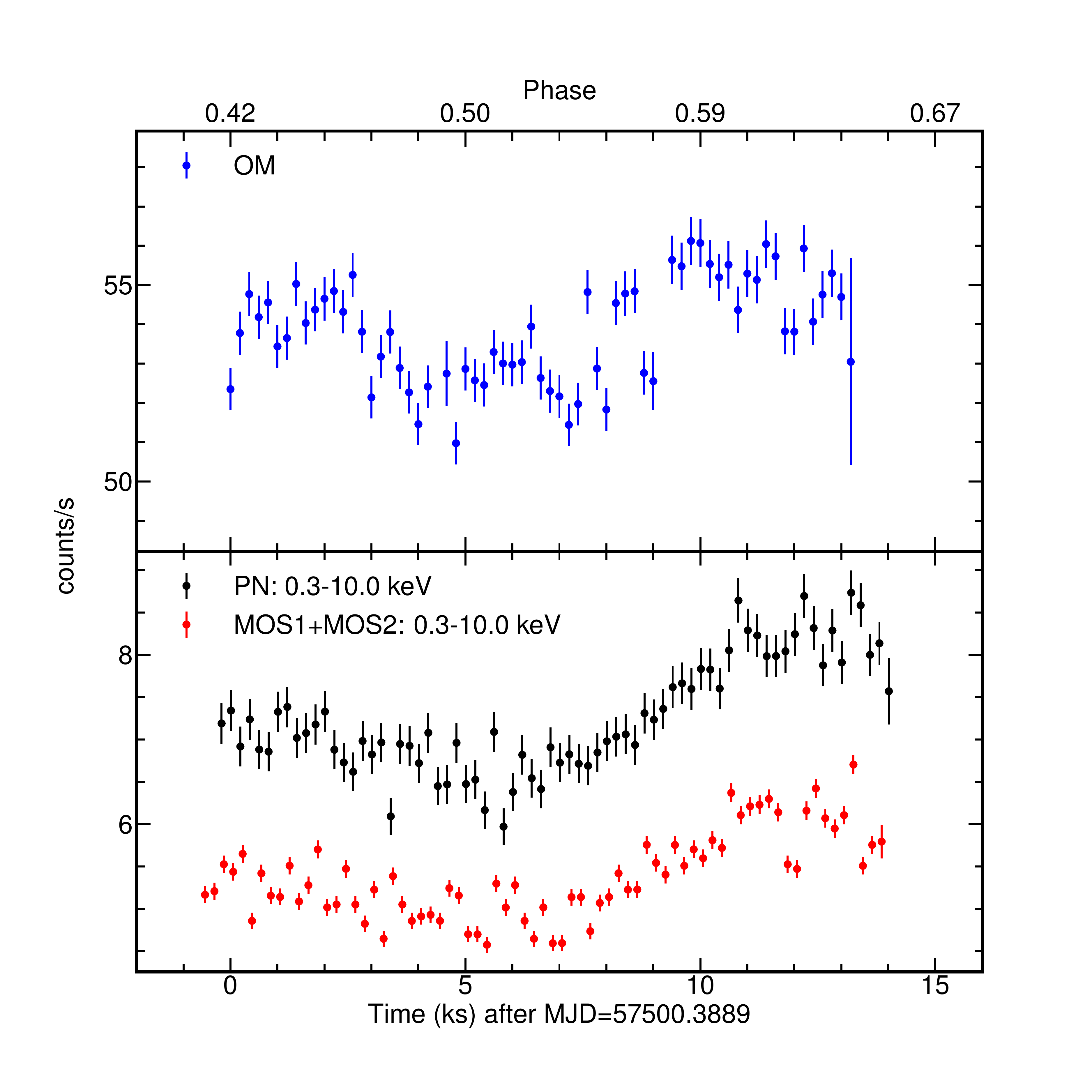}}
\subfigure[XY UMa]{\includegraphics[height=6.5cm,width=0.45\linewidth]{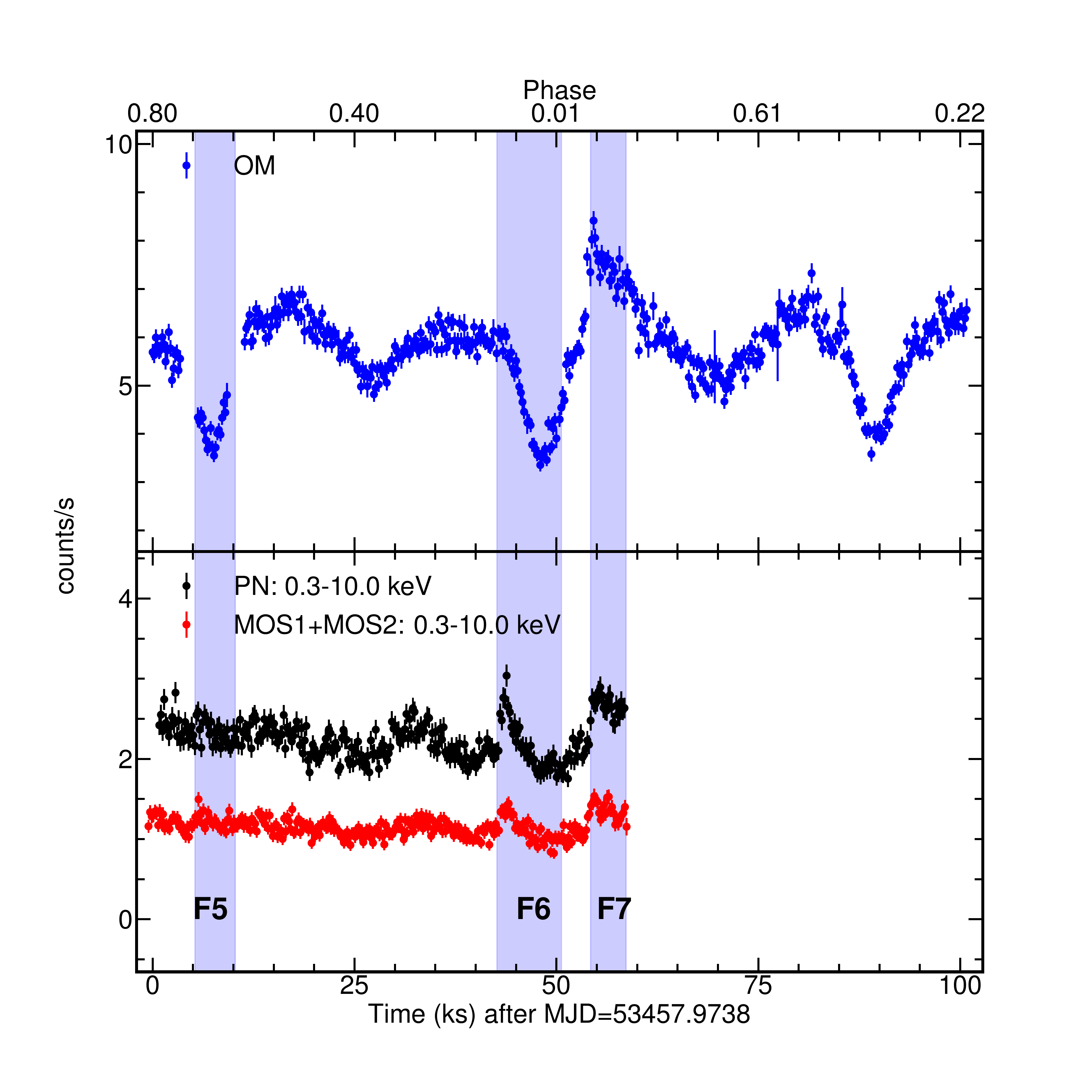}}
\subfigure[TX Cnc]{\includegraphics[height=6.5cm,width=0.45\linewidth]{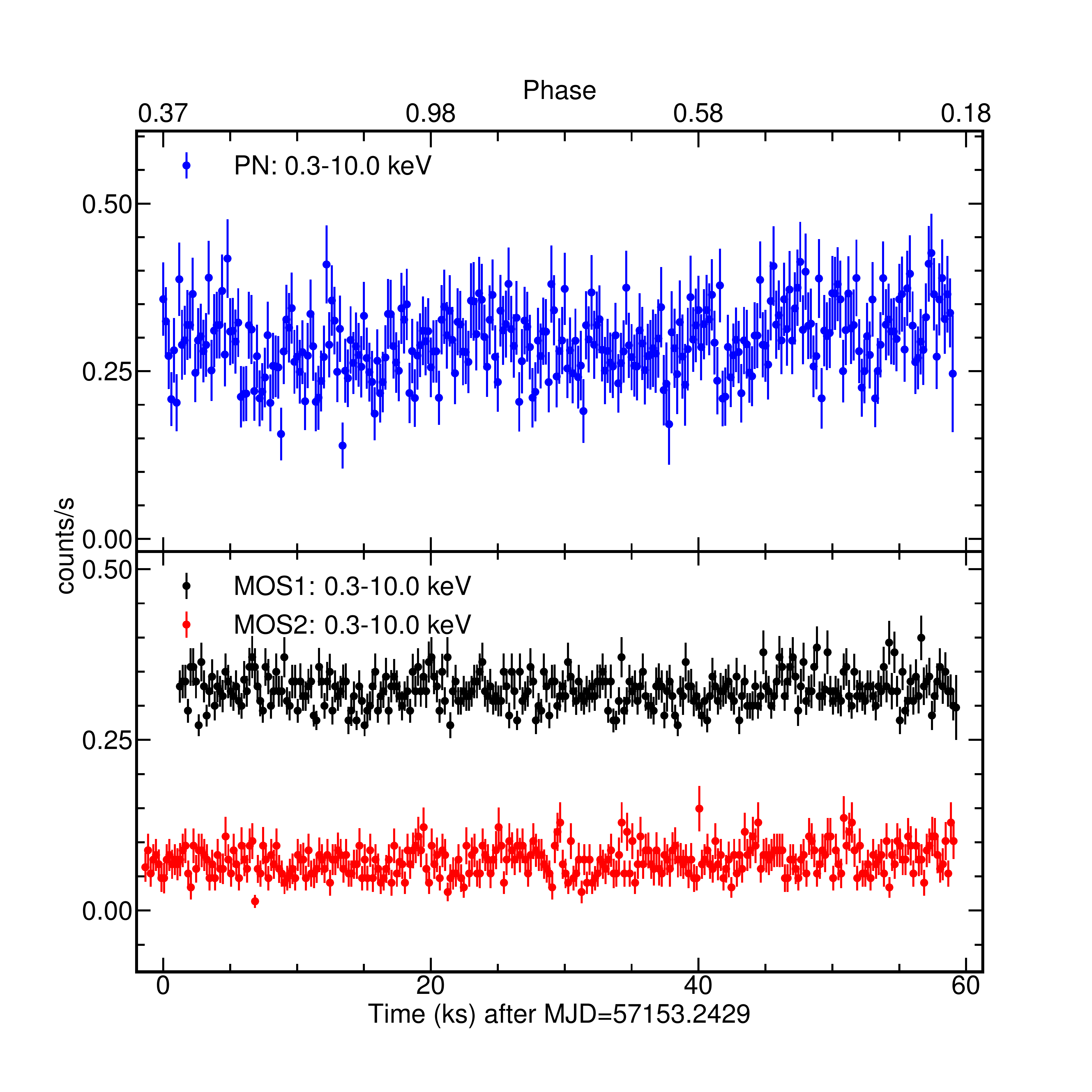}}
    \caption{UV and X-ray light curves of all target stars as observed by the OM (blue), PN (black), and MOS (red) detectors of XMM-Newton. The time bin size of each X-ray light curve is 200 s, whereas each observation from the OM instrument is constituted by each data point of the OM light curves. The energy bin for X-ray is 0.3 - 10.0 keV. The phase values as shown opposite to the time axis are calculated using the ephemeris given in Table  \ref{tab:basic_parameters_of_sample}. The blue shaded regions in Figures 1 (a-c) and (e) refer to flare duration and are marked by FN, where N = 1,2...,7.}
    \label{fig:1}
\end{figure*}

The  X-ray light curves as observed from the \textit{EPIC} instrument in the energy band of 0.3 - 10.0 keV and OM light curves in the different UV filters for all target stars are shown in Figure \ref{fig:1}. All the X-ray light curves were binned with 200 s, whereas the OM light curve is constituted with each OM observation.  The time corresponding to each data point of the light curve was converted to the orbital phase using the ephemeris given in Table  \ref{tab:basic_parameters_of_sample} and shown in the opposite to the time axis of the same figure. Excluding the star TX Cnc,  the X-ray observations for all target stars suffering from flaring events are depicted by the blue shaded areas in Figure \ref{fig:1} (see section \ref{sec:orb} for a detailed description of the flare detection method).

\subsection{Flare detection and orbital modulation }\label{sec:orb}
\begin{figure*}
\centering
\subfigure[44 Boo]{\includegraphics[height=6.5cm,width=0.45\linewidth]{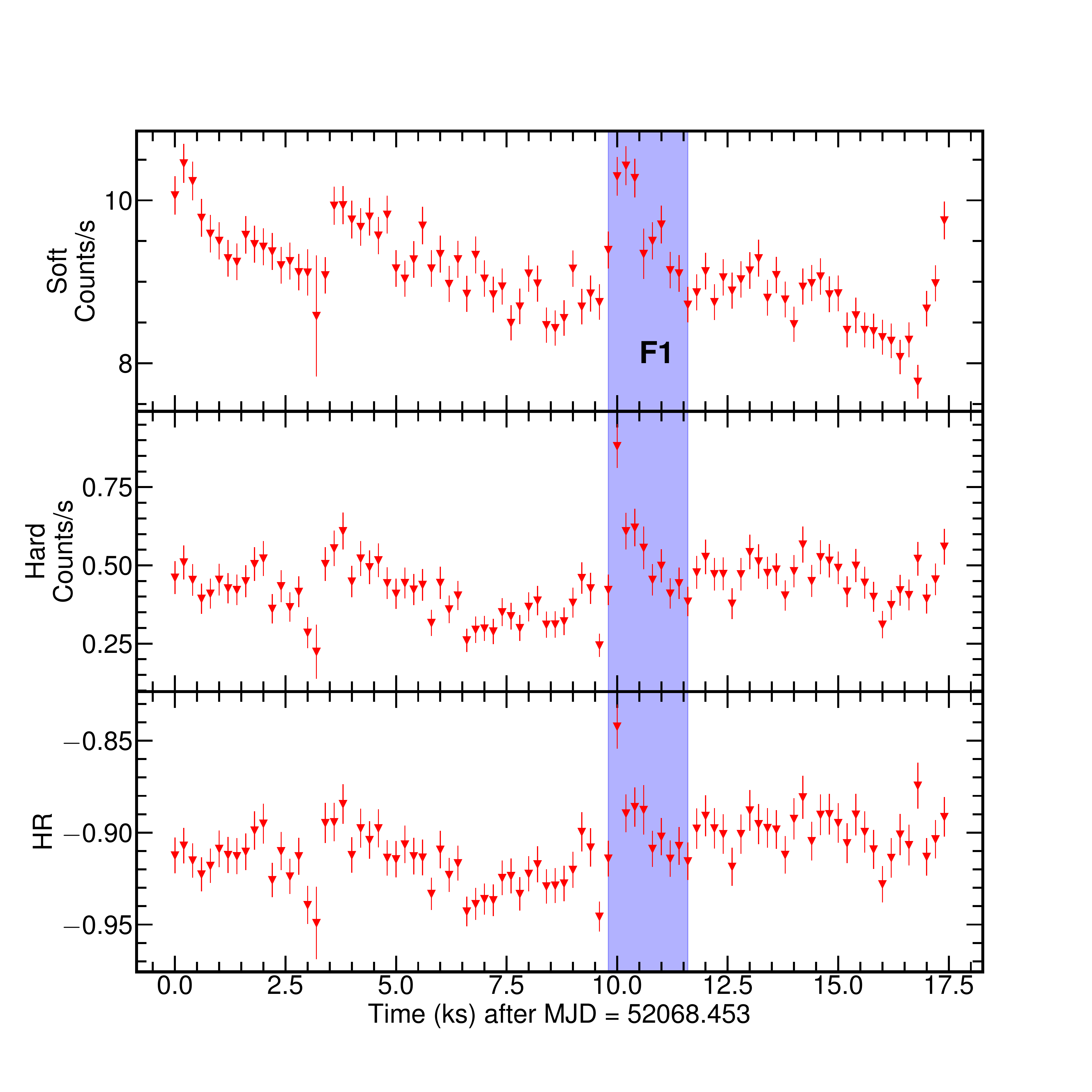}}
\subfigure[DV Psc]{\includegraphics[height=6.5cm,width=0.45\linewidth]{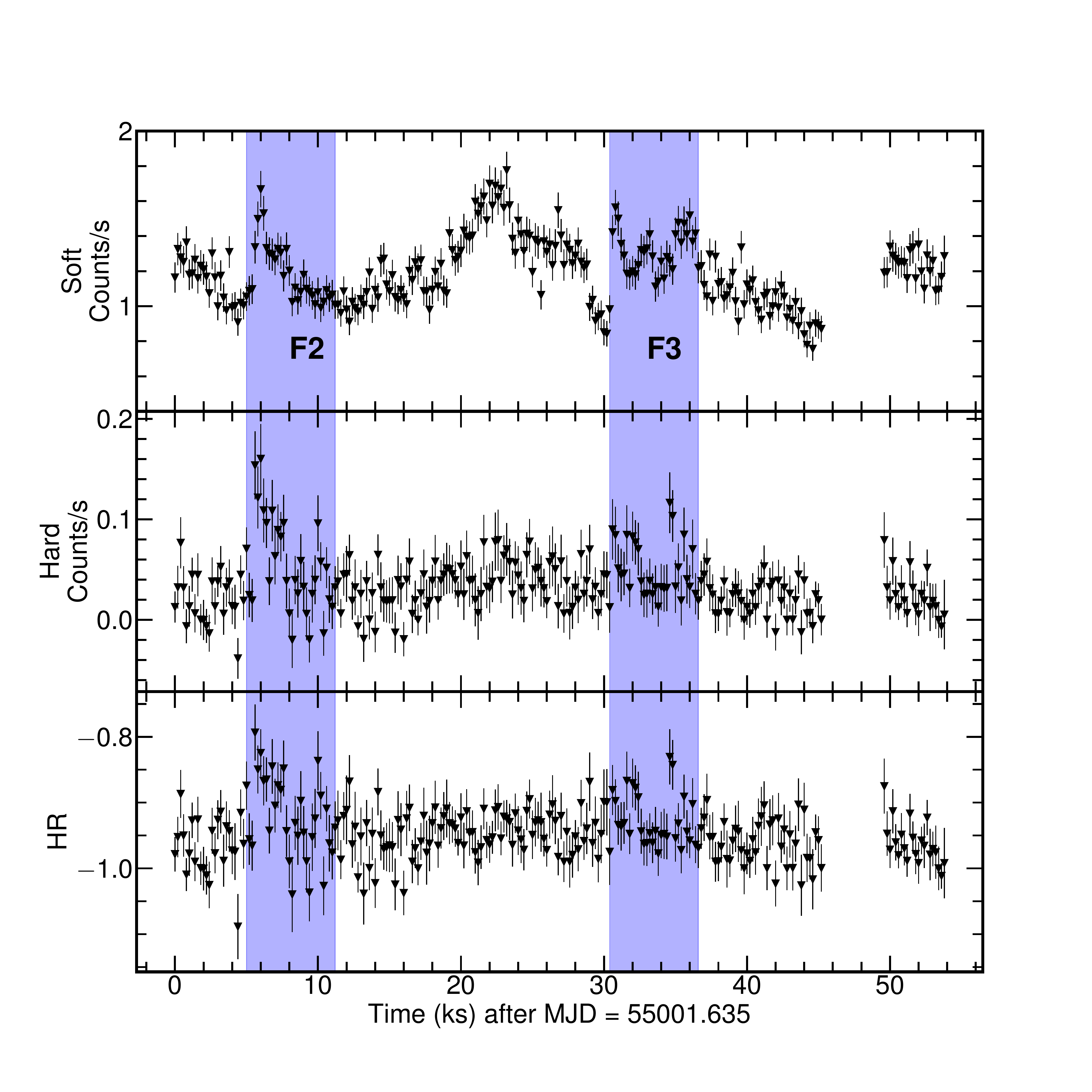}}
\subfigure[ER Vul (obs. ID: 0781500101)]{\includegraphics[height=6.5cm,width=0.45\linewidth]{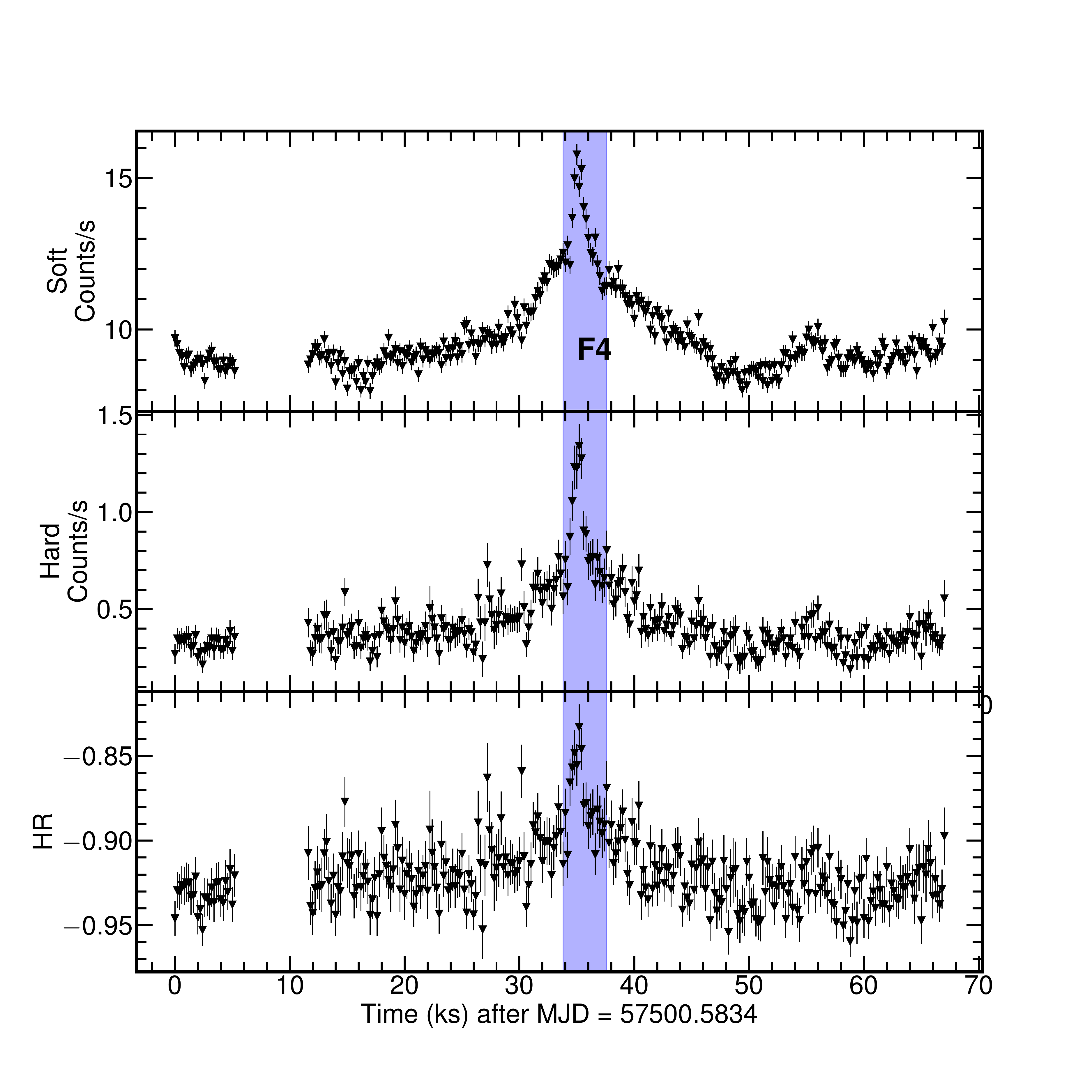}}
\subfigure[ER Vul (obs. ID: 0785140601)]{\includegraphics[height=6.5cm,width=0.45\linewidth]{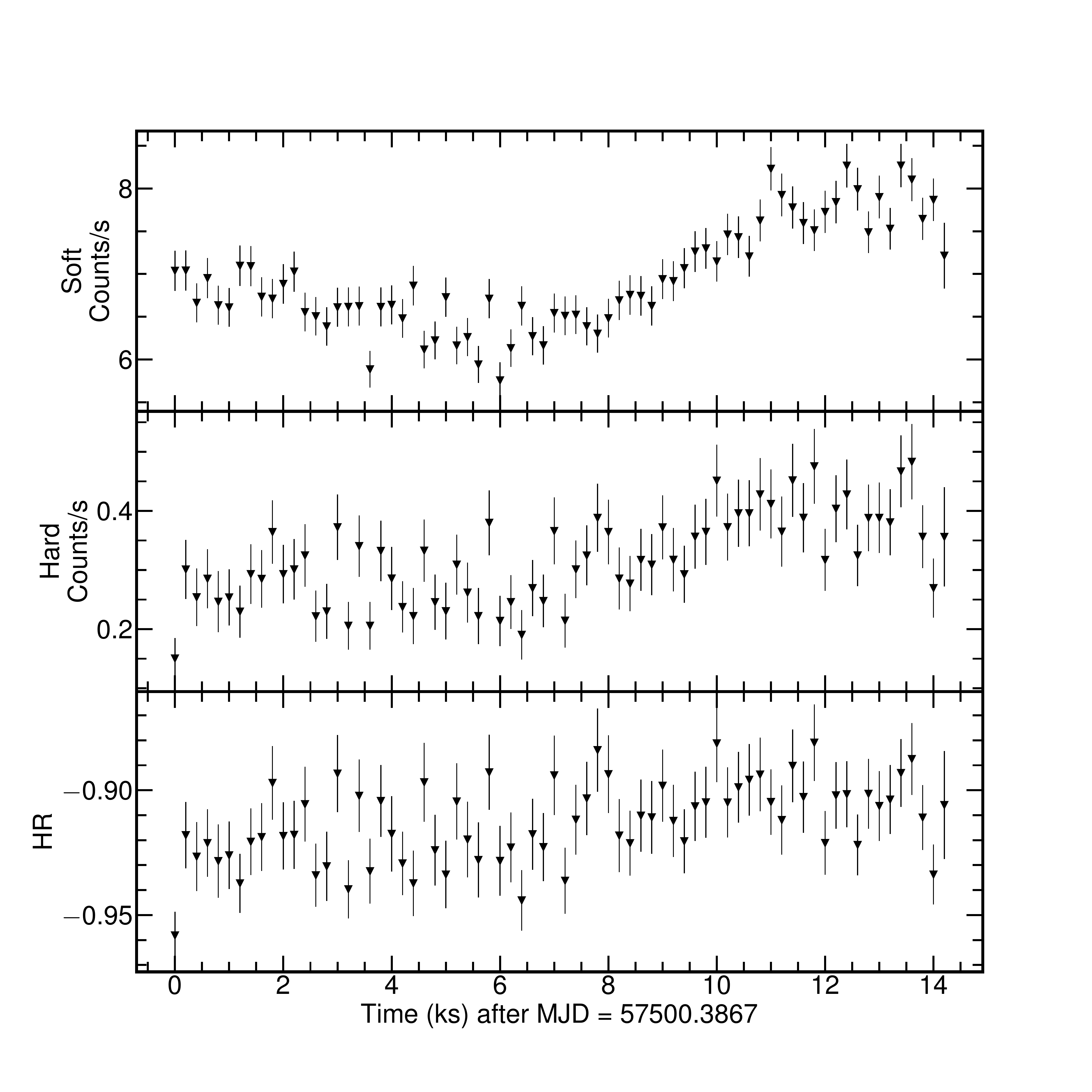}}
\subfigure[XY UMa]{\includegraphics[height=6.5cm,width=0.45\linewidth]{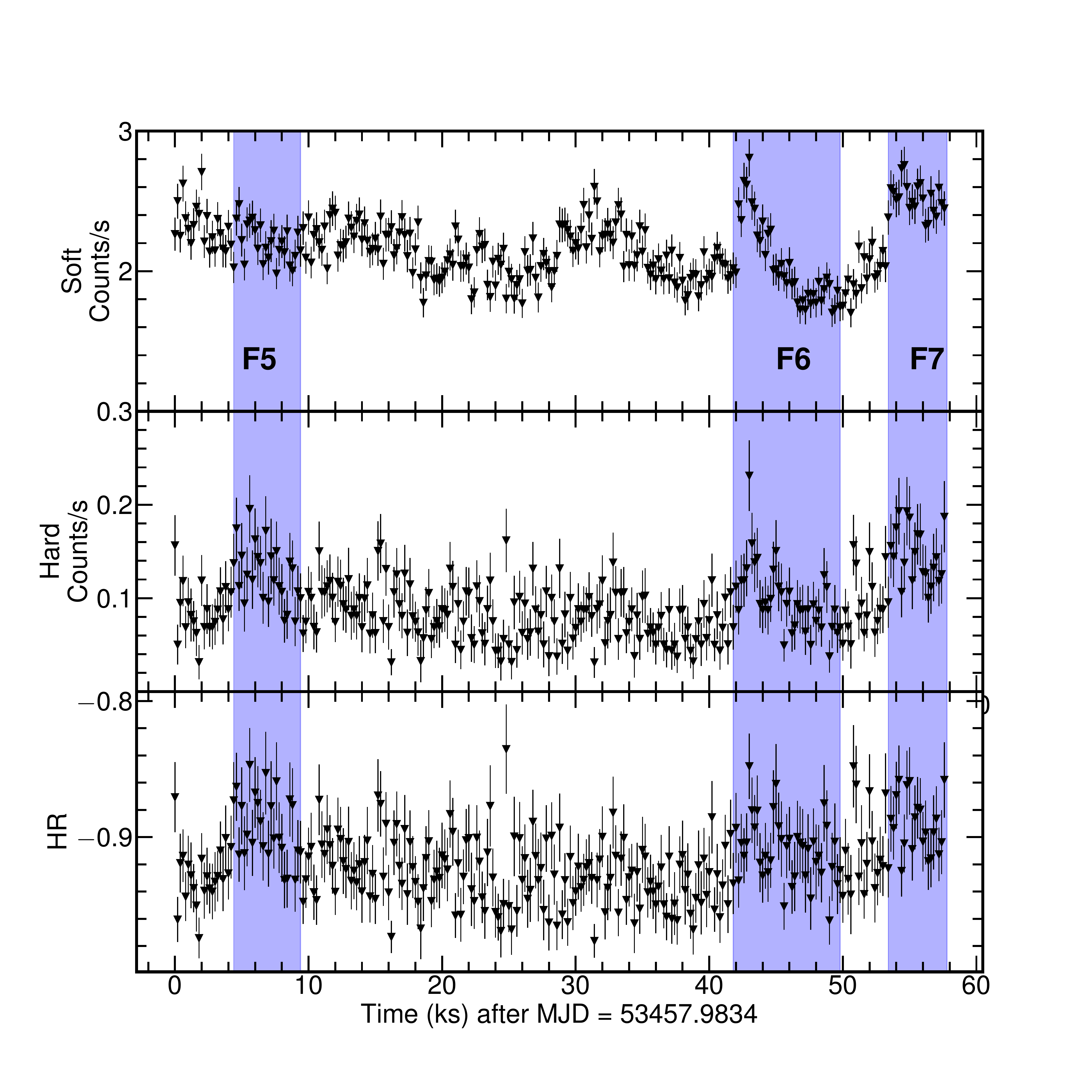}}
\subfigure[TX Cnc]{\includegraphics[height=6.5cm,width=0.45\linewidth]{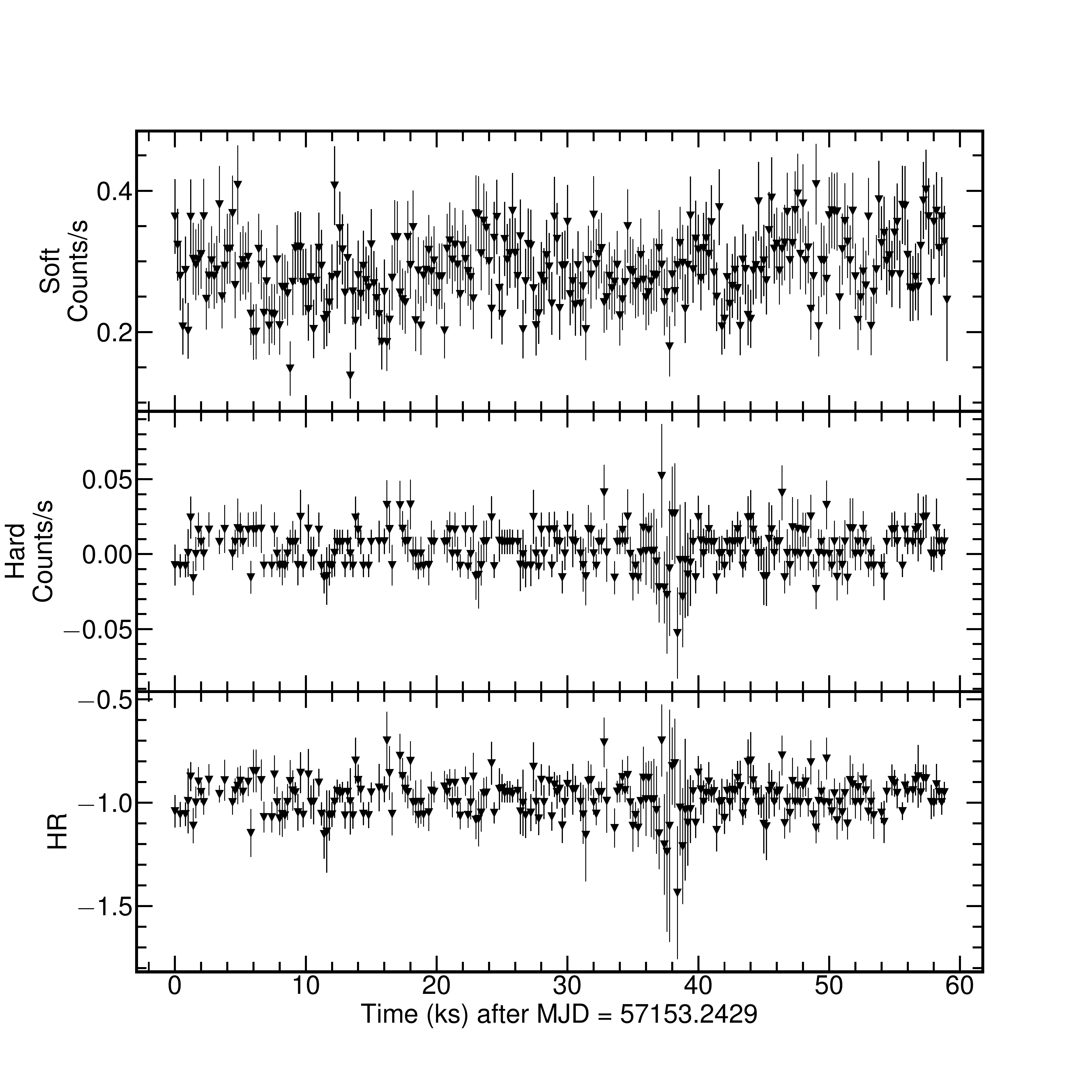}}
    \caption{Soft (0.3-2.0 keV) and hard (2.0-10.0 keV)  bands X-ray light curves along with the HR plots for the stars in the sample. The time bin size is 200 s for both the PN (black) and MOS (red) detectors. }
\label{fig:2}
\end{figure*}

We identified the probable flaring events by inspecting the hardness ratio (HR) curve.  The HR is defined as  $HR = \frac{H-S}{H+S}$, where  H and S are count rates in the hard (2.0-10.0 keV) and soft (0.3 - 2.0 keV) energy bands,  respectively.  X-ray light curves in the soft and hard energy  bands together with the HR curve are shown in Figure \ref{fig:2}.  We identified flaring episodes as times when HR increases rapidly, mimicking the X-ray light curves. Such episodes are shown as blue-shaded regions in Figures \ref{fig:1} and \ref{fig:2}. We identified seven such flaring episodes and refer to them as FN, where N is 1,2,3....7. The  HR value for all stars reached close to -0.9, indicating that the sample stars are very soft X-ray sources. 


\begin{figure*}
\centering
\subfigure[44 Boo]{\includegraphics[height=6.5cm,width=0.45\linewidth]{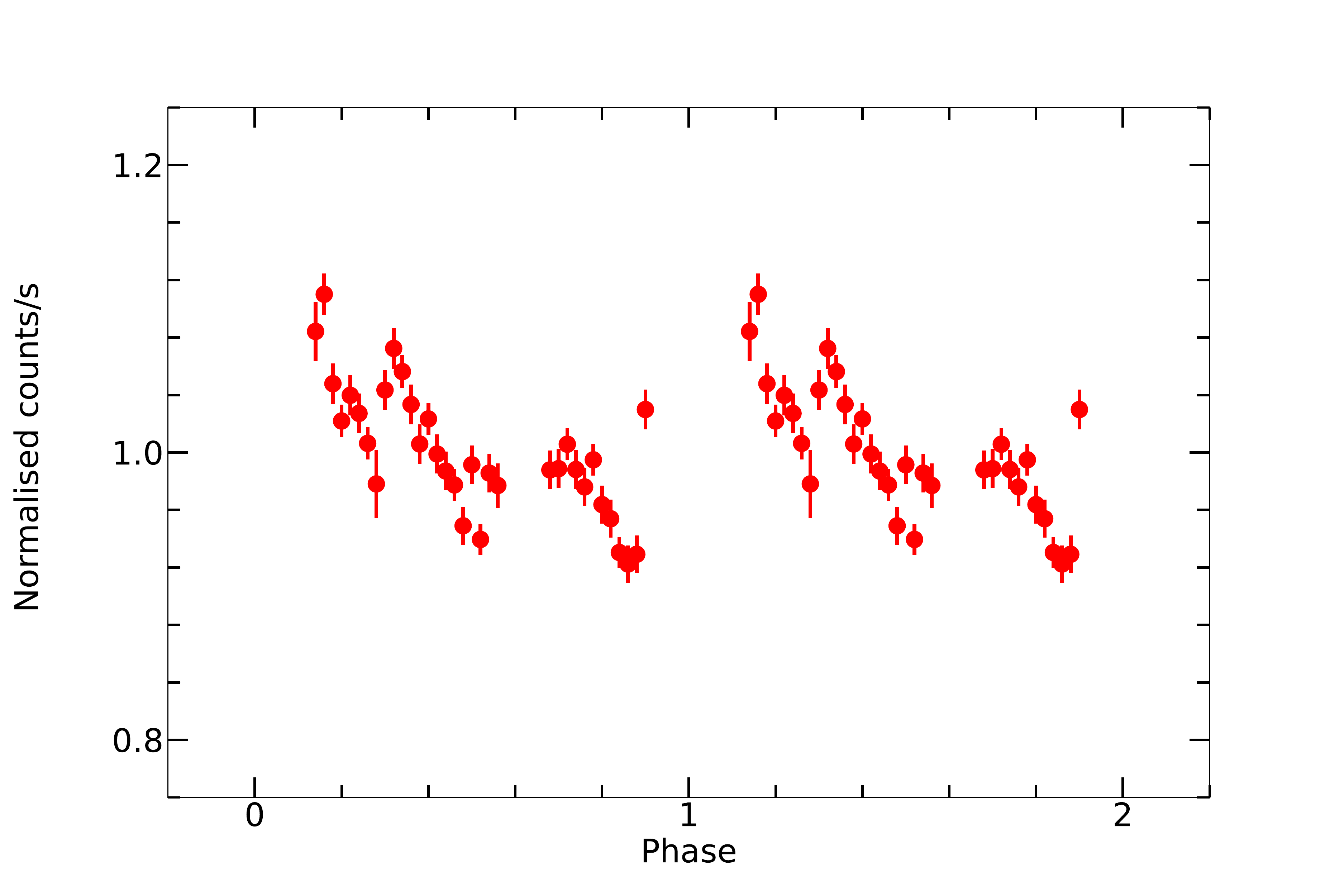}}
\subfigure[DV Psc]{\includegraphics[height=6.5cm,width=0.45\linewidth]{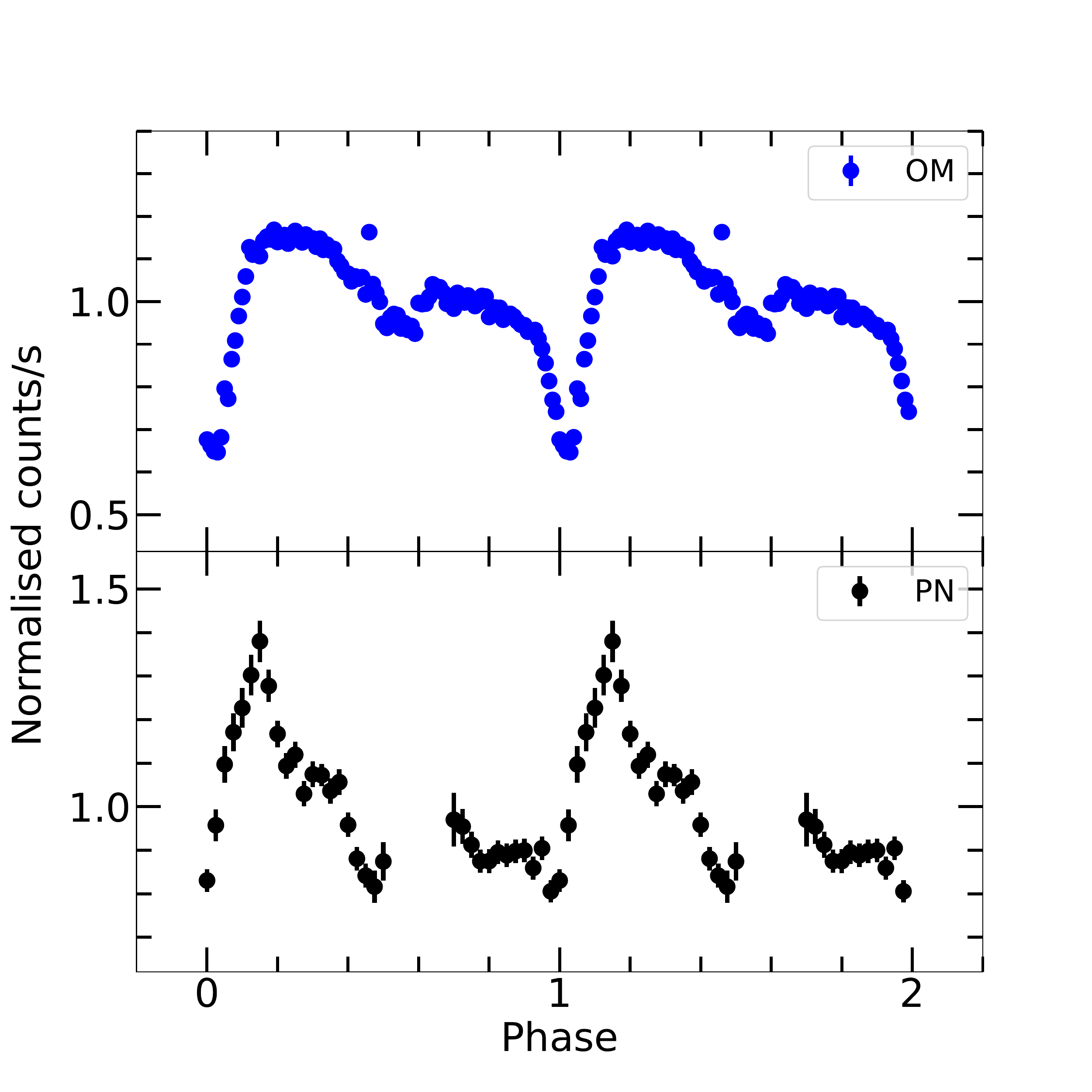}}
\subfigure[ER Vul (obs.ID: 0781500101)]{\includegraphics[height=6.5cm,width=0.45\linewidth]{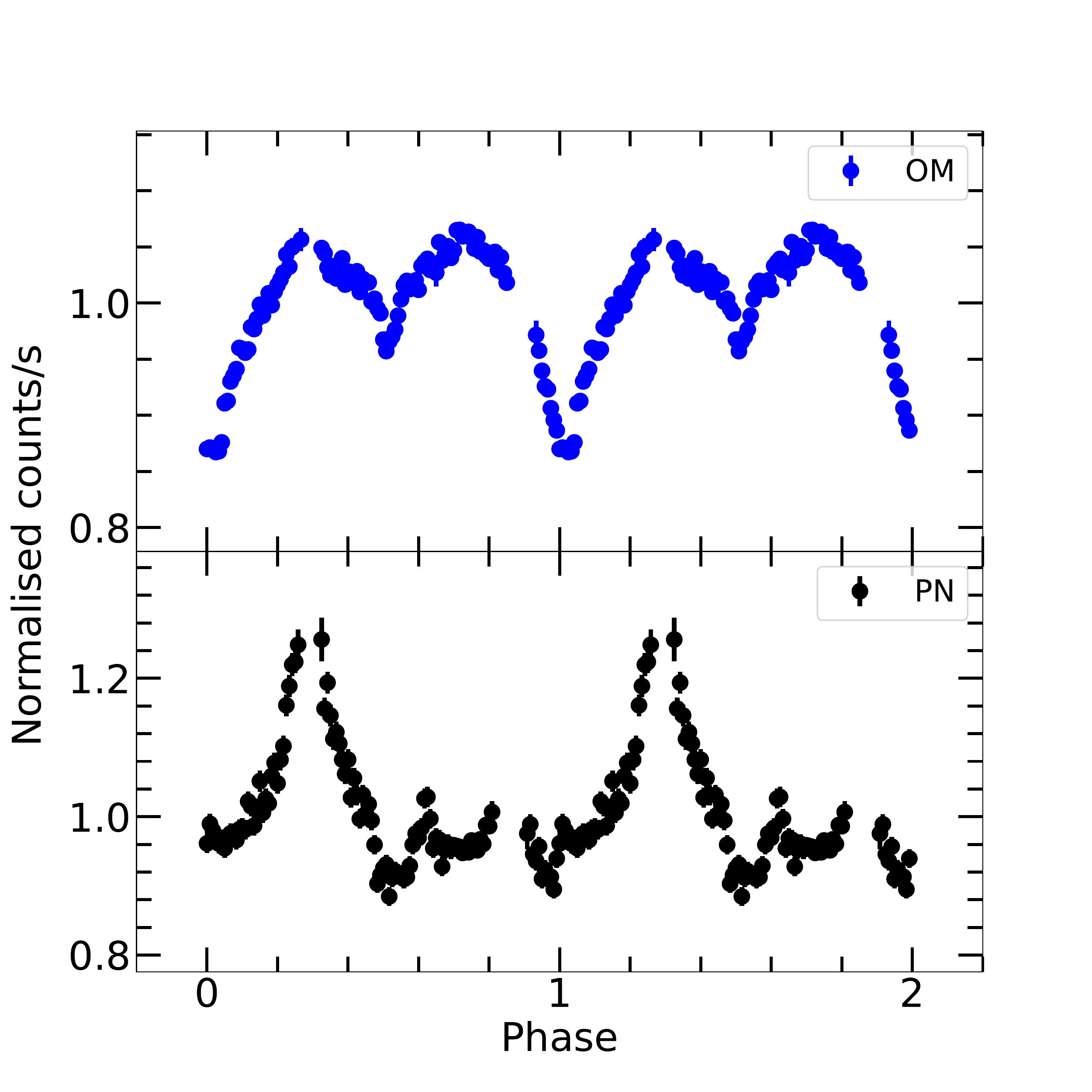}}
\subfigure[ER Vul (obs. ID: 0785140601)]{\includegraphics[height=6.5cm,width=0.45\linewidth]{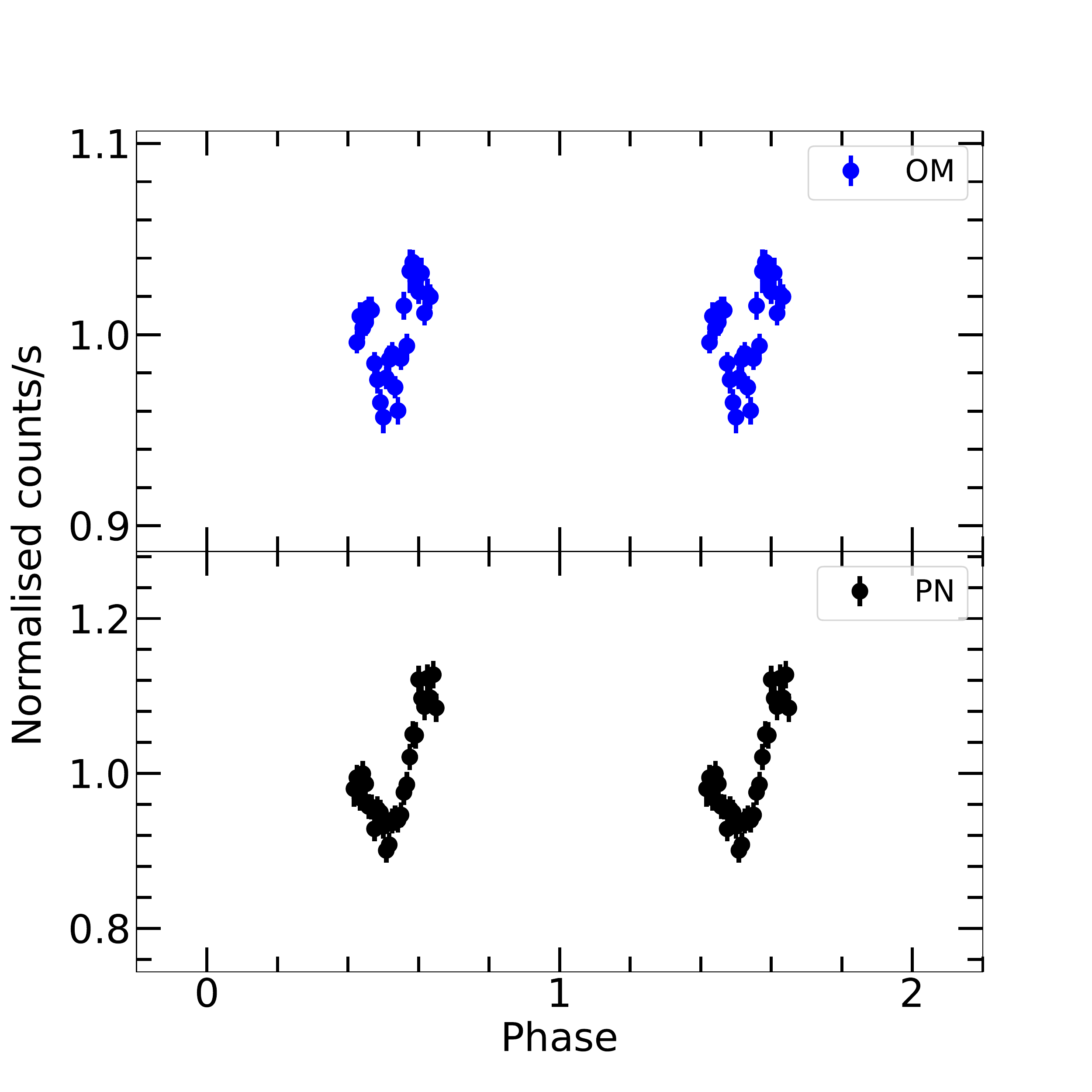}}
\subfigure[XY UMa]{\includegraphics[height=6.5cm,width=0.45\linewidth]{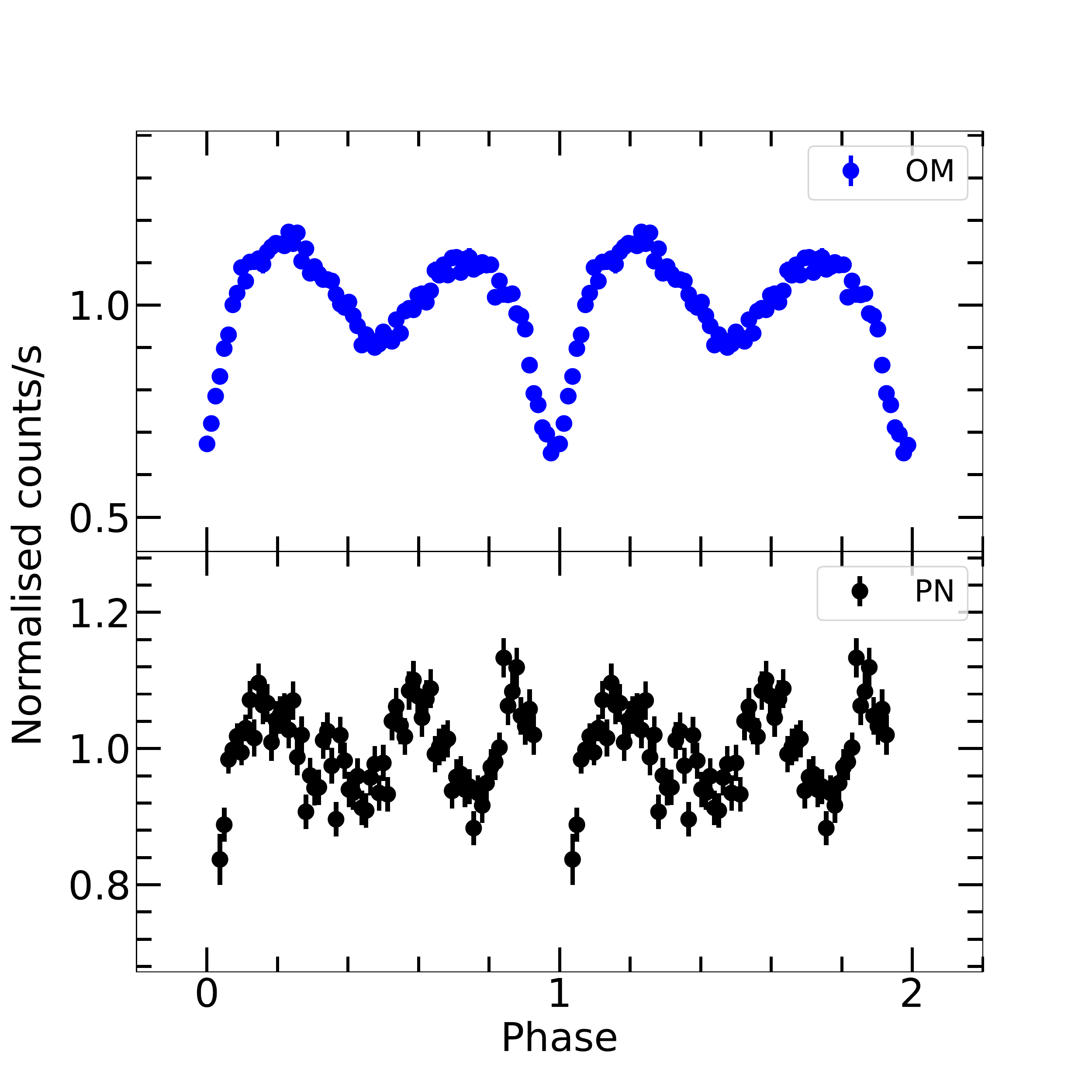}}
\subfigure[TX Cnc]{\includegraphics[height=6.5cm,width=0.45\linewidth]{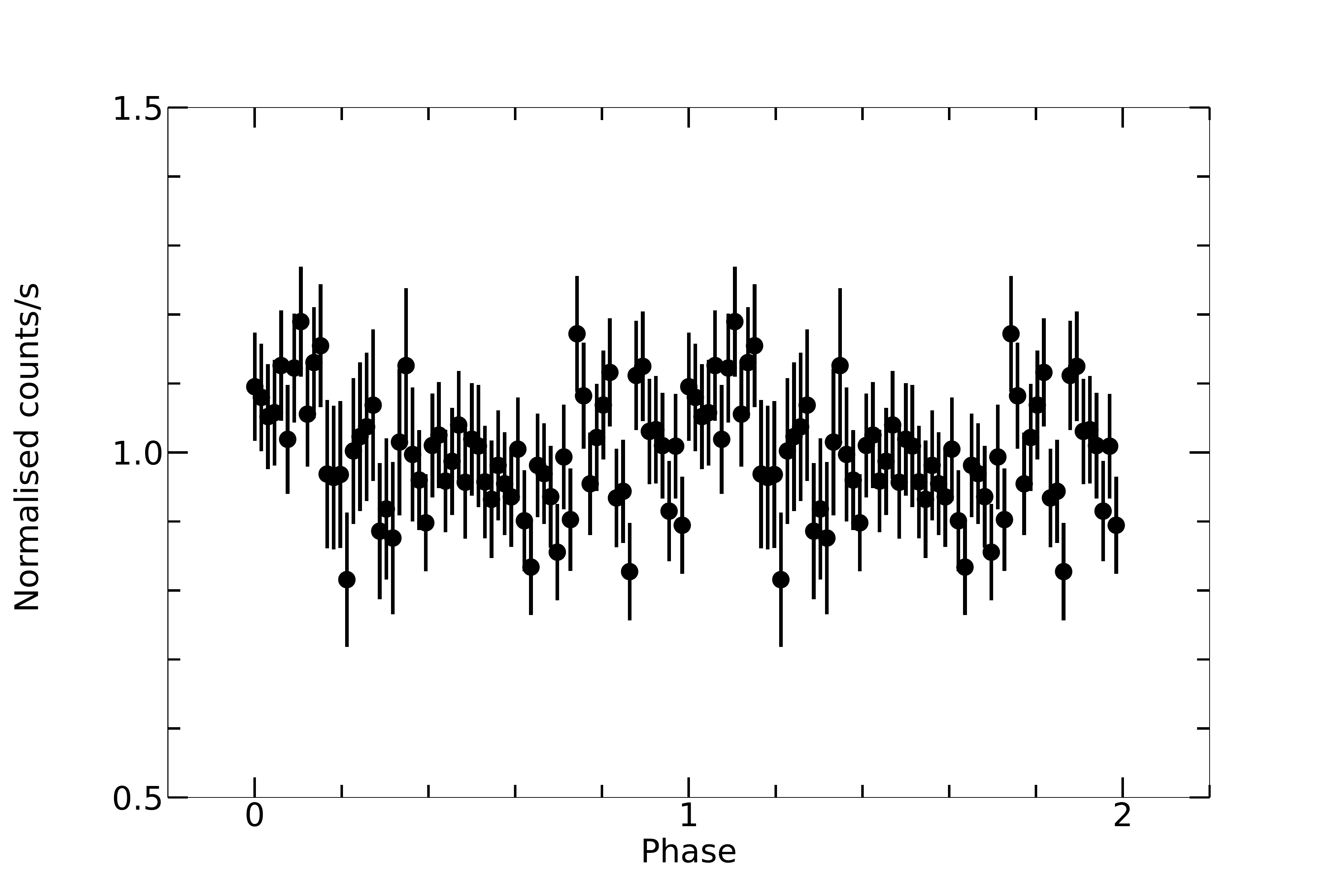}}
    \caption{ OM (blue) and EPIC [PN (black) and MOS (red)] folded quiescent state light curves of sample stars.  The ephemerides used are given in Table \ref{tab:basic_parameters_of_sample}}.
    \label{fig:3}
\end{figure*}

We used the PN light curve for further analysis for all stars, excluding 44 Boo, as MOS observations have relatively poor count statistics compared with the PN observations.  We excluded the flaring events from the X-ray and UV data of each system to make quiescent state light curves.  The quiescent state light curves were phase folded according to the ephemeris given in Table \ref{tab:basic_parameters_of_sample} and shown in Figure \ref{fig:3}. The X-ray light curves of the target stars are found to show phase-locked variation  even if the stars 44 Boo, ER Vul, and XY UMa were only  observed for approximately one orbital cycle.

The phase-folded X-ray light curve of 44 Boo shows a highly variable nature during the entire binary phase. A tendency of decreasing X-ray flux before the primary eclipse is seen in the X-ray light curve, even if the observations during the primary eclipse (0.92-0.10) were lacking. A dip near the secondary eclipse was observed as well. Additionally, brightening is seen near phase 0.35 indicating the existence of some hot plasma.  The presence of both eclipses in the light curves shows that both the components of binary are active. 

The phase-folded light curves of the stars DV Psc and ER Vul show a similar trend, with a brighter X-ray around the first quarter phase than that from the third quarter phase. This kind of variation shows that one side of binary components is being crowded by more active regions than the other side. Based on the presence of both primary and secondary eclipses, one can infer that both primary and secondary components are active in both binary systems.

XY UMa, on the other hand, shows an odd phase-folded light curve with the presence of three dips during the entire orbital period. Two dips are at the phases of primary and secondary eclipses, indicating both components are active. The third dip is seen near phase 0.75. TX Cnc was found to be brighter during the primary eclipse.

Simultaneous UV and X-ray  observations were available for three EBs, ER Vul, DV Psc, and XY UMa.  UV light curves of DV Psc and XY UMa show variation outside the eclipse, mimicking the  O’Connell effect. We then seek correlations between UV and X-ray phase-folded light curves of these binaries. The results of the correlation are summarized in Table \ref{tab:correlations_xray_uv}. The X-ray emission is found to be positively correlated with the UV. In the case of DV Psc and one observation of ER Vul, this correlation was found to be stronger with correlation coefficients of more than 0.7, whereas in the case of XY UMa a marginal correlation was found.

\begin{deluxetable}{ccccc}
    
    \tablecaption{Correlation coefficient between X-ray and UV  fluxes of the stars in the sample along with a probability of No correlation.
    \label{tab:correlations_xray_uv}}
    \tablehead{
         \colhead{Star Name} & \colhead{Spearman's $\rho$}&\colhead{ p-value$^a$ }&\colhead{ Kendall $\tau$} &\colhead{ p-value$^a$}\\}
    \startdata
         DV Psc & 0.760 & $2\times10^{-8}$ & 0.538 & $1\times10^{-6}$\\
         ER Vul (1)& 0.283 & 0.00288& 0.199 & 0.00219\\
         ER Vul (2)& 0.82 & $3\times10^{-7}$& 0.59 & $7\times10^{-6}$\\
         XY UMa & 0.233& 0.0457& 0.165 & 0.0378\\
      \enddata
      \tablecomments{$^a$The p-value is the probability of no correlation. (1) and (2) for ER Vul corrospond to the observation IDs: 0781500101 and 0785140601, respectively.} 
\end{deluxetable}

\subsection{Photospheric contribution in the UV band}
The flux in the UV band from the late-type stars could be a combination of flux values arising from both the photosphere and chromosphere. Therefore,  we calculated the proportion of photospheric emission in the overall measured UV flux. For calculating the photospheric contribution,  we used the synthetic spectra from the  \textsc{BT$-$Settl (CIFIST)} spectral models of \cite{2015A&A...577A..42B}. These models are accounted for the solar photospheric abundances only \citep[]{2011SoPh..268..255C}. 
 To extract template spectra, the temperatures of primary and secondary components of DV PSc, ER Vul, and XY UMa were taken as  4400 and 3700 K, 5900 and 5800 K, and 5300 and 3900 K, respectively, whereas the log-g was taken as 4.5 for all stars. These values of temperature and log-g are very close to their well-estimated values listed in Table  \ref{tab:basic_parameters_of_sample}. 
The theoretical model spectra for the primary and secondary components were multiplied by their respective stellar surfaces and further folded over the OM effective area. At the quarter phases, when both  stars of the system are visible, the theoretical flux prediction is obtained using the following relation:

\begin{equation}
    F=\frac{\int_{0}^{\infty}(S_p(\lambda)A_p+S_s(\lambda)A_s)\times\Phi_{eff}d\lambda}{4\pi d^2\int_{0}^{\infty}\Phi_{eff}d\lambda}\sigma_{eff}
\end{equation}

\noindent
where $A_p=4\pi R_p^2$ and $A_s=4\pi R_s^2$ are the surface area of the primary, and secondary stars, respectively, $S_p(\lambda)$ and $S_s(\lambda)$ are the model spectra of the primary and secondary, respectively, $\phi_{eff}$
is response function of OM filter, d is distance of star, and $\sigma_{eff}$ is the effective width of the OM filter. The observed OM count rates were converted to flux using conversion factors given by the SAS team\footnote{\url{https://www.cosmos.esa.int/web/xmm-newton/sas-watchout-uvflux}}. 

This approach for estimating the photospheric flux does not account for Roche geometry and surface heating effects. Therefore, the estimated photospheric contribution to the observed UV flux is an approximate lower value as accounting for the Roche geometry and surface heating effects will increase the predicted flux. The results of the photospheric contribution are summarized in Table   \ref{tab:photospheric_contribution}. Considering the parameters of the stars as mentioned above, the observed photospheric contribution was found to be 78 \% for DV Psc, 53\%  for  ER Vul, and 68\% for  XY UMa.  However, the effective temperature of double-lined binaries is not precisely measured as for a single star and a small change in the temperature results in a large change in the predicted flux.  Therefore, the lower and upper limits of the predicted flux were estimated by changing the effective temperature of each component of the system by $\pm$ 100 K. We also included the errors in radii of both components and the distance of the system while estimating the upper and lower limits of the predicted flux. Considering this we estimated the minimum PC as  51\%, 31\%, and 43\% for DV Psc, ER Vul, and XY UMa, respectively.  Due to a strong dependence of UV flux on temperature, we could not constrain the upper limit of the predicted flux for DV Psc and XY UMa. 
\begin{table}
    \caption{Photospheric contribution to total observed UV flux. 
    }
    \label{tab:photospheric_contribution}
    \begin{tabular}{cccc}
    \hline
    \hline
         Star&Observed flux&Predicted flux&PC (\%)\\
         \hline
         DV Psc&$1.04_{-0.1}^{+0.1}$&$0.81_{-0.27}^{+0.42}$& $ >$51 \\
         ER Vul$^a$&$17.98_{-1.8}^{+1.8}$&$9.57_{-3.85}^{+5.63}$&53 (31 - 85)\\
         XY UMa&$0.68_{-0.07}^{+0.07}$&$0.46_{-0.16}^{+0.24}$& $>$ 43 \\
         \hline
    \end{tabular}
~\\

\textsc{Note--} Observed flux is calculated by averaging the flux from phase 0.25 and 0.75, and predicted UV flux is estimated from equation (1). Photospheric contribution is calculated as PC=$\frac{predicted flux}{observed flux}\times100\%$.\\
{\scriptsize  
$^a$For the observation ID: 0781500101. Here, both observed and predicted flux are in the units of $10^{-11} erg ~s^{-1}cm^{-2}$. The error in the observed flux is calculated by adding a maximum relative error of 10\% to the error obtained from the count rate. Whereas, the error in predicted flux is estimated by propagating the  error in radius, distance, and temperature (100 K error). The values in the parenthesis in the last column are the range of minimum to maximum values of PC. }
\end{table}

\subsection{X-ray light-curve modeling: Coronal imaging}
The coronal structure of each component in an eclipsing binary system can be constrained if the light curve spans nearly the entire orbital period of the system. We used the 3D deconvolution method of \cite{1992MNRAS.259..453S} and \cite{1996ApJ...473..470S}  to reconstruct the 3D coronal structures of each binary system, 
since the inversion of a 1D light curve to 3D emitting structures is a mathematically ill-posed problem and leads to multiple solutions. However, as shown by \citet{1992MNRAS.259..453S}, with the given physically reasonable priors, solutions converge to physically sensible structures.
The stellar surrounding is divided into uniform cubical volume bins occupied by variable emission density [$f_{em}(x,y,z)$]. If we consider optically thin plasma, then the total flux ($F(\phi)$) observed at any phase $\phi$ is the sum of the contribution from all those bins that are not occulted by either star, i.e.\\
$$F(\phi)=\sum_{x,y,z}f_{em}(x,y,z)\textbf{M}(\phi,x,y,z)dxdydz$$
where $\textbf{M}(\phi,x,y,z)$ is the occultation matrix with values ranging from 0 (totally occulted) to 1 (fully visible). 
The occultation matrix is calculated with the assumption that the system rotates rigidly and the distance to the system is much larger than the binary separation, so shadows can be considered cylindrical shapes rather than conical. These assumptions allow our model to take care of the inclination angle when computing the occultation matrix, so, it is an extension of the \cite{1992MNRAS.259..453S} model. 
We start with uniform emission density (i.e. $f_{em}^1(x,y,z)=1 \quad\forall\quad x,y,z \in X,Y,Z$ where $X,Y,Z$ means the physical domain of computation). Based on this $f_{em}(x,y,z)$ and occultation matrix $M(\phi,x,y,z)$, the light curve $F_c$ is computed.
Then for each phase, $\phi_i$, a correction factor is determined based on observed flux ($F_o$) and distributed to all visible volume bins. This correction factor updates the existing emission density estimations as \\
$$f_{em}^{n+1}(x,y,z)=f_{em}^n(x,y,z)\frac{\sum_i\frac{F_o(\phi_i)}{F_c(\phi_i)}\textbf{M}(\phi_i,x,y,z)}{\sum_i\textbf{M}(\phi_i,x,y,z)}$$ 
This updated emission density is then used to recompute the light curve. The discrepancy between modeled and observed light curves was checked using $\chi^2$ statistics. The iteration process was stopped when the reduced $\chi^2$ approaches 1.
We used a cell size of $0.1\times0.1\times0.1 R^3_\odot$ with the physical constraint that the corona of each binary is extended up to  $1 R_*$ above the photosphere. Based on the solution obtained, we plotted $f_{em}$ distribution with the axis (i.e. $f_x,f_y,f_z$). 
Here, $f_x=\Sigma_{y,z}f_{em}(x,y,z)$, provides information about how emitting regions are varying with the variation of longitude, and $f_y=\Sigma_{x,z}f_{em}(x,y,z)$ shows the variation of X-ray emitting plasma along the line connecting the binary system. 
This distribution shows the intra-binary plasma component  and provides the relative brightness of the primary with respect to the secondary. The $f_z=\Sigma_{x,y}f_{em}(x,y,z)$ distribution shows whether emitting regions are uniformly distributed in both the upper and lower hemispheres of the binary or not. 

Results obtained from the coronal imaging are shown in Figures 4 - 8.
In each figure, panel (a) is the top view: a view of the upper hemisphere as seen by an observer at the top of the orbital plane, panel (b) is the bottom view:   a view of the lower hemispheres as seen by an observer at the bottom of the orbital plane, panel (c) is the front view: 0$^\circ$-180$^\circ$ longitudinal view of the orbital plane from the front side, panel (d) shows back view: 180$^\circ$-360$^\circ$ longitudinal view of the orbital plane from the backside, panel (e) represents the distributions of  $f_{em}$, where R = 0 corresponds to a 0 longitude for the $f_x$ distribution, the center of mass of the binary system for the $f_y$ distribution, and the equator for the $f_z$ distribution, and panel (f) shows the best-fit modeled light curve along with the data. 
The results obtained from eclipse modeling for different systems are described below. 

\subsubsection{44 Boo}
\label{iboo_eclipse_modelling_results}
\begin{figure*}
\centering
\subfigure[]{\includegraphics[scale=0.3]{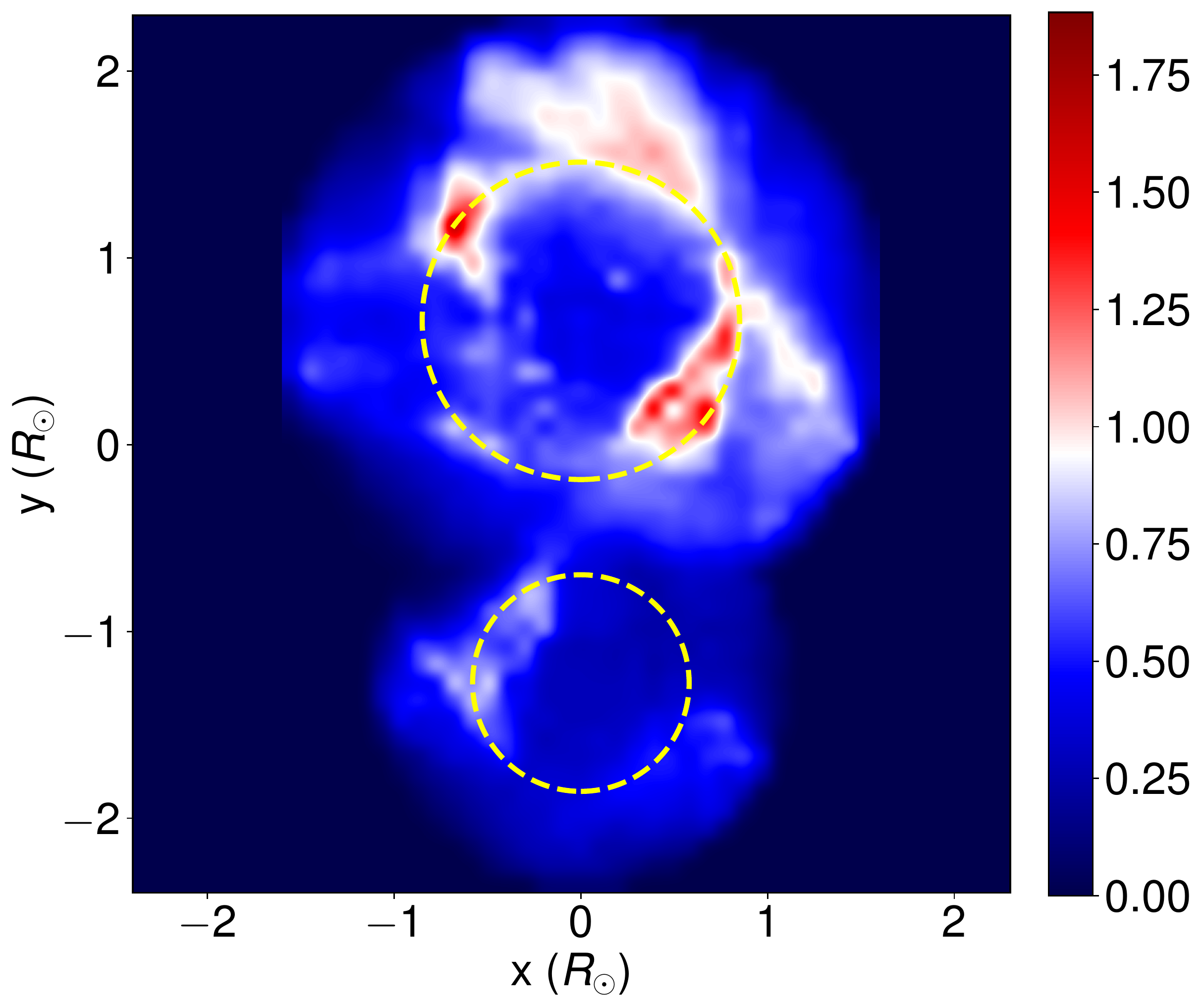}}
\subfigure[]{\includegraphics[scale=0.3]{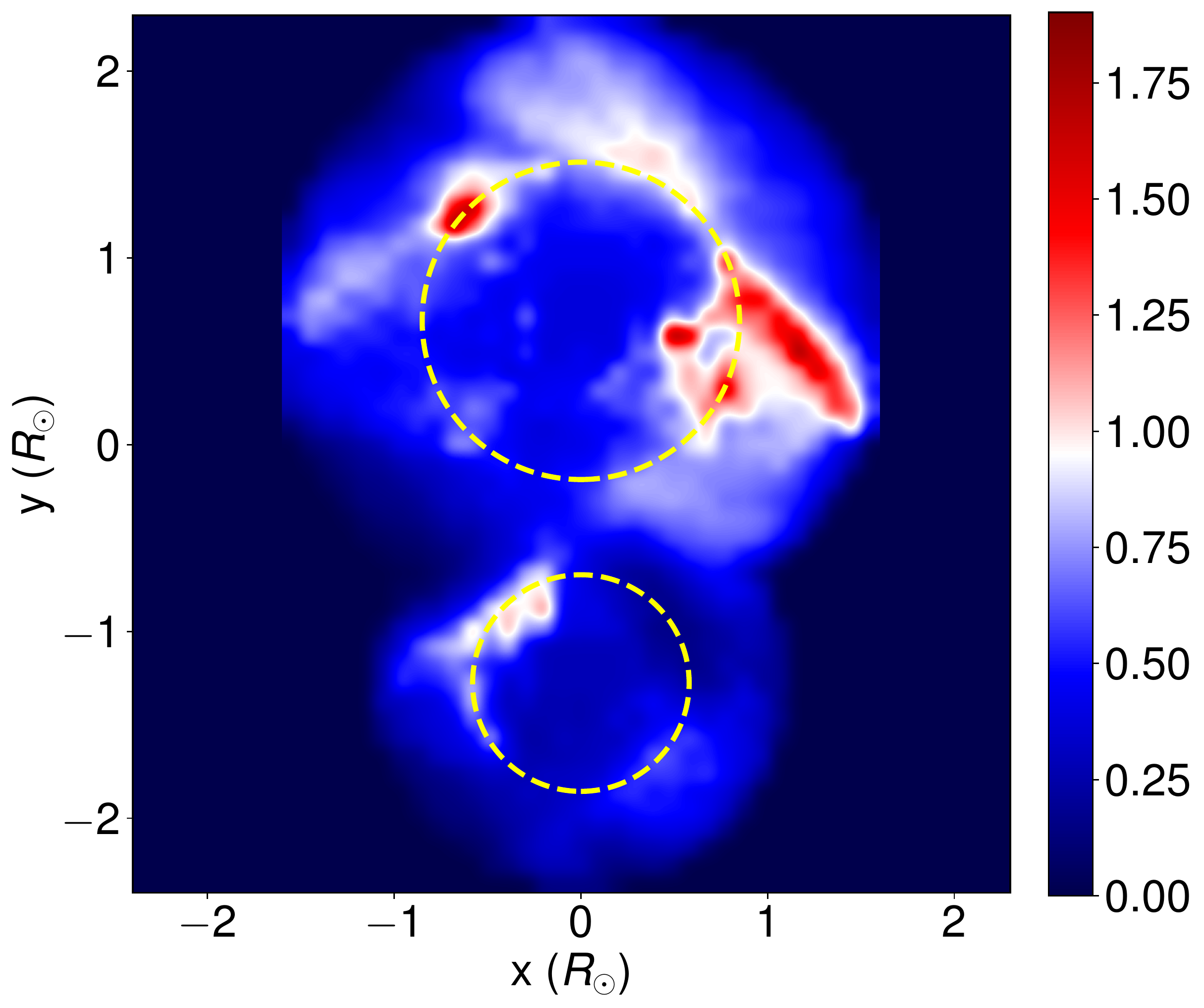}}
\subfigure[]{\includegraphics[scale=0.3]{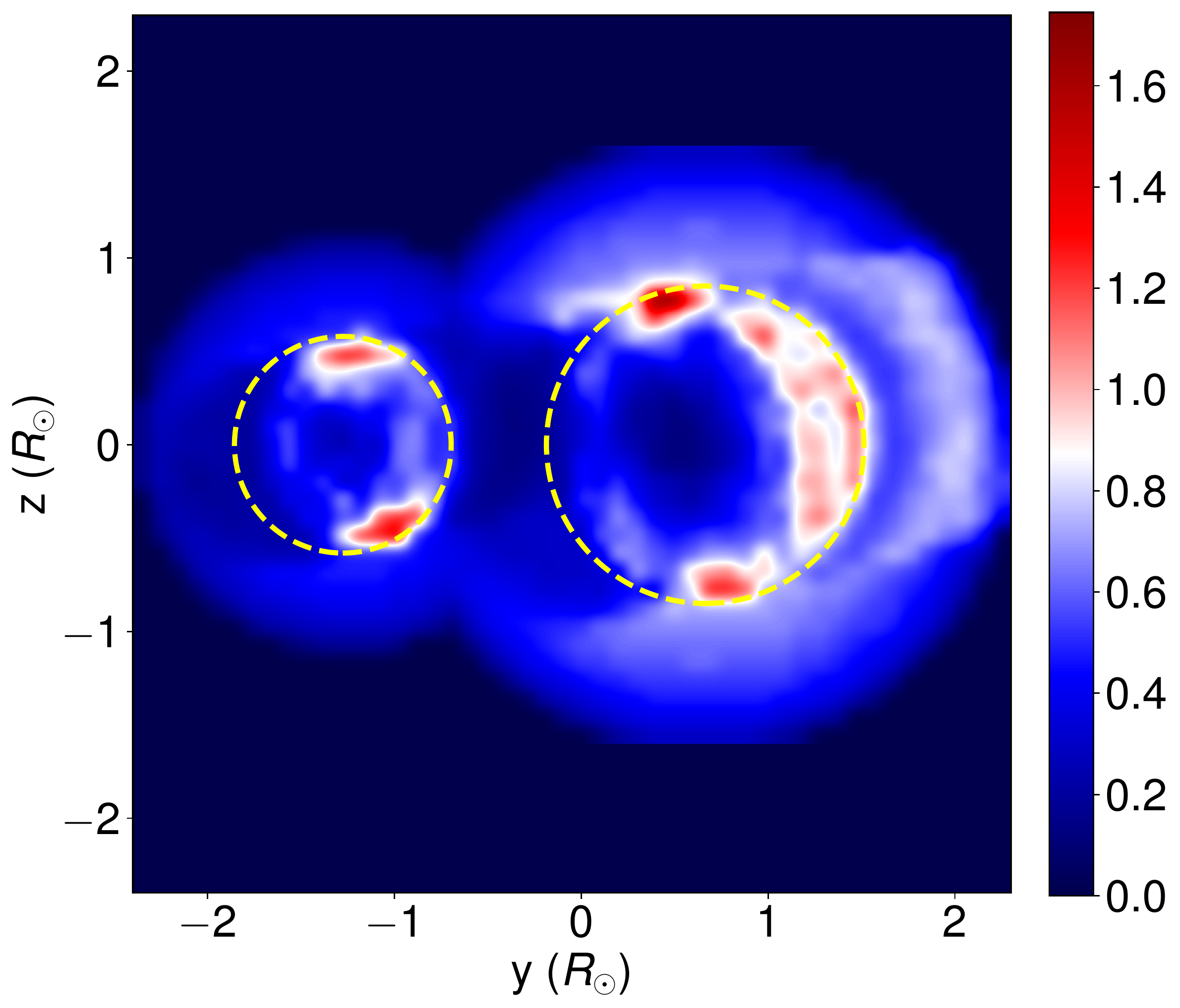}}
\subfigure[]{\includegraphics[scale=0.3]{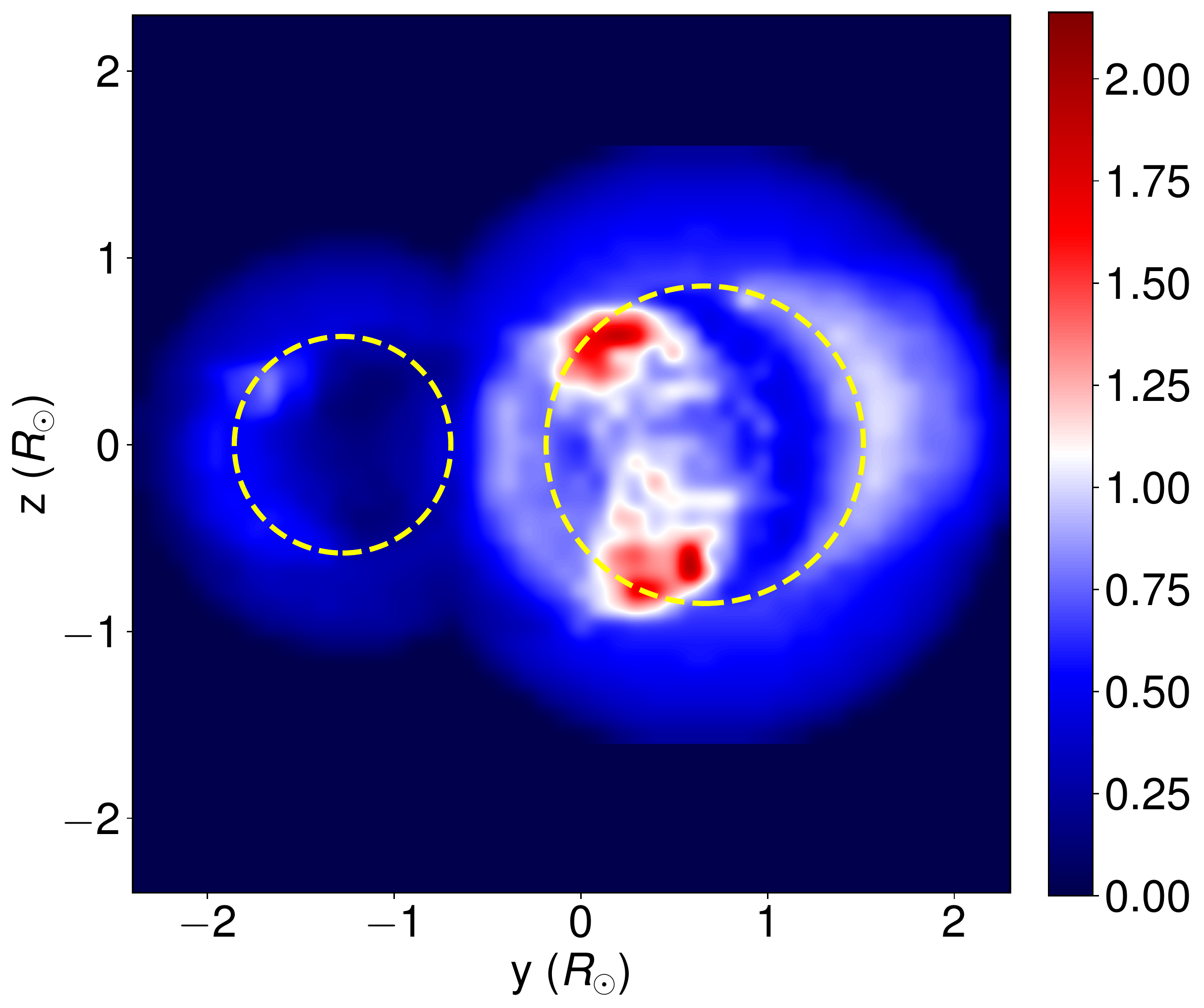}}
\subfigure[]{\includegraphics[width=0.8\columnwidth]{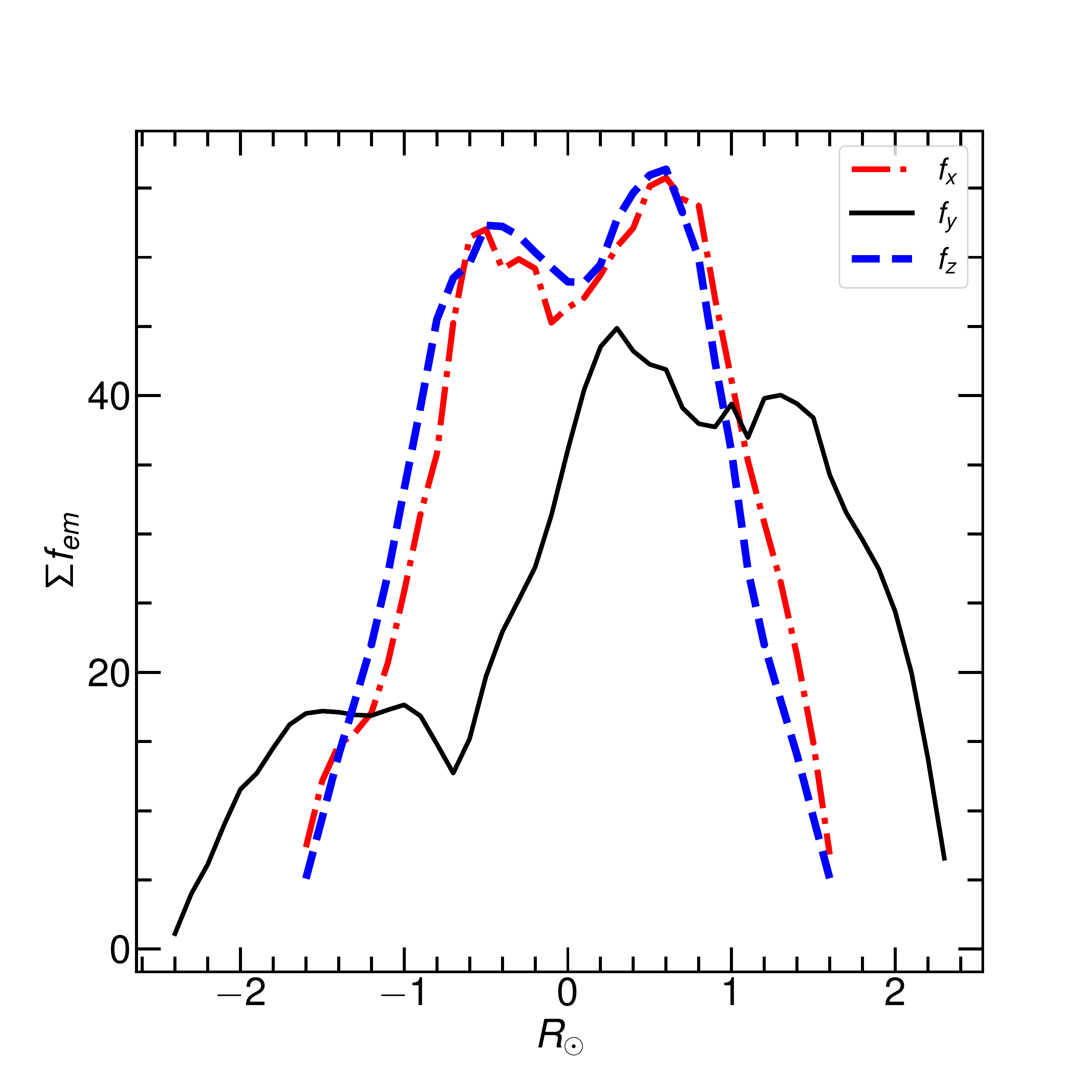}}
\subfigure[]{\includegraphics[width=0.8\columnwidth]{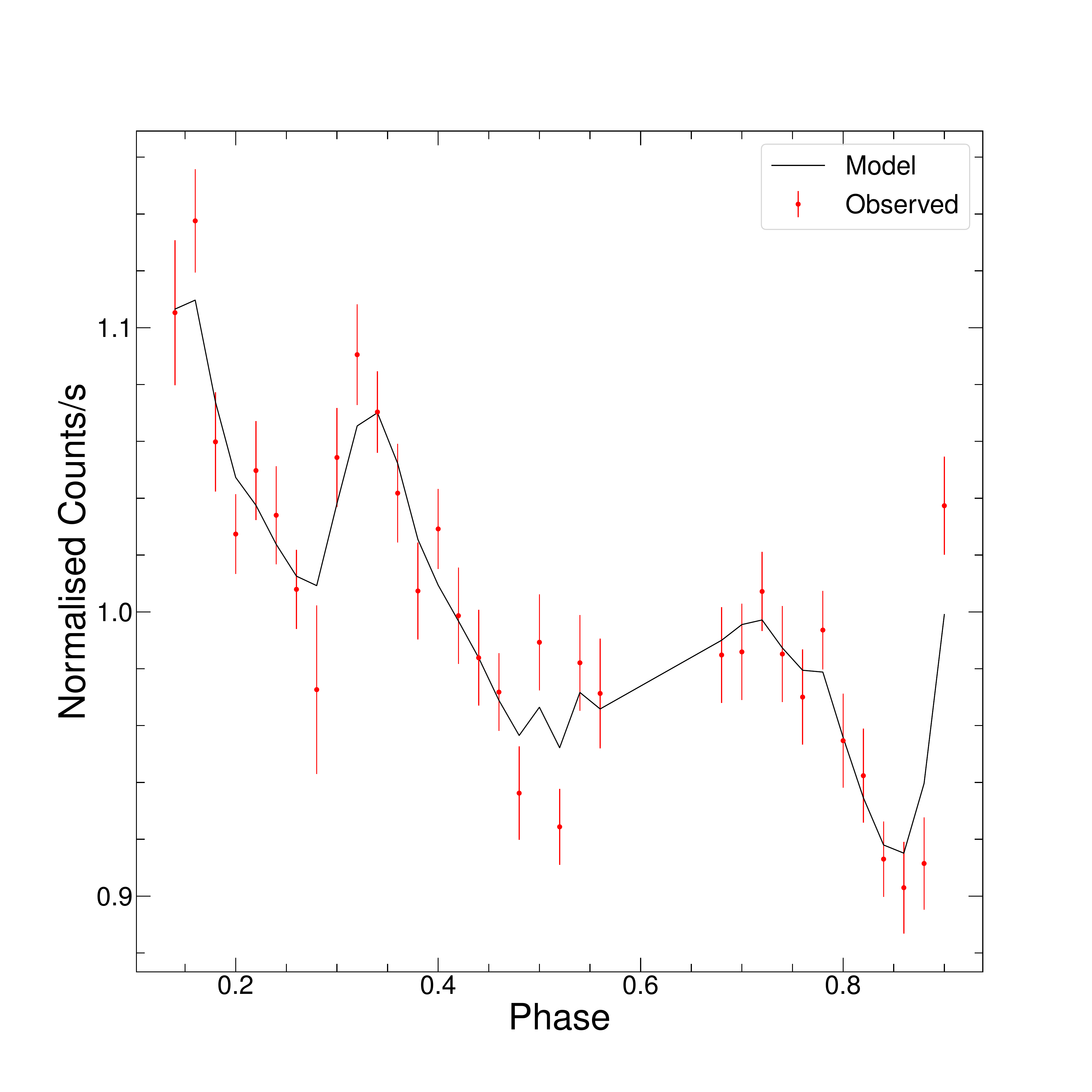}}
    \caption{Results of the X-ray light-curve modeling for the system 44 Boo. (a) Top view: a view of the upper hemisphere as seen by an observer at the top of the orbital plane, (b) bottom view:   a view of the lower hemispheres as seen by an observer at the bottom of the orbital plane, (c) front view: 0$^\circ$-180$^\circ$ longitudinal view of the orbital plane from the front side (d) back view: 180$^\circ$-360$^\circ$ longitudinal view of the orbital plane from the backside,  (e) the distributions of  $f_{em}$, where R = 0 corresponds to a 0 longitude for $f_x$ distribution, the center of mass of the binary system for the $f_y$ distribution, and equator for $f_z$ distribution, (f) the best-fit modeled light curve along with the observed light curve. The yellow dashed circles show the extent of the photosphere of  primary and secondary. }
    \label{fig:iboo_emr}
\end{figure*}

Figures \ref{fig:iboo_emr}(a)-(d), show that both components of 44 Boo  have highly inhomogeneous coronae.   The identical structure of the top and bottom views (see Figure \ref{fig:iboo_emr}(a), (b)) indicate that the upper and lower hemispheres contain an almost similar distribution of active regions in both components of 44 Boo.  The presence of two peaks in the  distribution of $f_z$ also suggests the presence of  two active latitudes (see Figure \ref{fig:iboo_emr} e). The majority of the X-ray active regions are found to be concentrated toward the poles and very less active regions  are found near the equator. 
A similar trend is seen in the $f_x$ distribution as well, two active X-ray emitting regions exist along the longitude. It is evident from back and front views as shown in Figures \ref{fig:iboo_emr} (c) and (d) that the secondary component has more active regions on the one side than on the other while the primary has active regions on both longitudes.  The $f_y$ distribution as shown in Figure \ref{fig:iboo_emr} (e) indicates that the corona of the primary is approximately twice as bright as the secondary and the coronae of both stars overlap each other.  In terms of the brightness per unit area, both components are found to be almost at the same activity level.
Figure  \ref{fig:iboo_emr} (f) shows the best-fit modeled light curve over the real data. The brightening near phase 0.35 can now be attributed to the high active region on the secondary.

\subsubsection{DV Psc}
\begin{figure*}
\centering
\subfigure[]{\includegraphics[scale=0.3]{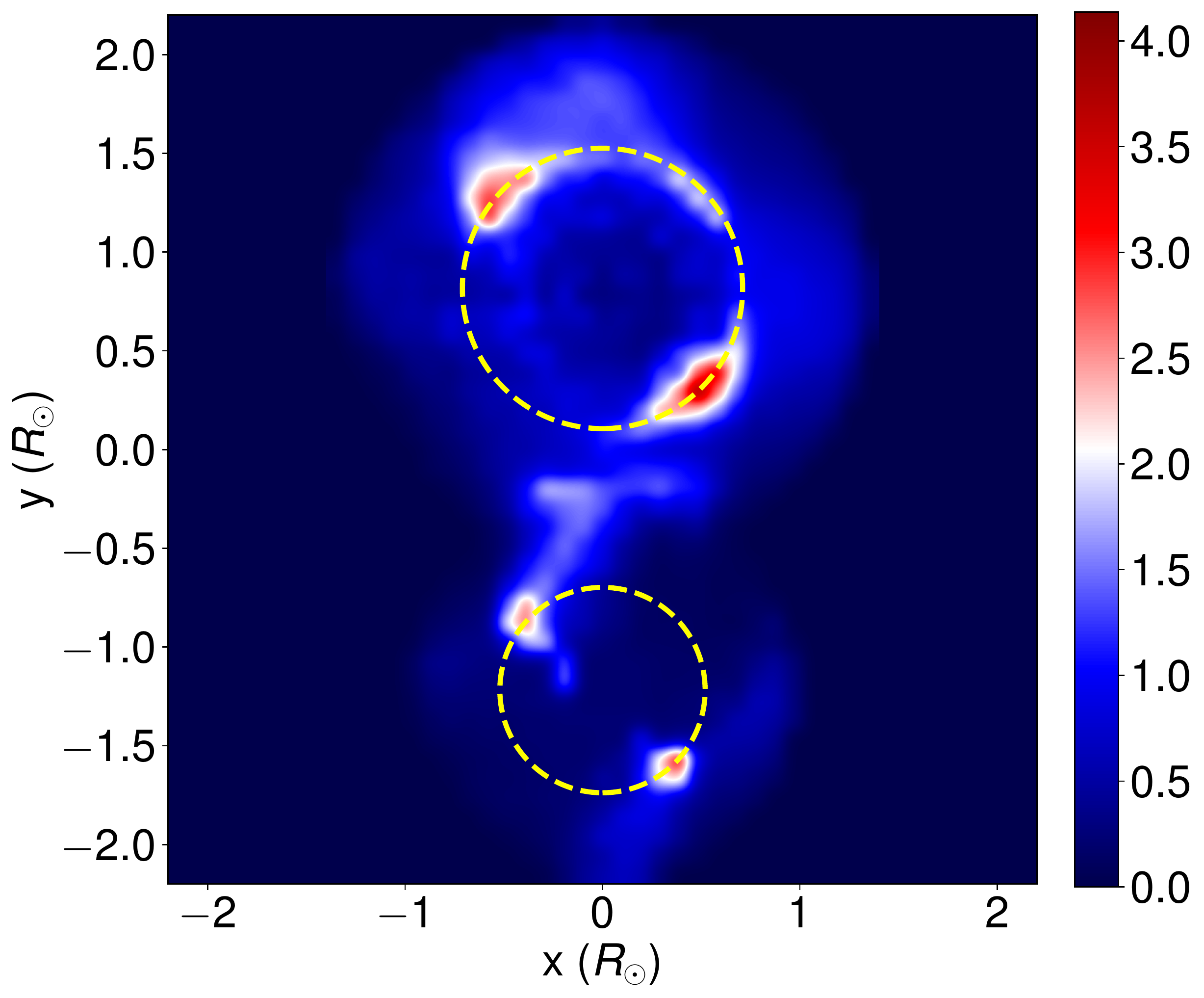}}
\subfigure[]{\includegraphics[scale=0.3]{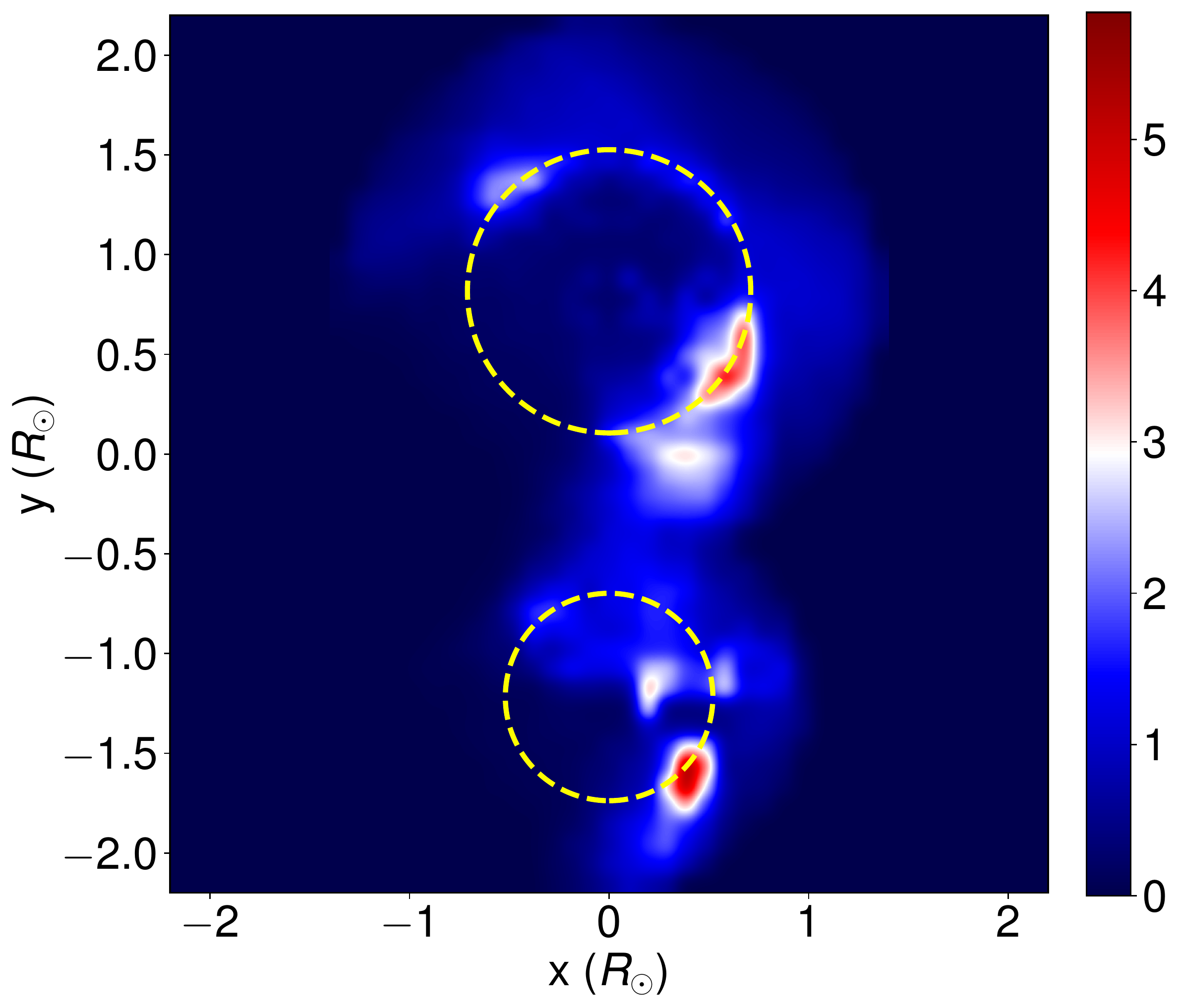}}
\subfigure[]{\includegraphics[scale=0.3]{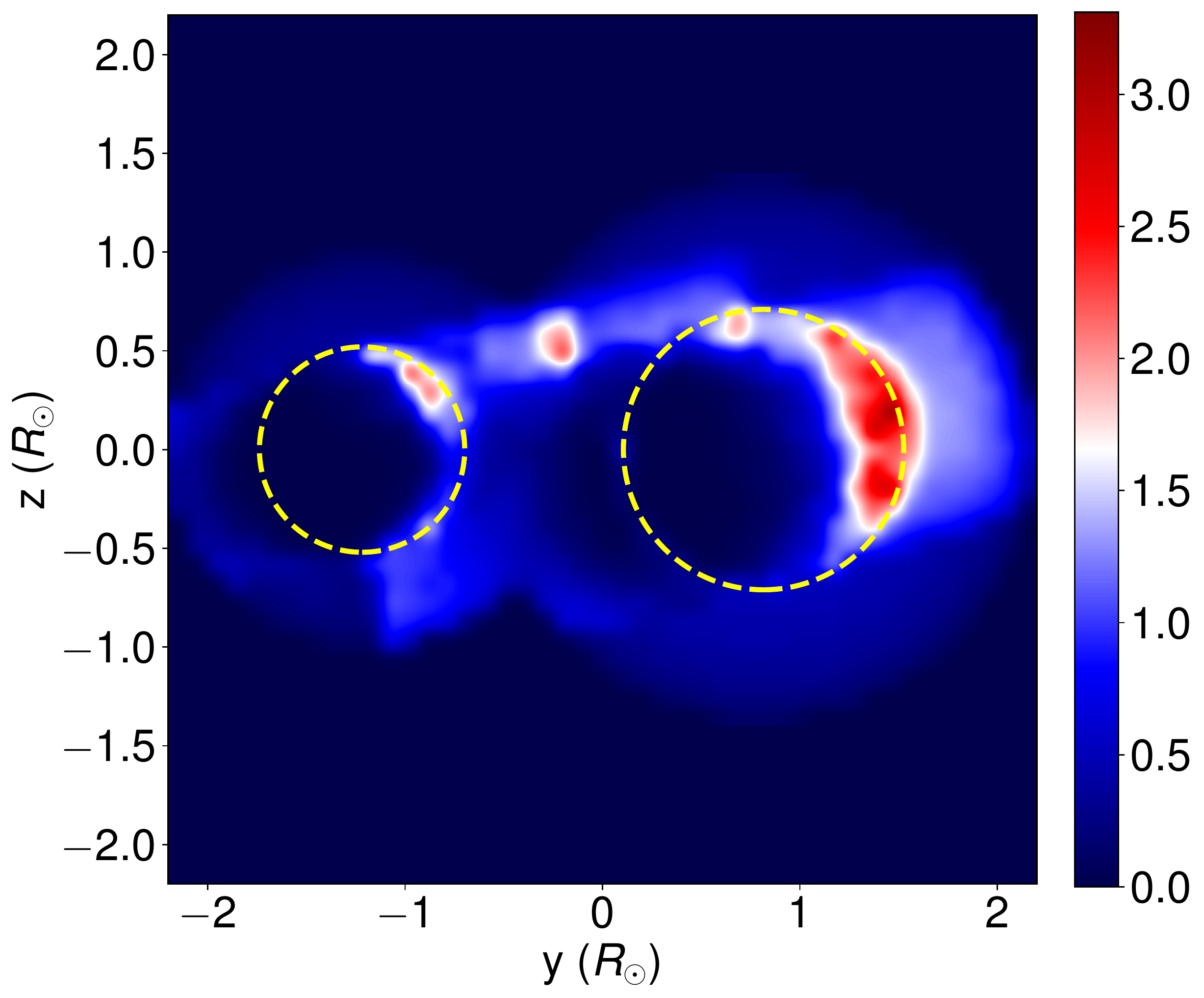}}
\subfigure[]{\includegraphics[scale=0.3]{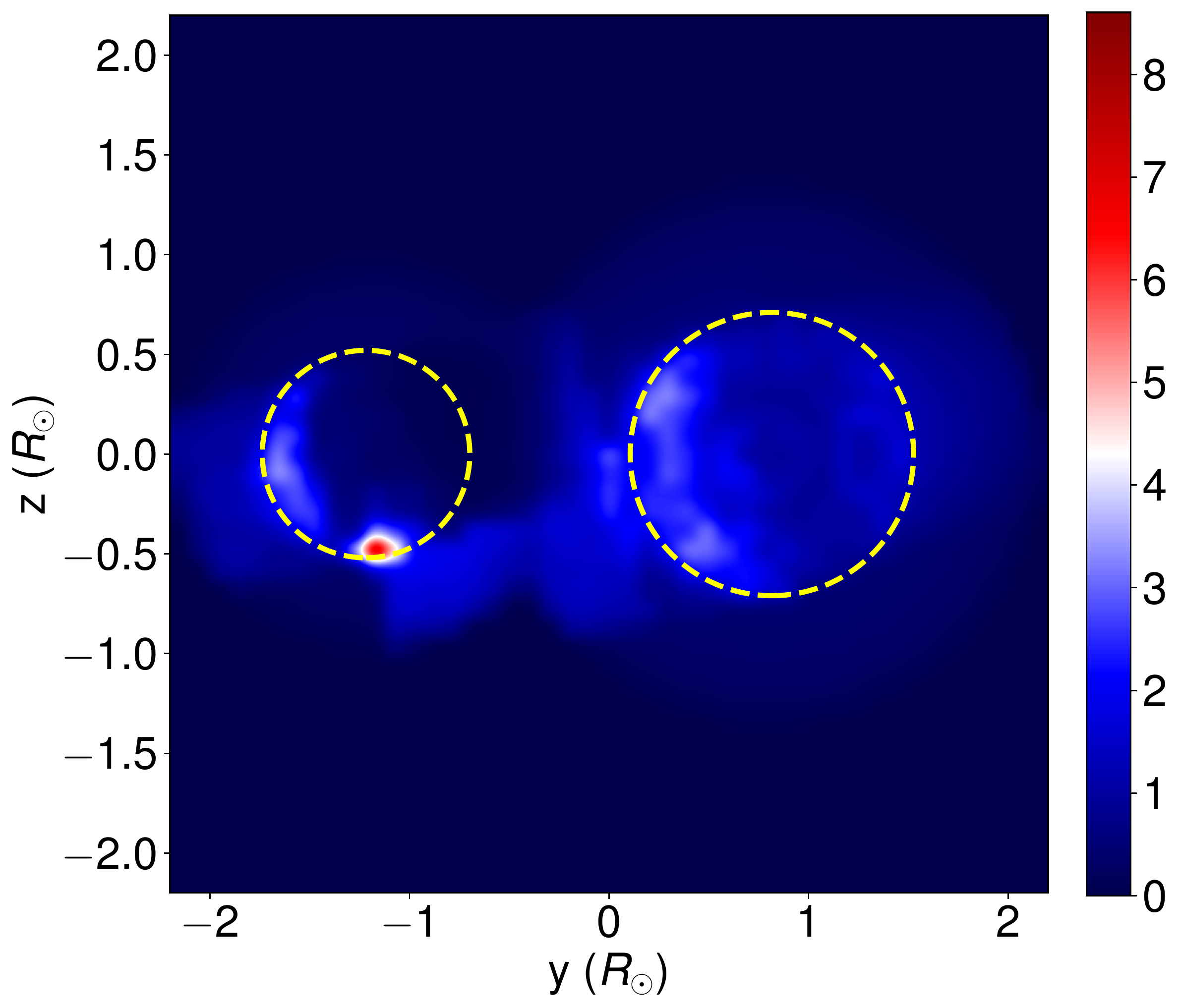}}
\subfigure[]{\includegraphics[width=0.8\columnwidth]{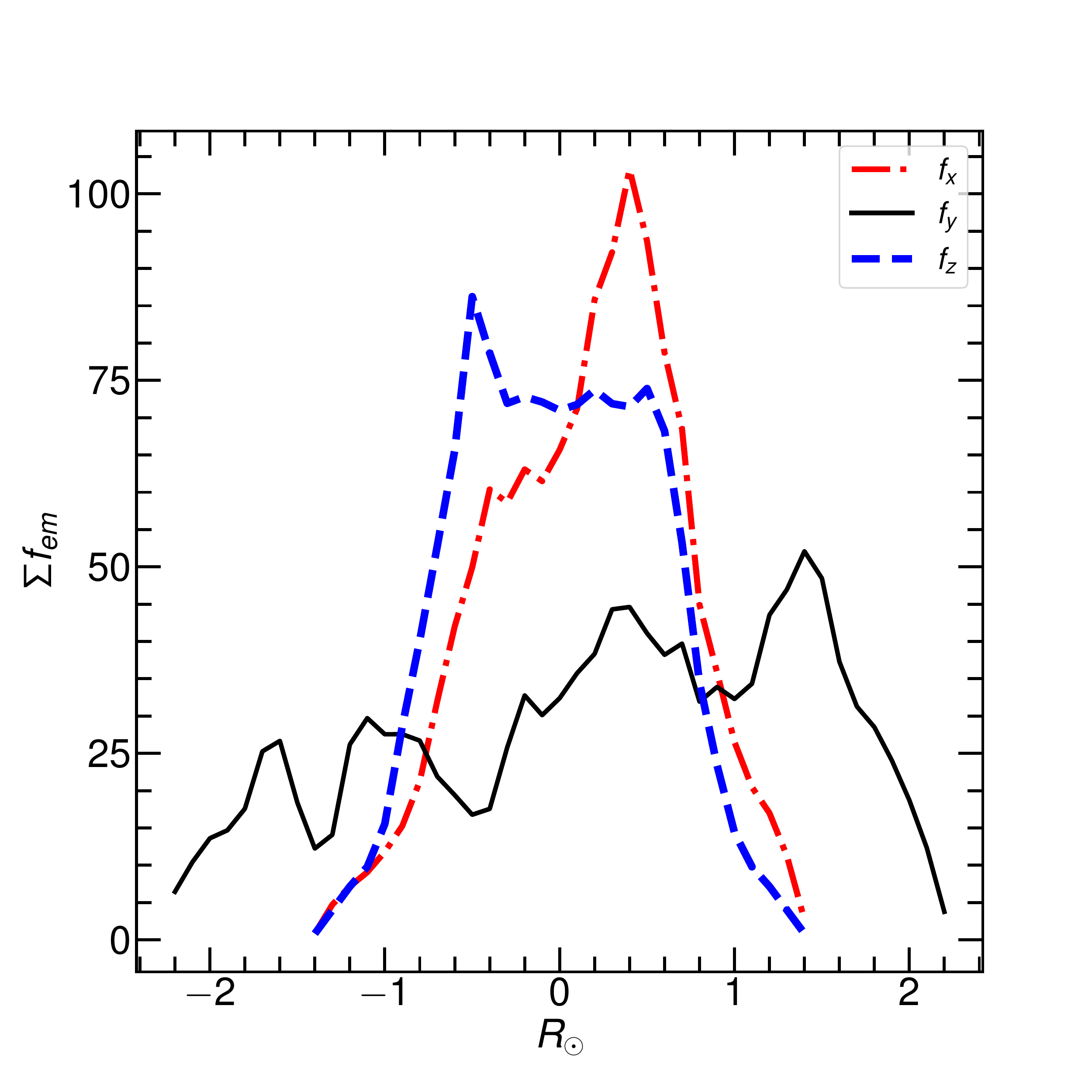}}
\subfigure[]{\includegraphics[width=0.8\columnwidth]{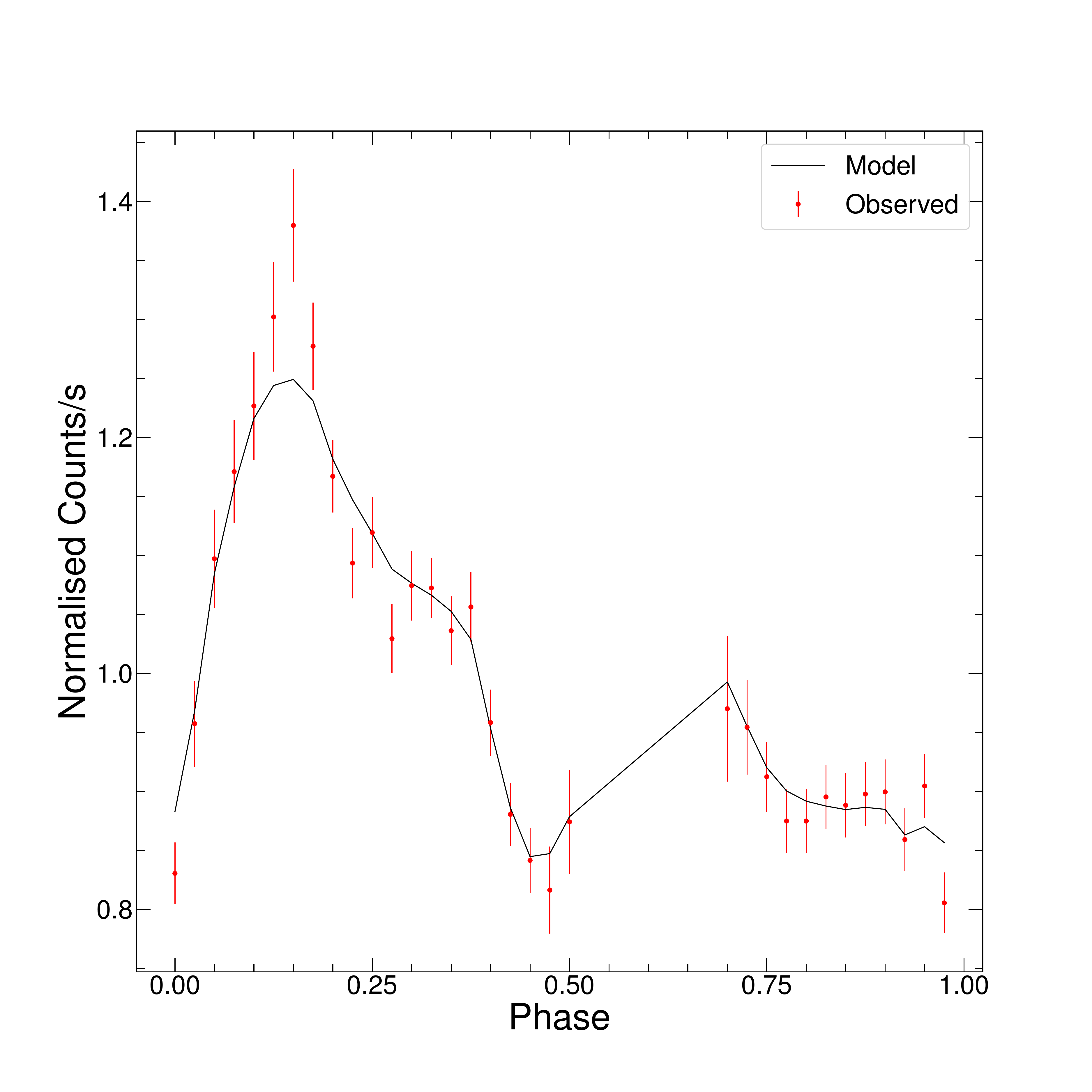}}
    \caption{Similar to Figure \ref{fig:iboo_emr} but  for DV Psc. }
    \label{fig:dvpsc_emr}
\end{figure*}

Figure \ref{fig:dvpsc_emr} shows the results of the X-ray light-curve modeling for DV Psc. Based on Figures \ref{fig:dvpsc_emr}(a) and (b), it is evident that the primary has more active regions in the upper hemisphere while secondary has them in the lower hemisphere. From Figure \ref{fig:dvpsc_emr} (c), a clear pole-to-pole connection of the active region for this system can be seen. The $f_z$ distribution in Figure \ref{fig:dvpsc_emr} (e) shows that the lower hemisphere is relatively more active than the upper one, which is probably due to the larger contribution of the  X-ray active region on the lower hemisphere of the secondary star. The $f_x$ distribution shows a skewed nature, which means that there exists only one active longitude; this can be also inferred from Figures \ref{fig:dvpsc_emr} (c) and (d). This further shows active regions are distributed uniformly from the pole to the equator of the primary star. The $f_y$ distribution shows a very dynamic trend, which also indicates that the relative brightness of the primary for this system is almost 1.7 times that of the secondary. We also noticed that both the components are equally active when X-ray flux is taken into account per unit surface area.

\subsubsection{ER Vul}
\begin{figure*}
\centering
\subfigure[]{\includegraphics[scale=0.3]{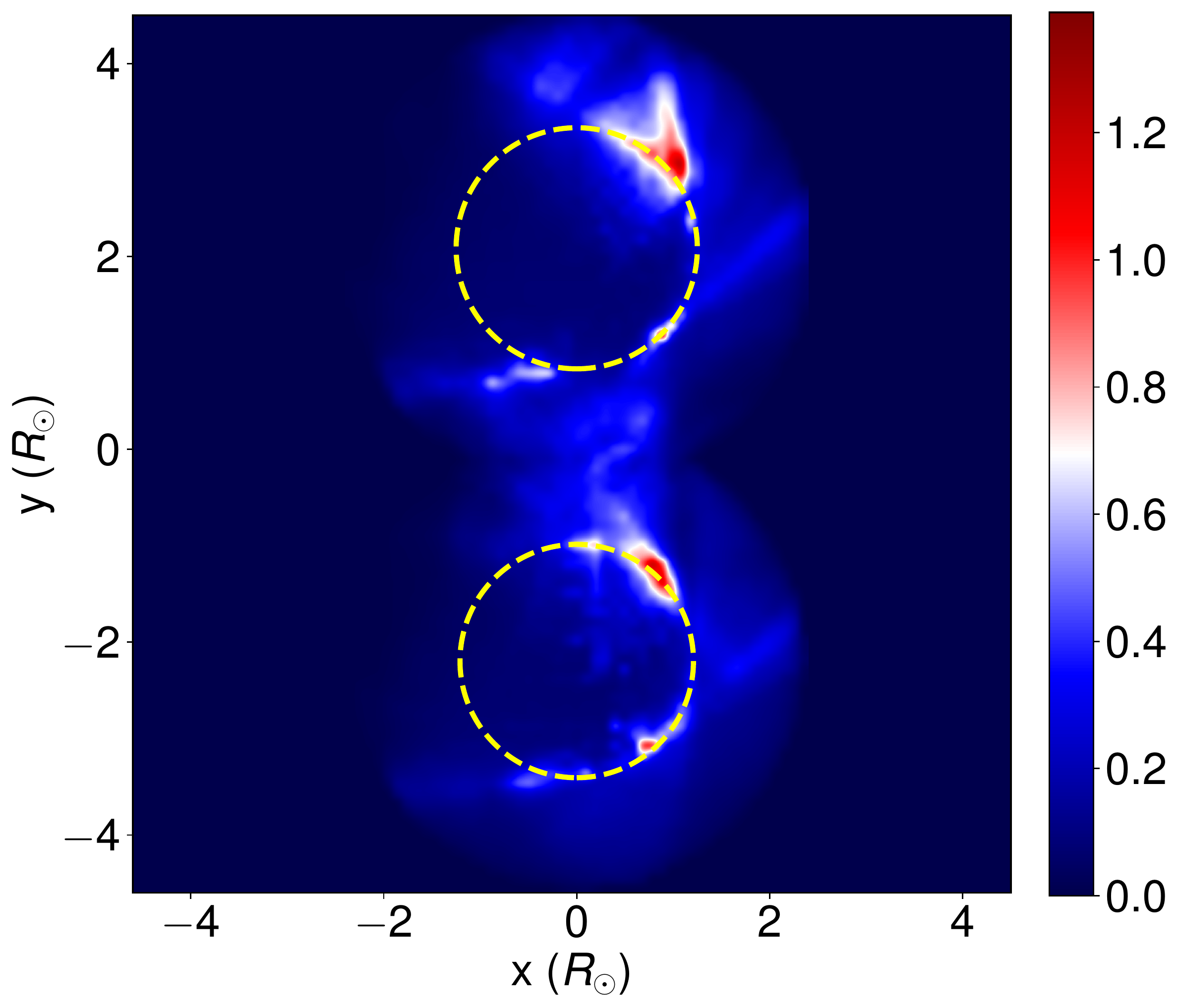}}
\subfigure[]{\includegraphics[scale=0.3]{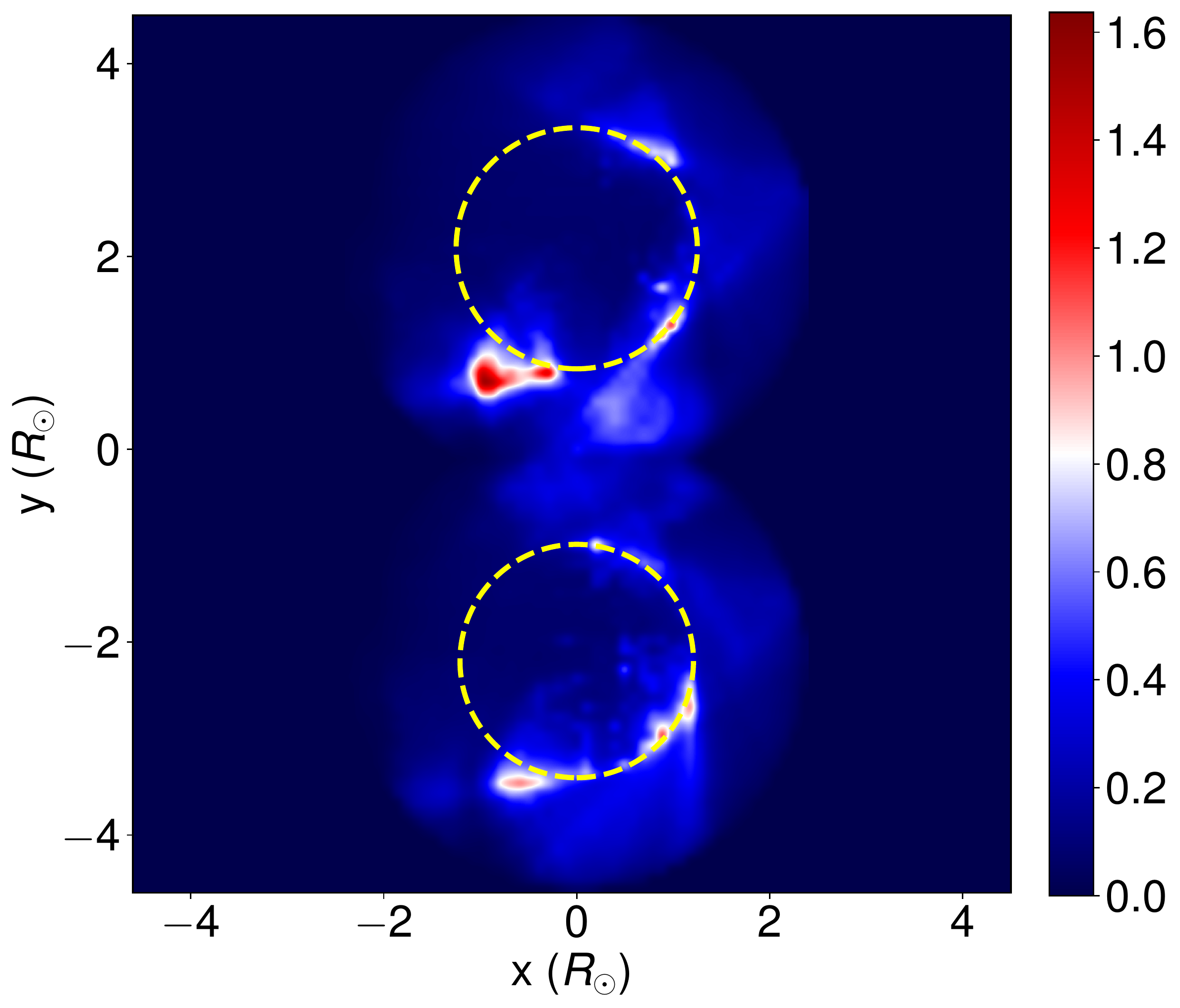}}
\subfigure[]{\includegraphics[scale=0.3]{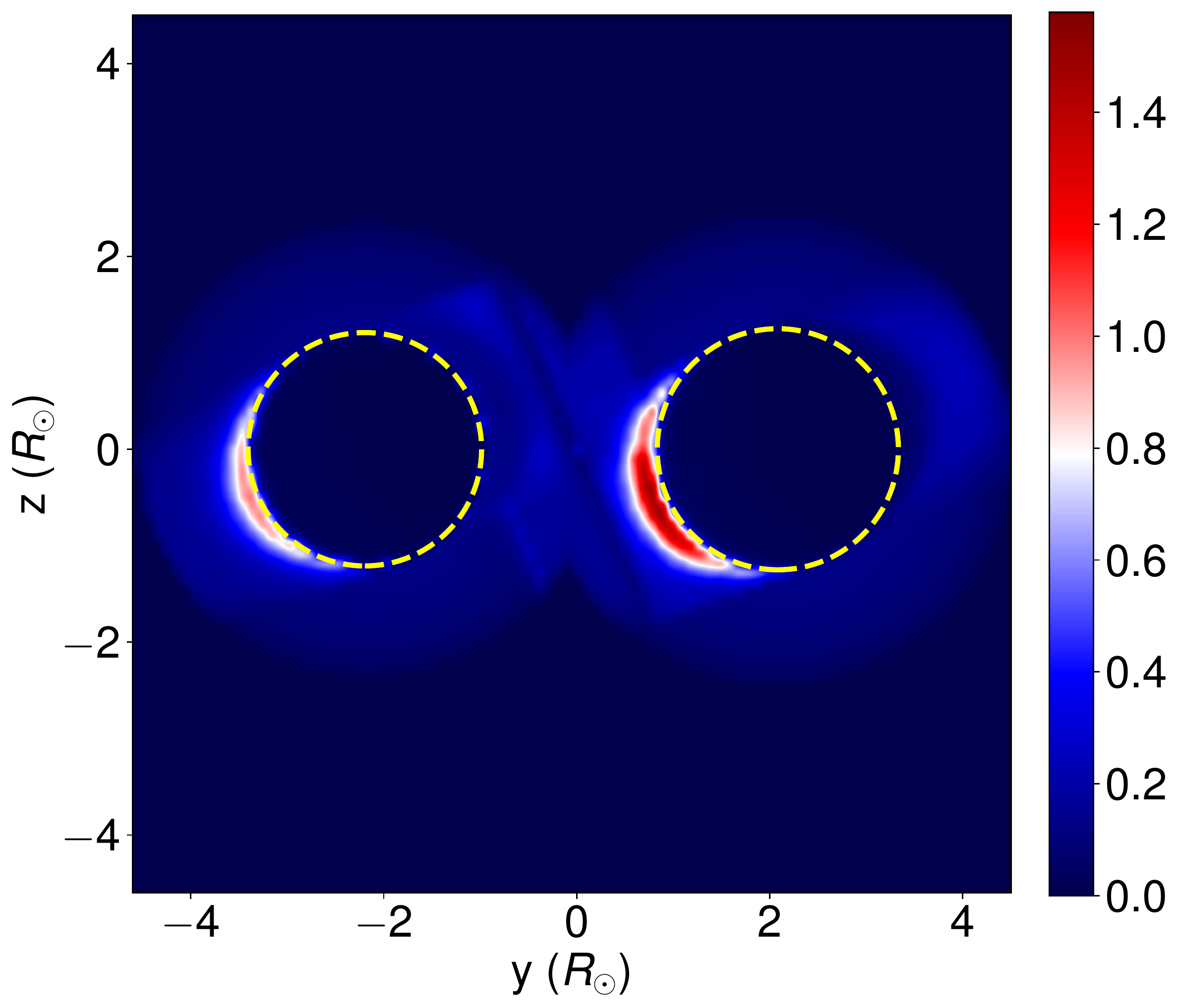}}
\subfigure[]{\includegraphics[scale=0.3]{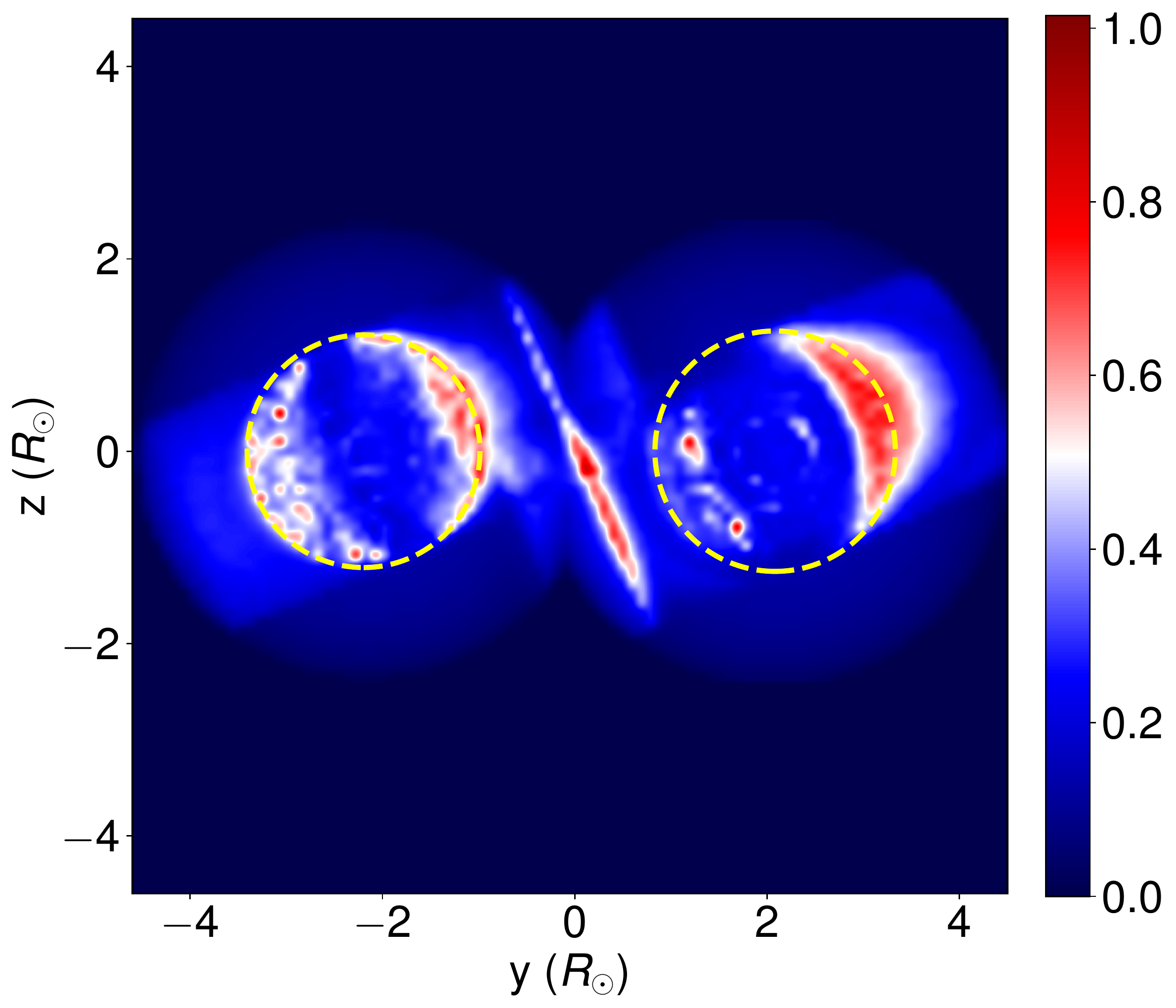}}

\subfigure[]{\includegraphics[width=0.8\columnwidth]{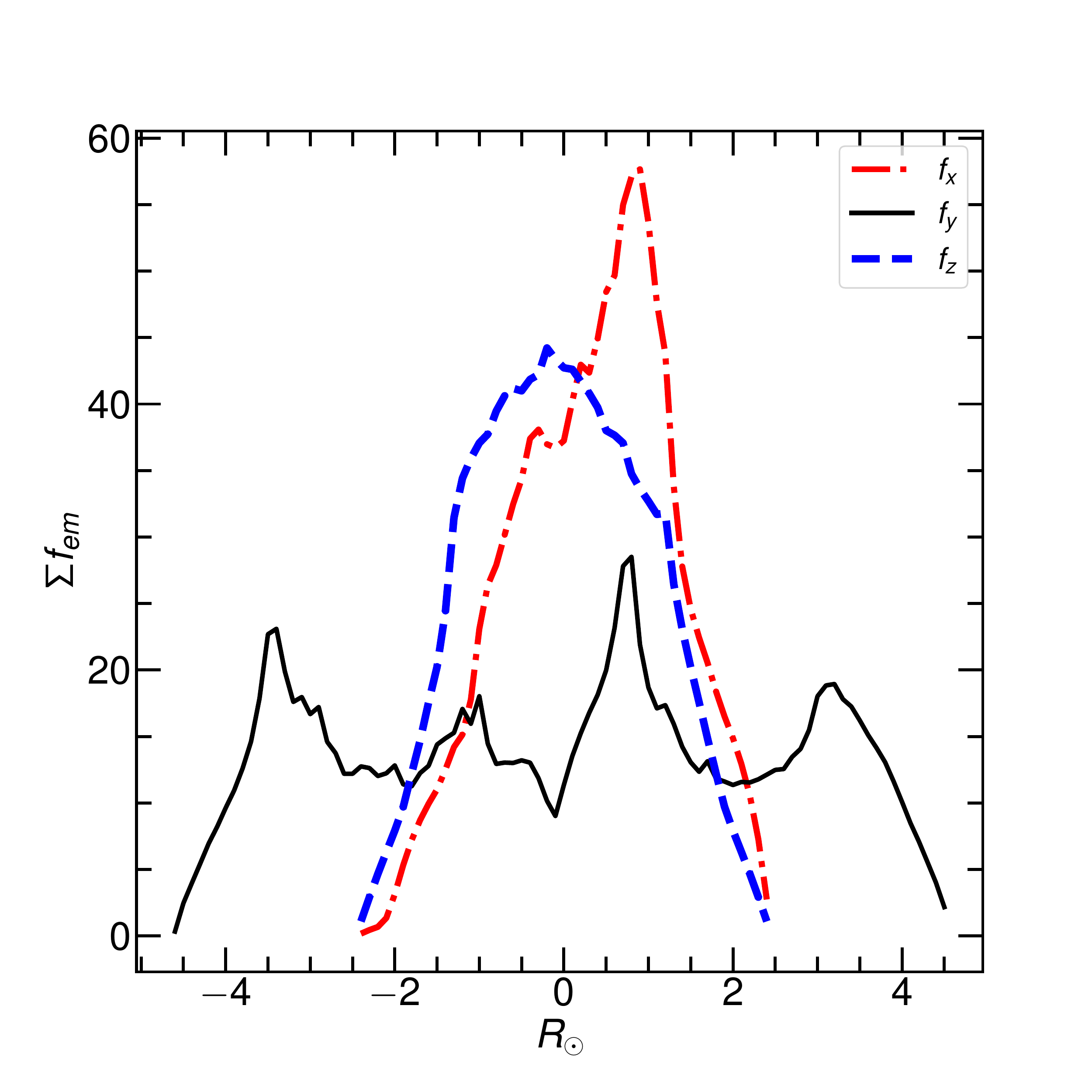}}
\subfigure[]{\includegraphics[width=0.8\columnwidth]{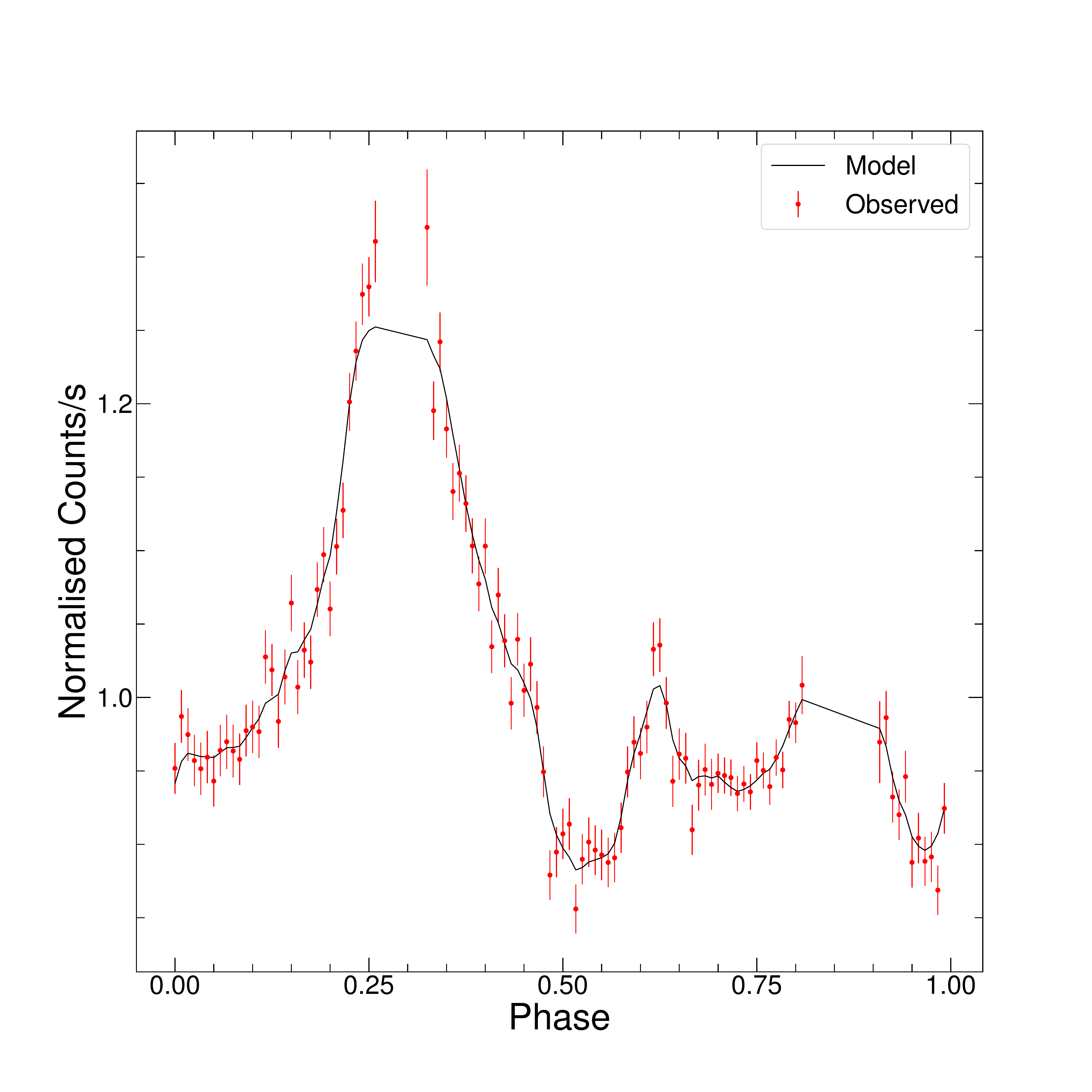}}
    \caption{Same as Figure \ref{fig:iboo_emr} but for ER Vul}
    \label{fig:ervul_emr}
\end{figure*}

We found very similar results for ER Vul to those found for  DV Psc. The presence of one active longitude and uniformly distributed active regions throughout the pole to the equator for both components of this system are found to be present. We also noticed the presence of an inter-binary active region in  ER Vul, which can be seen in Figure \ref{fig:ervul_emr} (d). The $f_y$ distribution as shown in  Figure \ref{fig:ervul_emr}(e), shows that X-ray emitting regions are concentrated toward one side of the binary components and the relative brightness of both coronae was the same. Both components are of almost similar sizes indicating  a similar activity level.

\subsubsection{XY UMa}
\begin{figure*}
\centering
\subfigure[]{\includegraphics[scale=0.3]{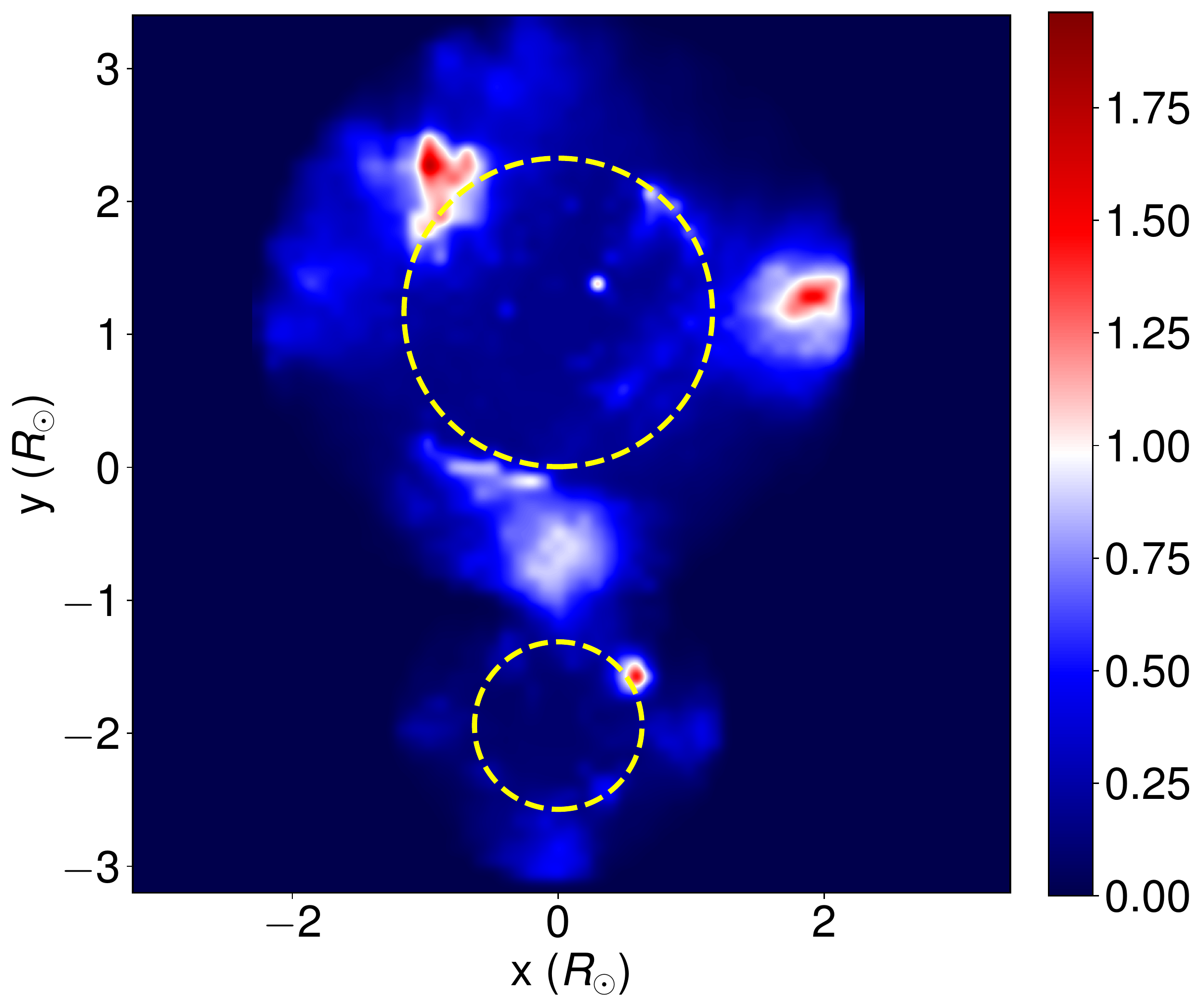}}
\subfigure[]{\includegraphics[scale=0.3]{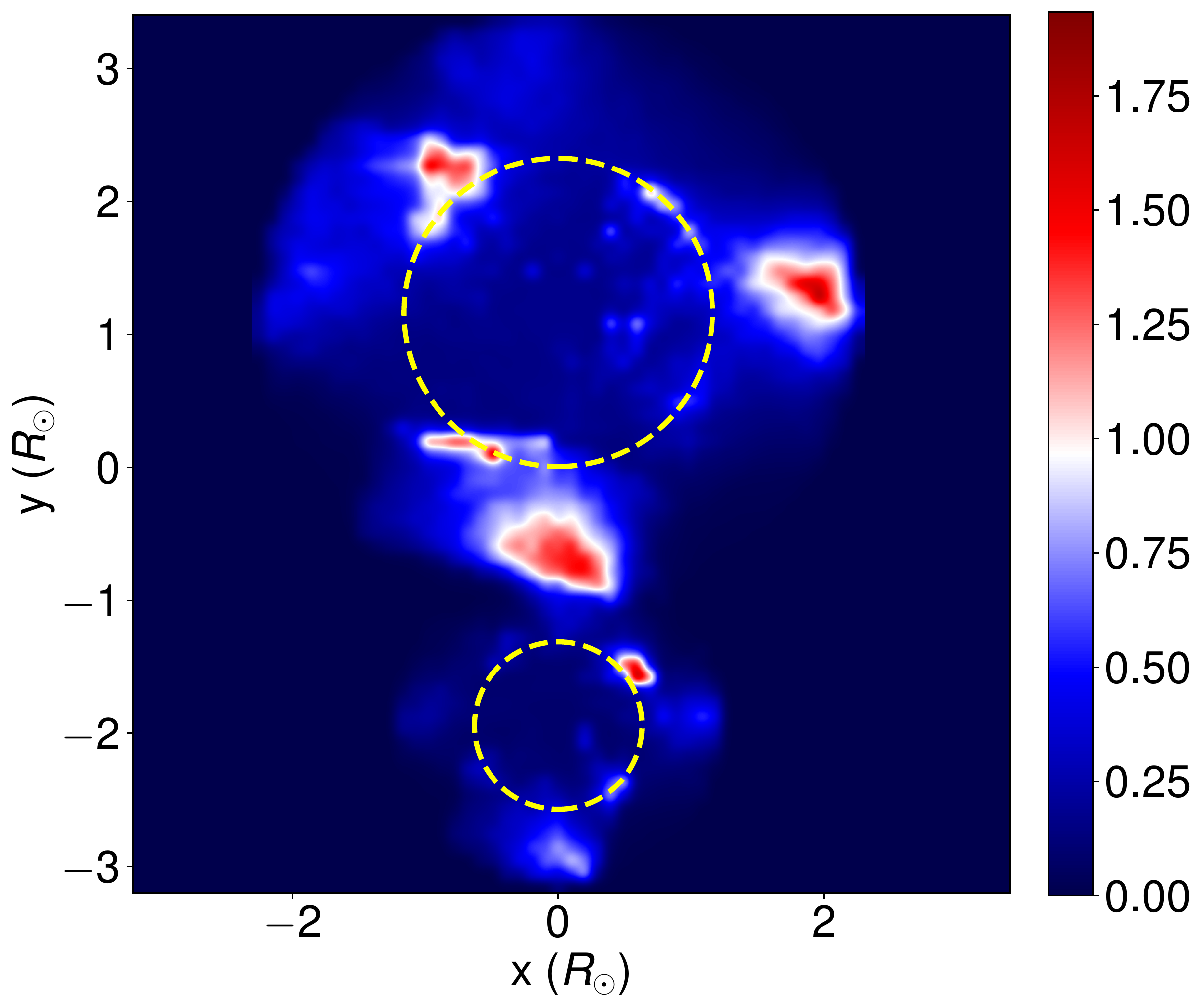}}
\subfigure[]{\includegraphics[scale=0.3]{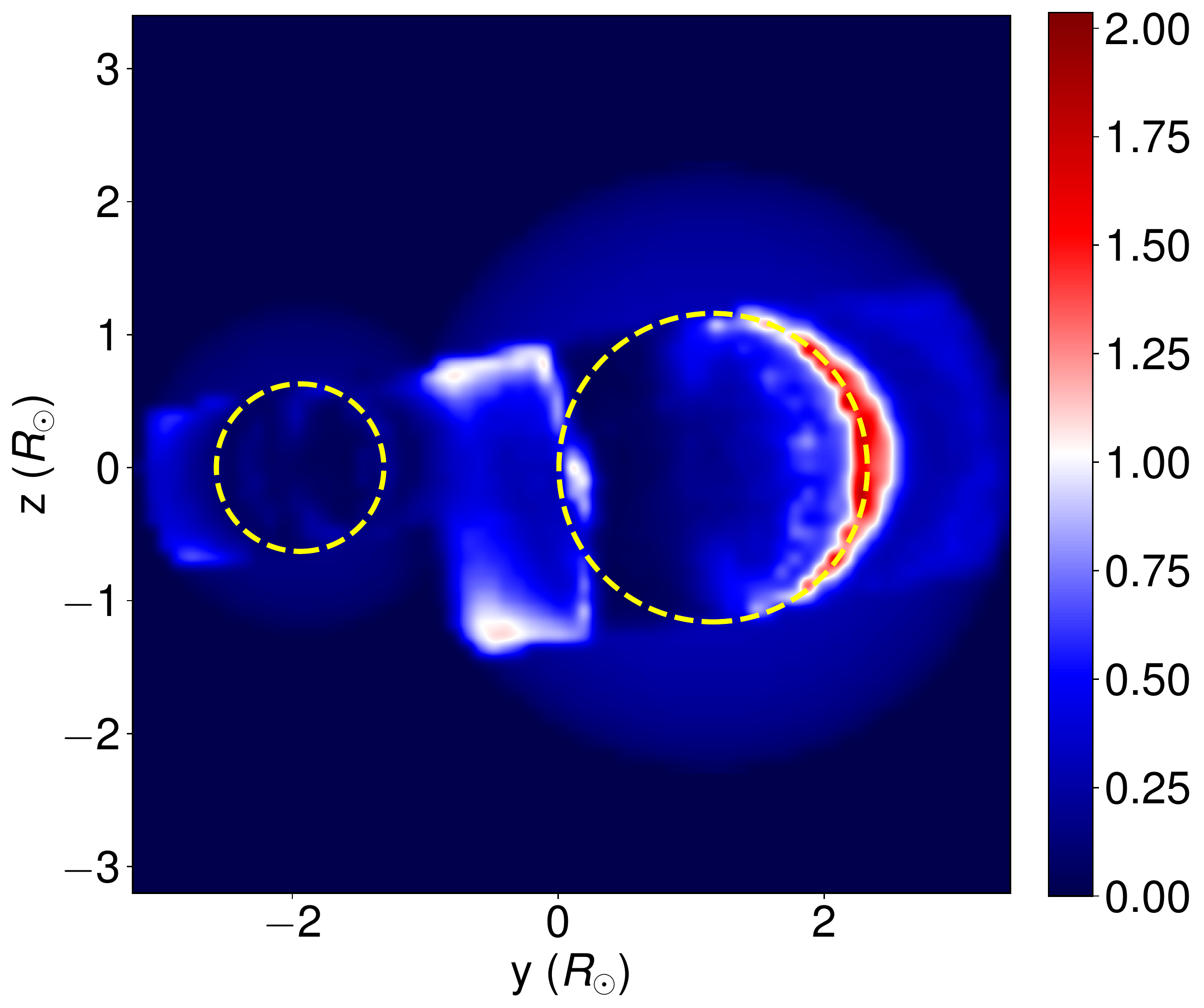}}
\subfigure[]{\includegraphics[scale=0.3]{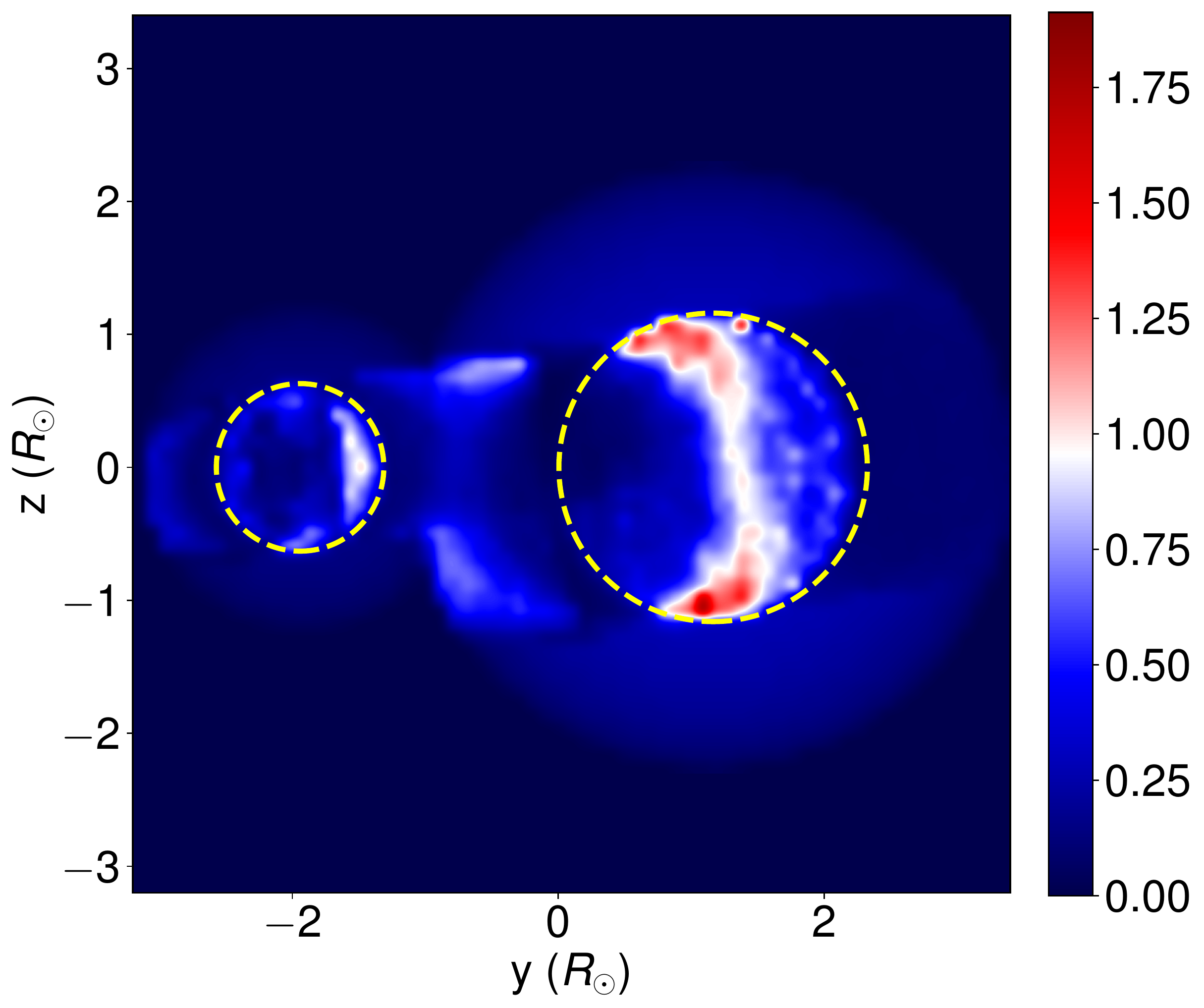}}
\subfigure[]{\includegraphics[width=0.8\columnwidth]{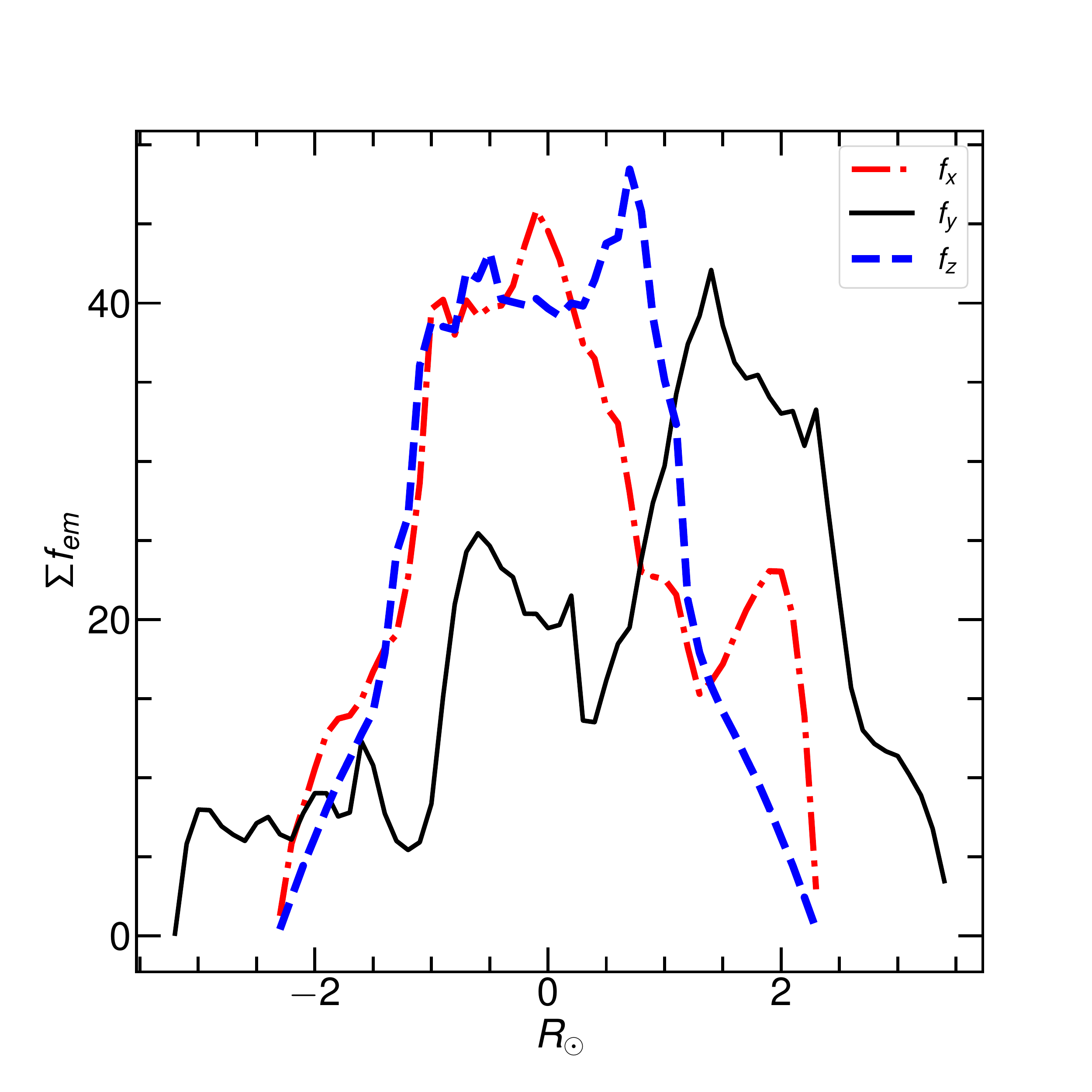}}
\subfigure[]{\includegraphics[width=0.8\columnwidth]{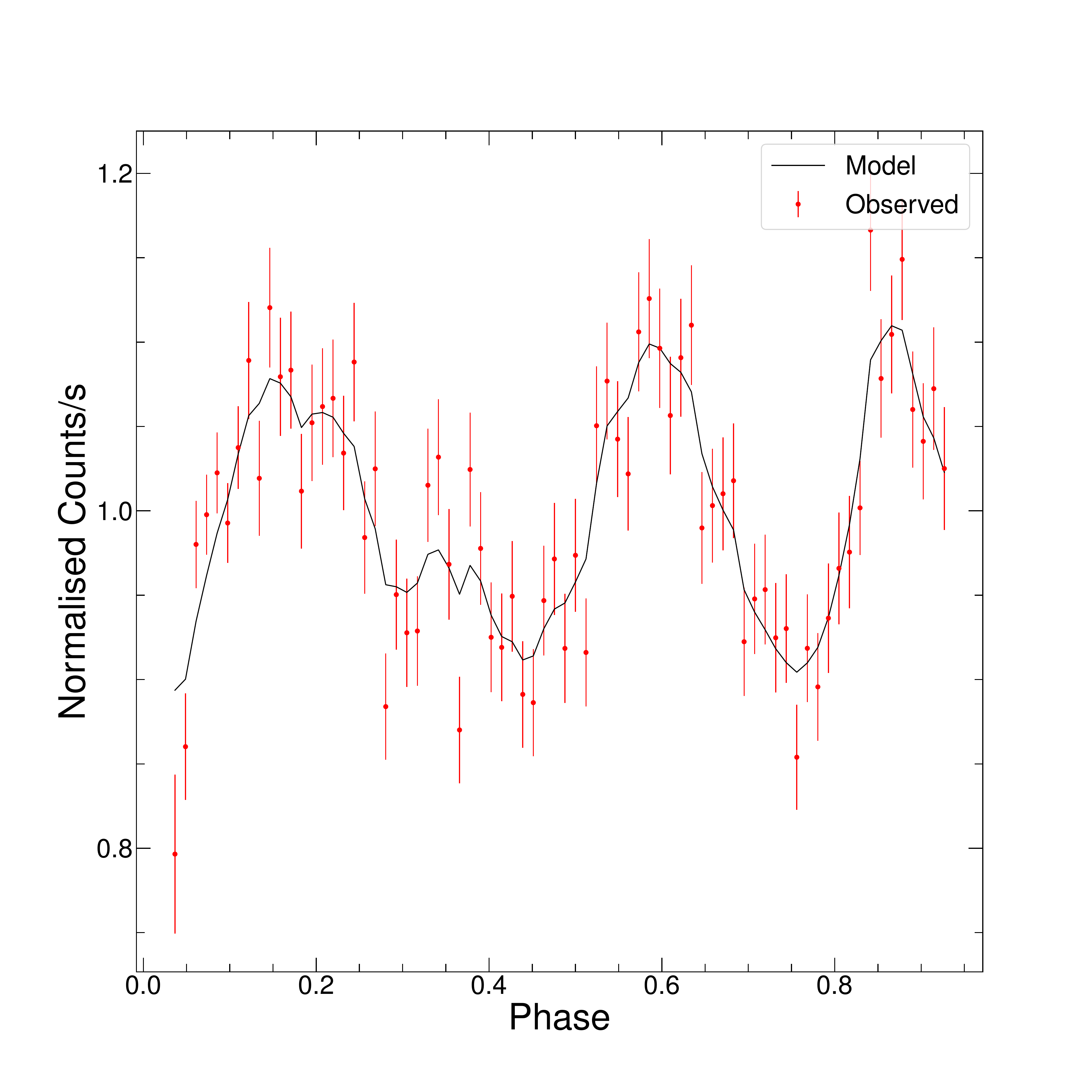}}
    \caption{Same as Figure \ref{fig:iboo_emr} but for  XY UMa.}
    \label{fig:xyuma_emr}
\end{figure*}

The corona of XY UMa shows more dynamic features than any other system in our current sample. In Figures \ref{fig:xyuma_emr} (a) and (b), three bright X-ray emitting regions  with two being on the primary facing away from the secondary and one being on the place where coronae of both components  are connected. The $f_z$ distribution shows X-ray emitting regions are not distributed homogeneously over both hemispheres for the primary as well as the secondary (see Figure \ref{fig:xyuma_emr}(a), (b)). From Figures \ref{fig:xyuma_emr} (c) and (d), we can infer that the coronae of both components are clearly connected. In fact, the  distribution of $f_y$, shows three peaks, which can be explained by the presence of three active regions in which  two active regions are on the primary whereas the third active region arises from  the X-ray emitting plasma near the contact point between the binary components. The relative brightness of the primary and coronal connection regions is about 4 and 2 times more than that of the secondary component. Two active longitudes were also visible in the $f_x$ distribution.

\subsubsection{TX Cnc}
\begin{figure*}
\centering

\subfigure[]{\includegraphics[scale=0.3]{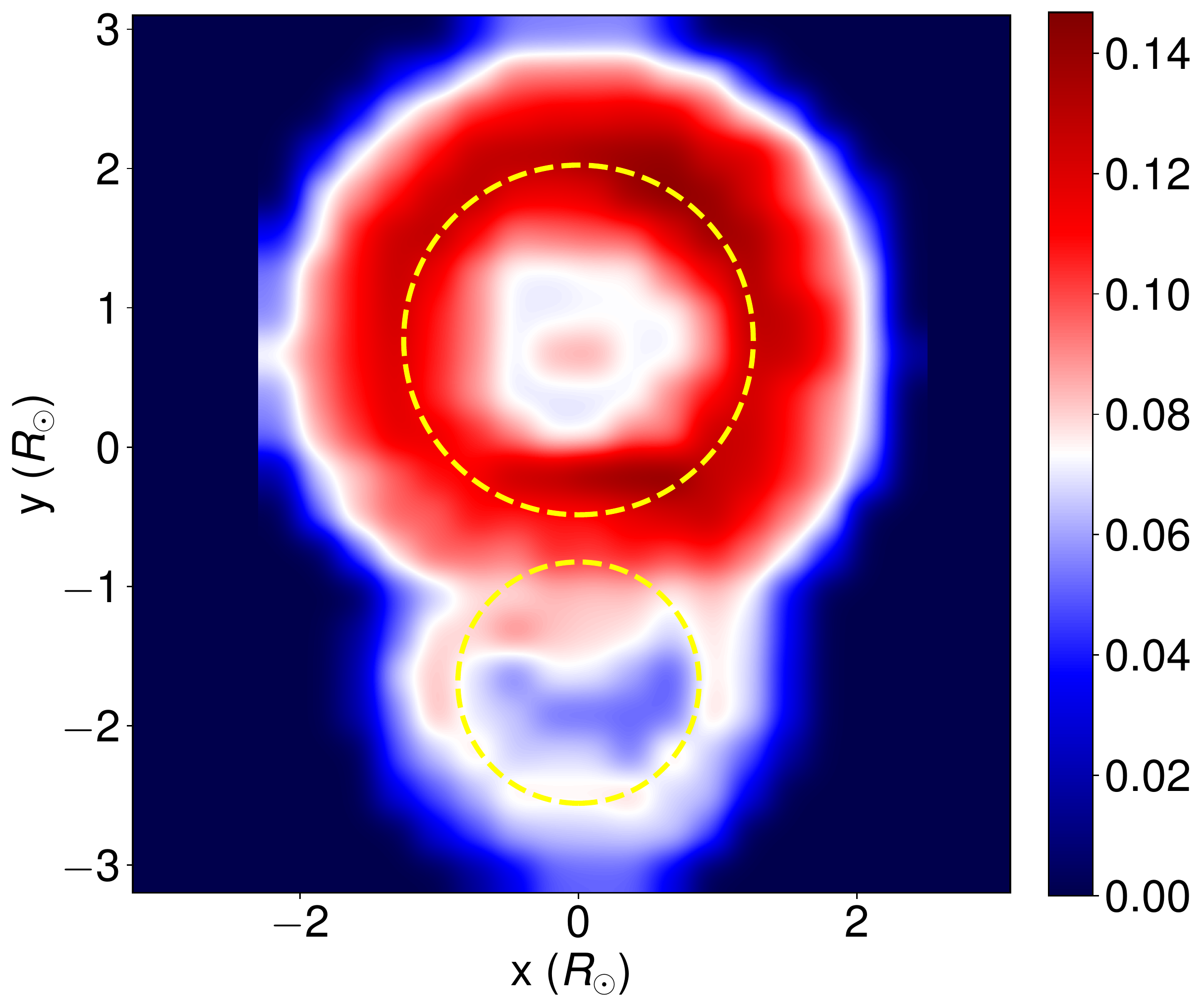}}
\subfigure[]{\includegraphics[scale=0.3]{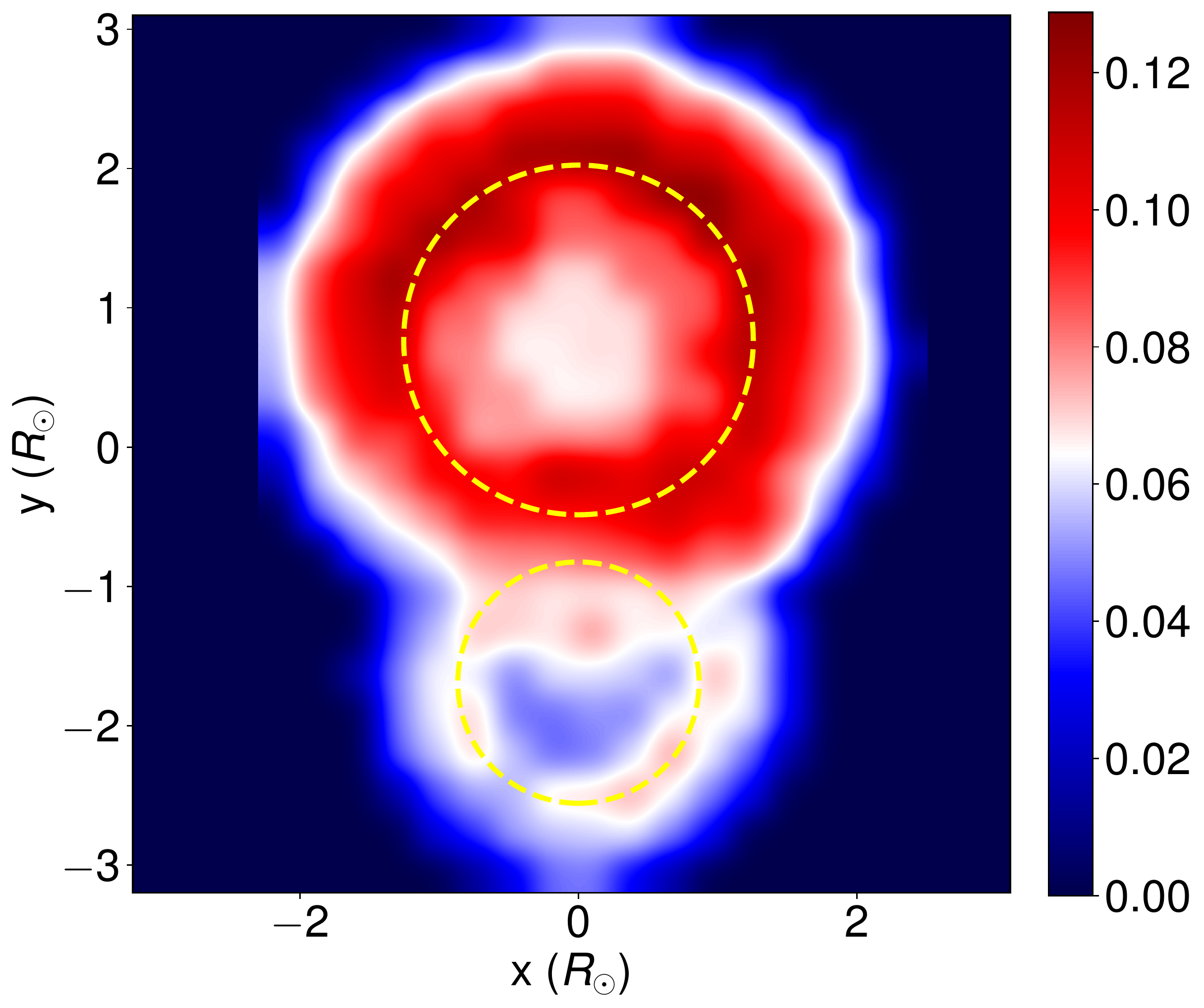}}
\subfigure[]{\includegraphics[scale=0.3]{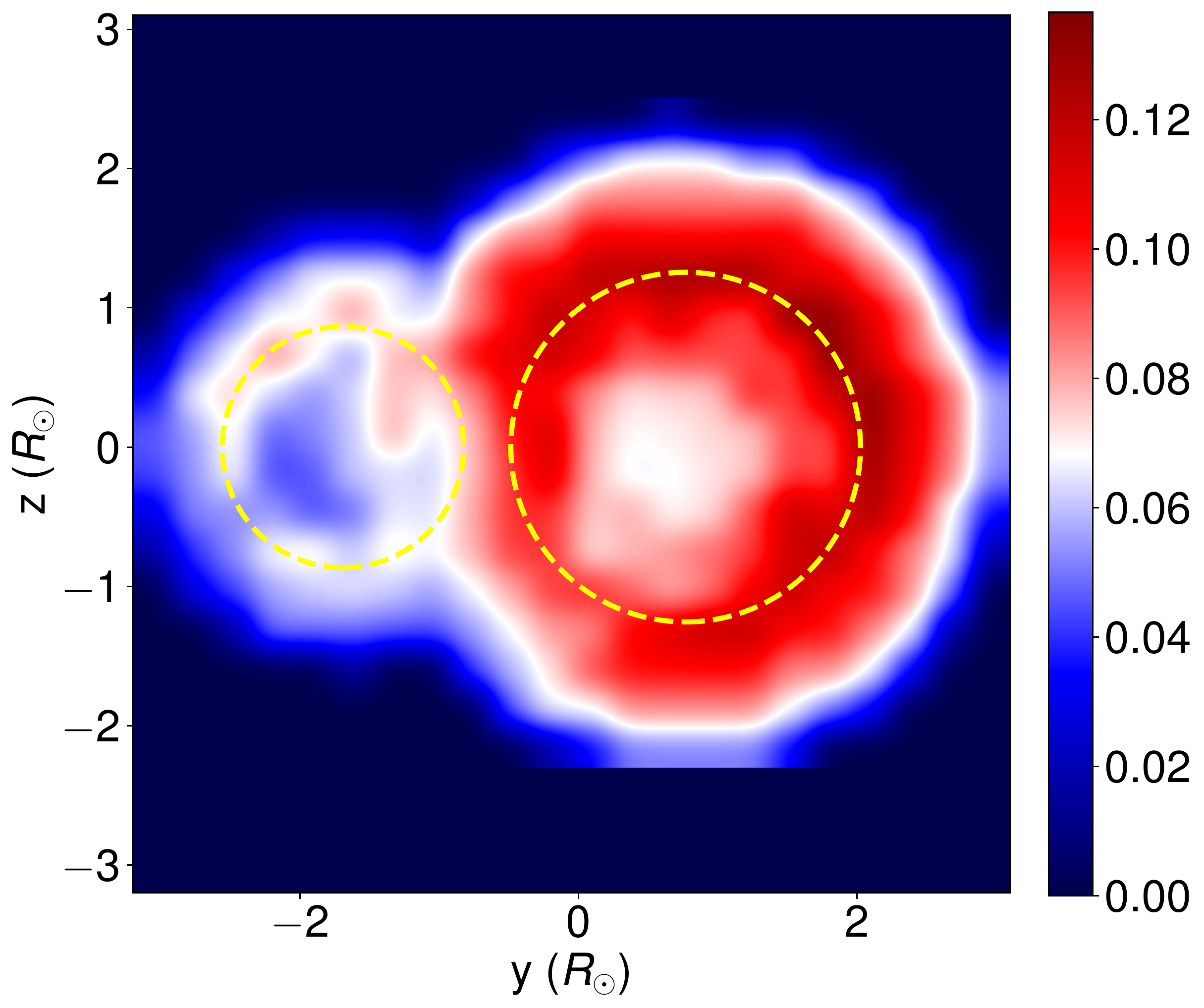}}
\subfigure[]{\includegraphics[scale=0.3]{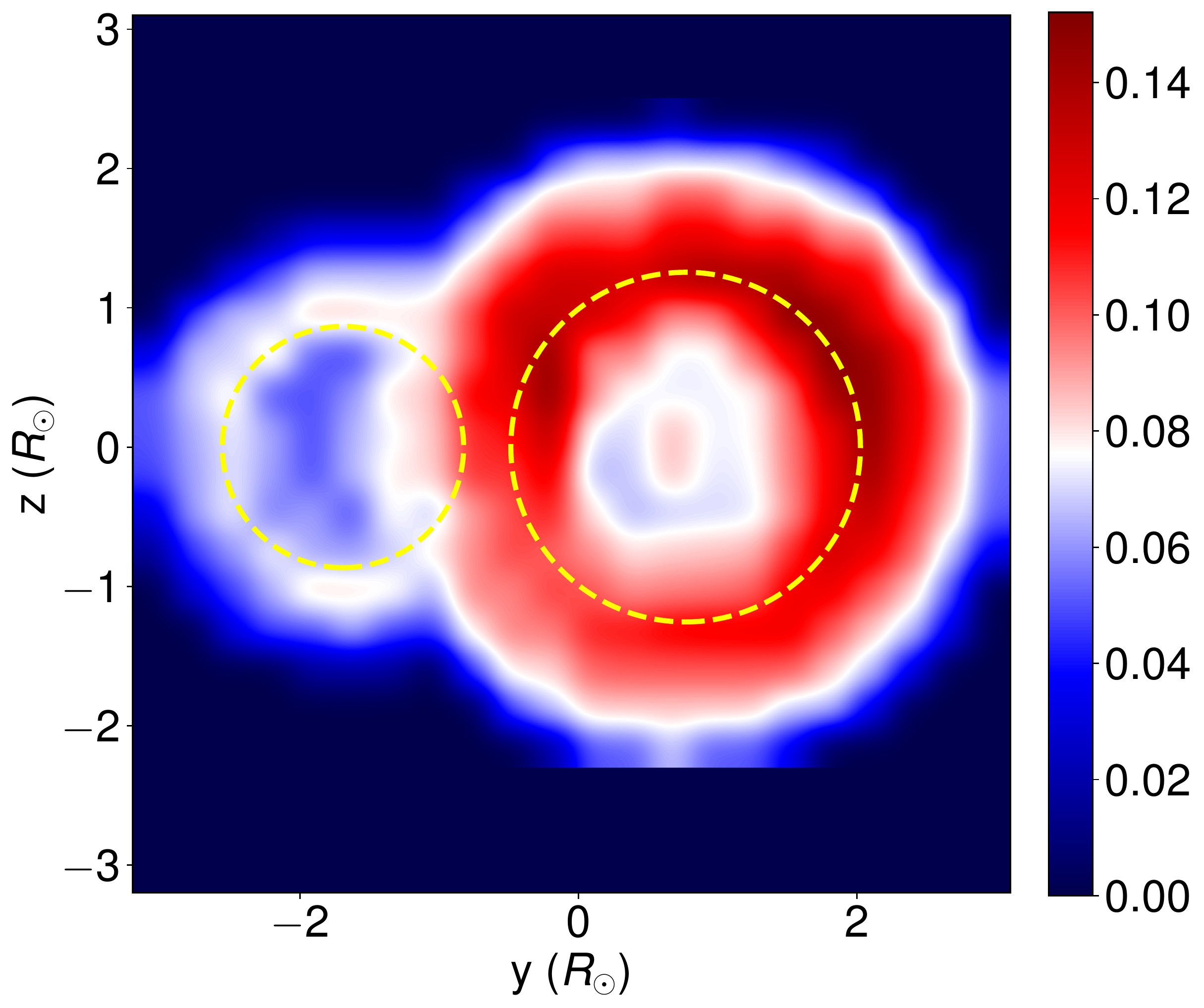}}
\subfigure[]{\includegraphics[width=0.8\columnwidth]{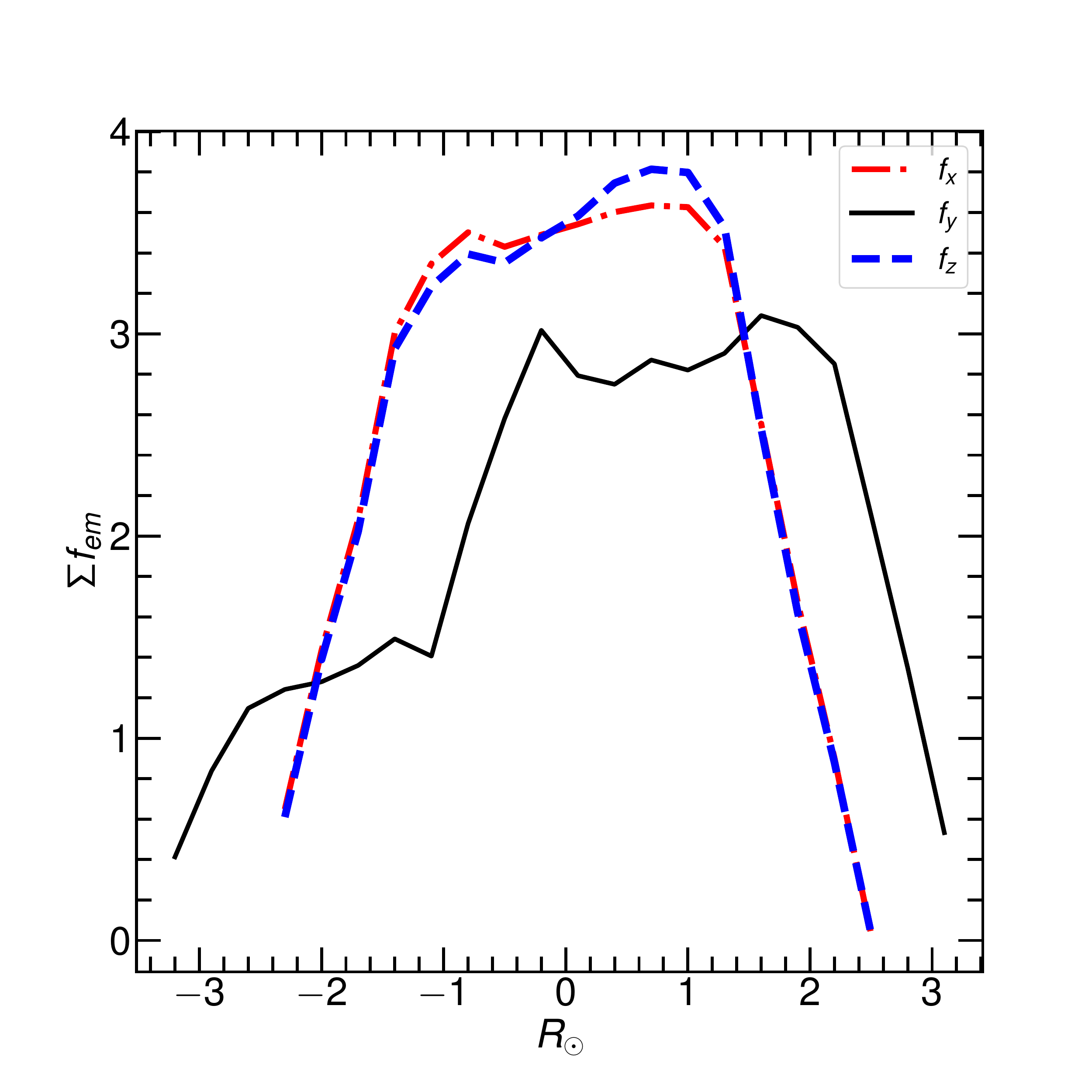}}
\subfigure[]{\includegraphics[width=0.8\columnwidth]{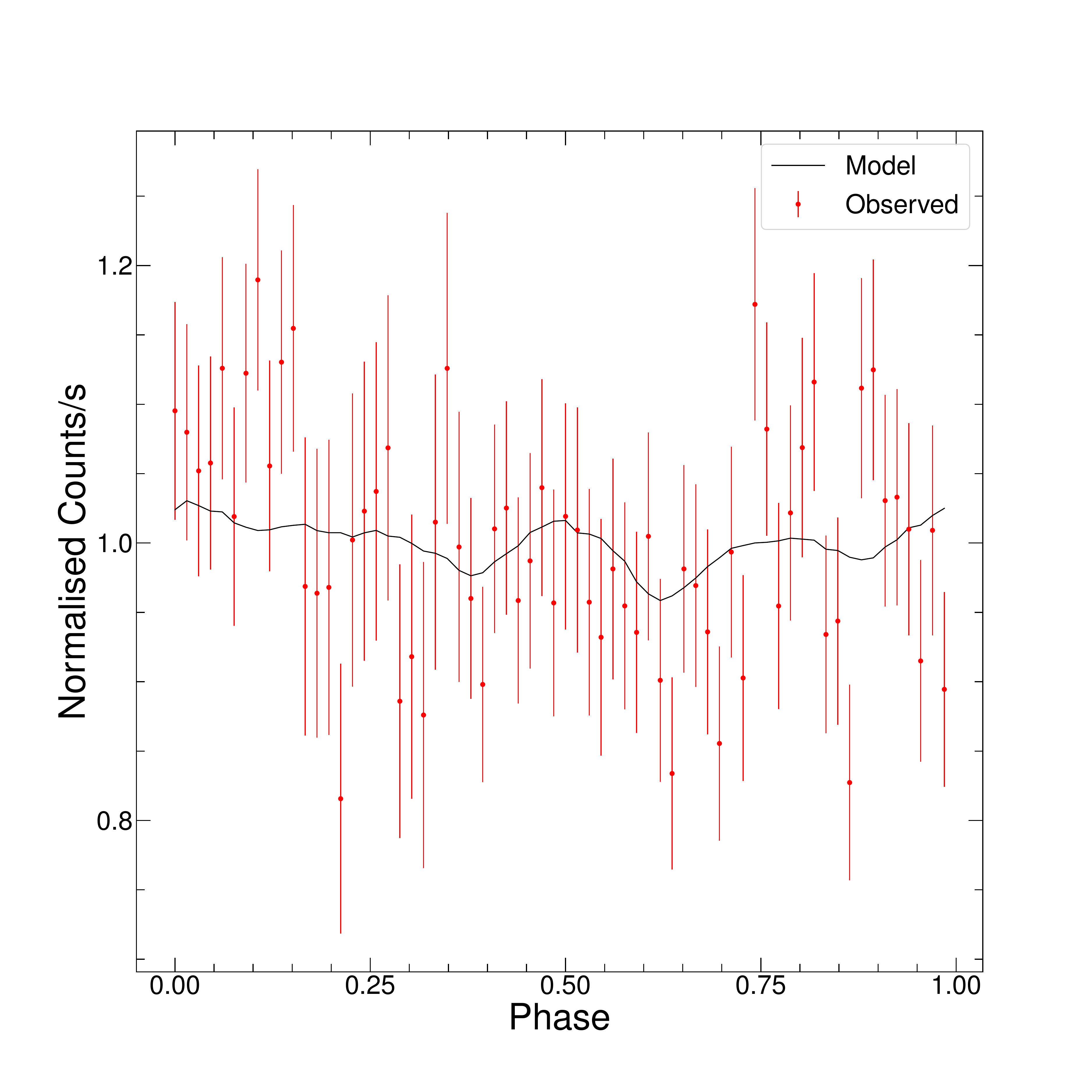}}
    \caption{Same as Figure \ref{fig:iboo_emr} but for TX Cnc.}
    \label{fig:txcnc_emr}
\end{figure*}

The results for TX Cnc are shown in Fig. \ref{fig:txcnc_emr}. The coronae of both components have uniformly distributed X-ray emitting regions, with a small excess in the upper hemispheres for both components. It appears that a common envelope of the X-ray emitting region exists around the binary. The relative brightness of the primary is nearly 3 times that of the secondary, but again both components become equally active  if we consider the relative brightness per unit of surface area. 

\subsection{X-Ray spectral analysis}
\begin{figure}
    \centering
    \includegraphics[width=\columnwidth]{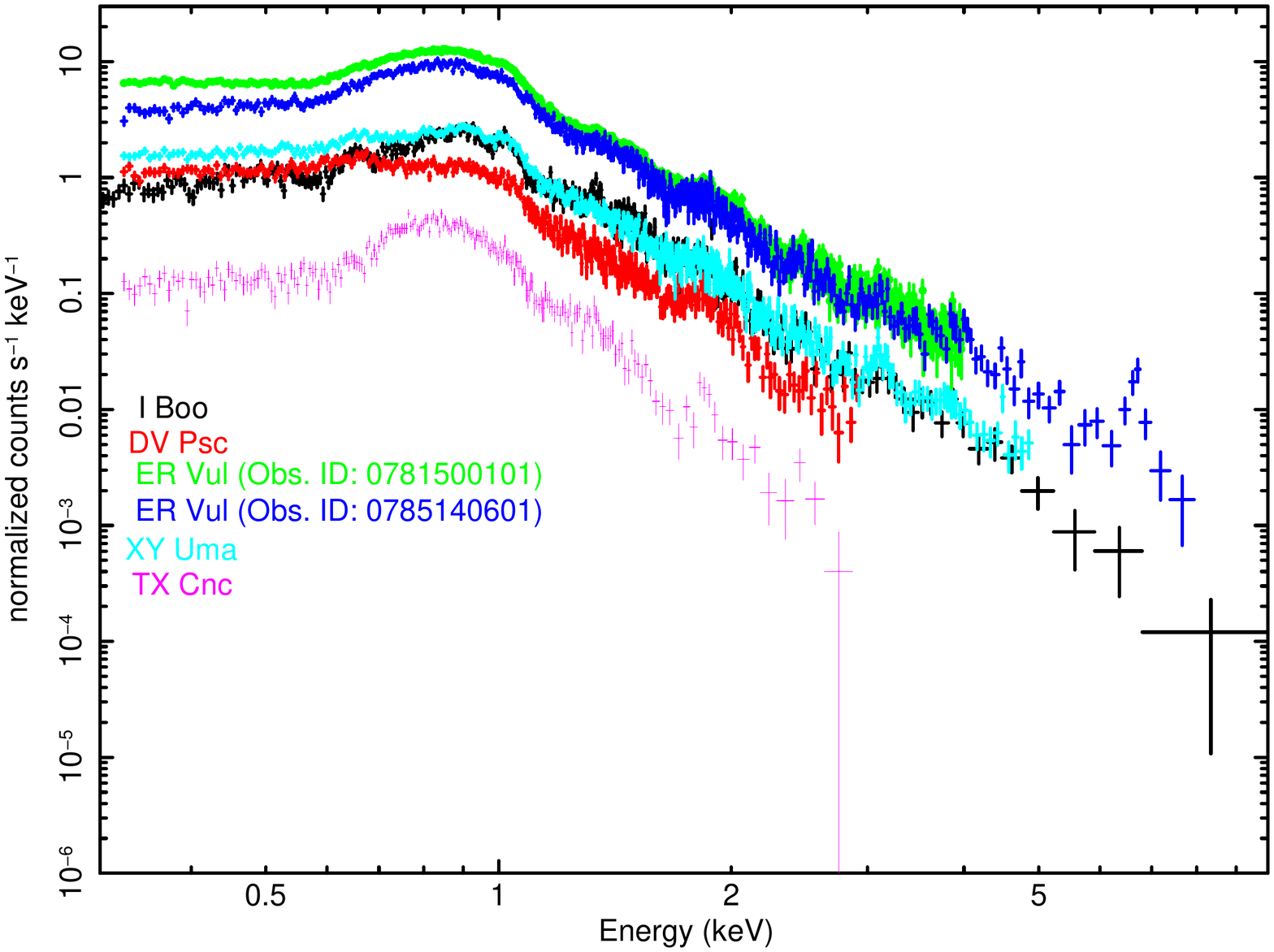}
    \caption{Quiescent spectra of target stars. Here, the color  scheme  is  as  follows:  black  for  I  Boo,  red  for  DV Psc,  green and blue for ER Vul with the observation IDs: 0781500101 and 0785140601, respectively, cyan for XY UMa, and Magenta for TX Cnc.}
    \label{fig:spectra}
\end{figure}

The X-ray spectra of coronally active stars are well modeled by collisionally ionized plasma. We used  \textsc{xspec} (version 12.11.0 \cite{1996ASPC..101...17A}) for X-ray spectral fitting with the absorbed {\sc apec} model \citep[][]{2001ApJ...556L..91S} considering the solar abundances from \cite{1989GeCoA..53..197A}. We analyzed the X-ray spectra at the quiescent and flaring states separately. The energy range for the spectral fitting was chosen to keep source spectra considerably above the background X-ray spectrum. Figure \ref{fig:spectra} shows the average quiescent state spectra of all five EBs. The quiescent state spectra from all three EPIC instruments were fitted simultaneously. The single temperature absorbed \textsc{apec} model resulted in a large value of $\chi^2$ for all systems. After adding one additional temperature component to the model, fitting was improved significantly. The abundances of both components of the two-temperature \textsc{apec} were tied and left free along with other parameters. However,  the parameter of the redshift was fixed to zero. 

The best-fit average spectral parameters for the quiescent state of all target stars are given in Table \ref{tab:phaseresolvedparams}.
The average cool temperature  is found to be  0.26-0.30 KeV for stars 44 Boo, DV Psc, and XY UMa and 0.56-0.64 keV for stars ER Vul and TX Cnc.  However, the average temperature of the hot component is found to be between 0.9  and 1.1 keV for all stars.  The ratio of emission measures corresponding to the hot and cool components is found to be in the range of 0.84 - 1.67. The abundances are found to be near 0.14 $Z_\odot$ for the majority of stars, whereas for TX Cnc, it was found to be  0.35 $Z_\odot$. As expected, the value of $N_H$ for each star is found to be less than the galactic value of $N_H$ \citep[]{1990ARAA..28..215D} toward the direction of that star. 

\subsubsection{Orbital Phase-resolved X-ray spectral analysis}
\begin{figure*}
\centering
\subfigure[44 Boo]{\includegraphics[height=6.5cm,width=0.45\linewidth]{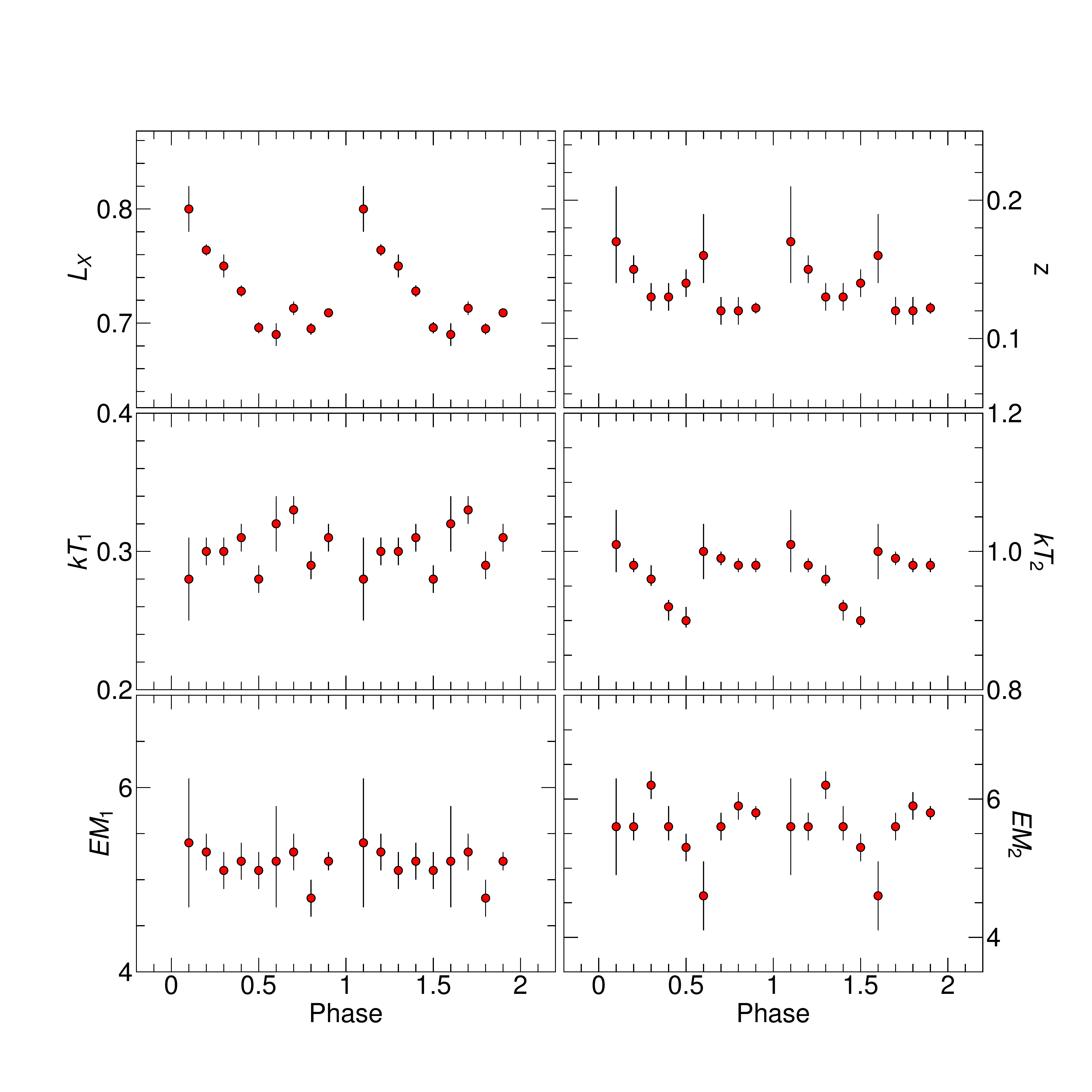}}
\subfigure[DV Psc]{\includegraphics[height=6.5cm,width=0.45\linewidth]{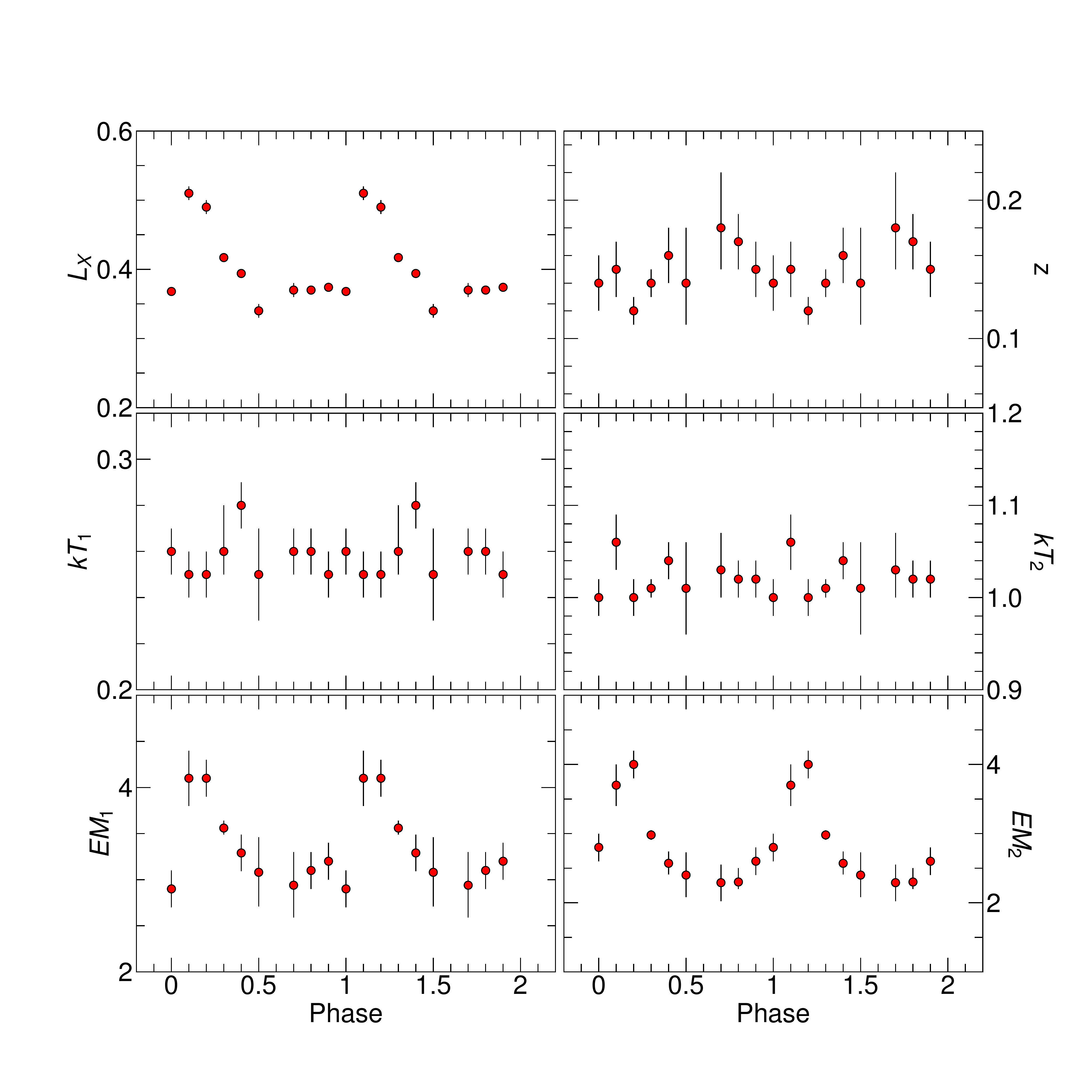}}
\subfigure[ER Vul (filled circle: Obs ID. 0781500101 and open circle: Obs ID. 0785140601)]{\includegraphics[height=6.5cm,width=0.45\linewidth]{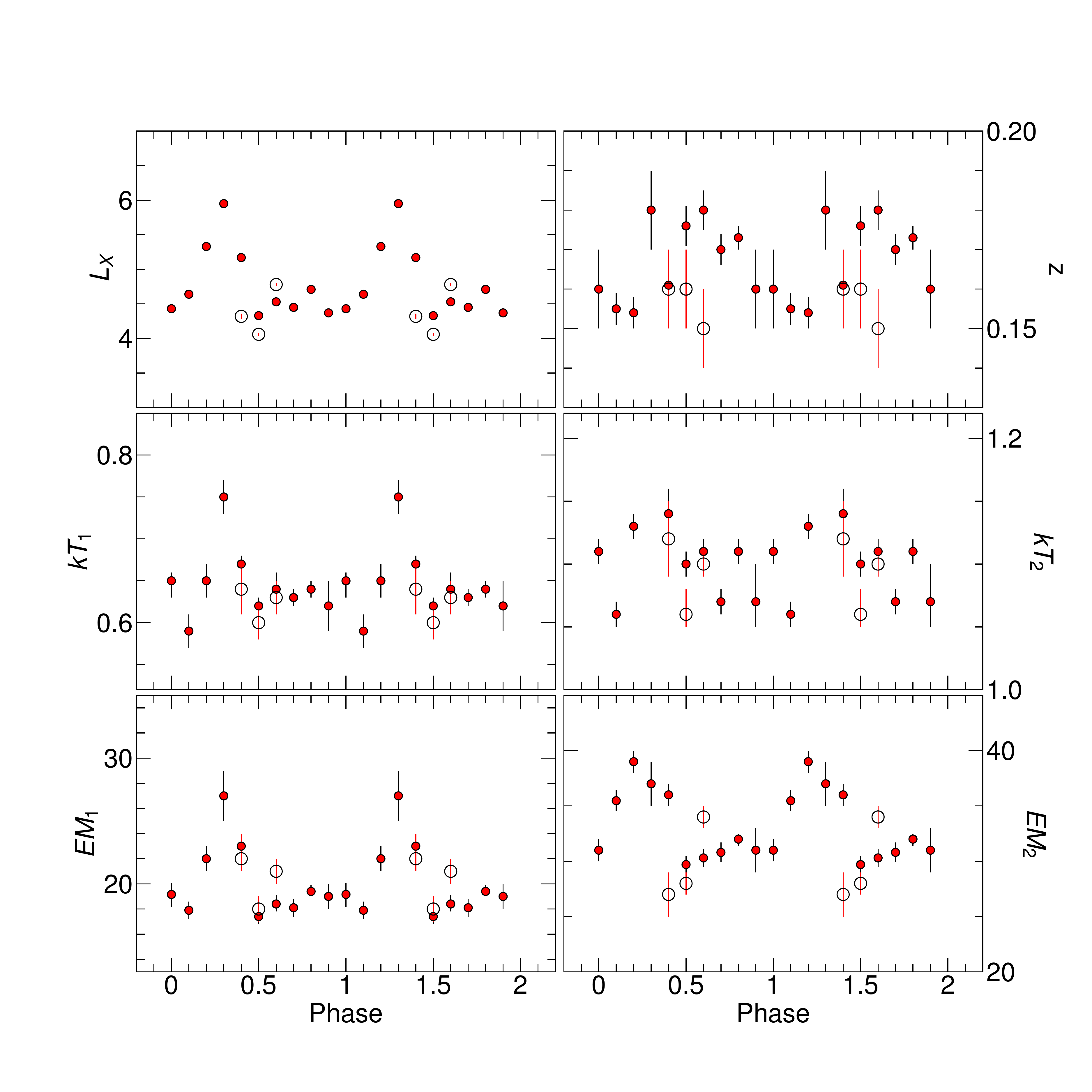}}
\subfigure[XY UMa]{\includegraphics[height=6.5cm,width=0.45\linewidth]{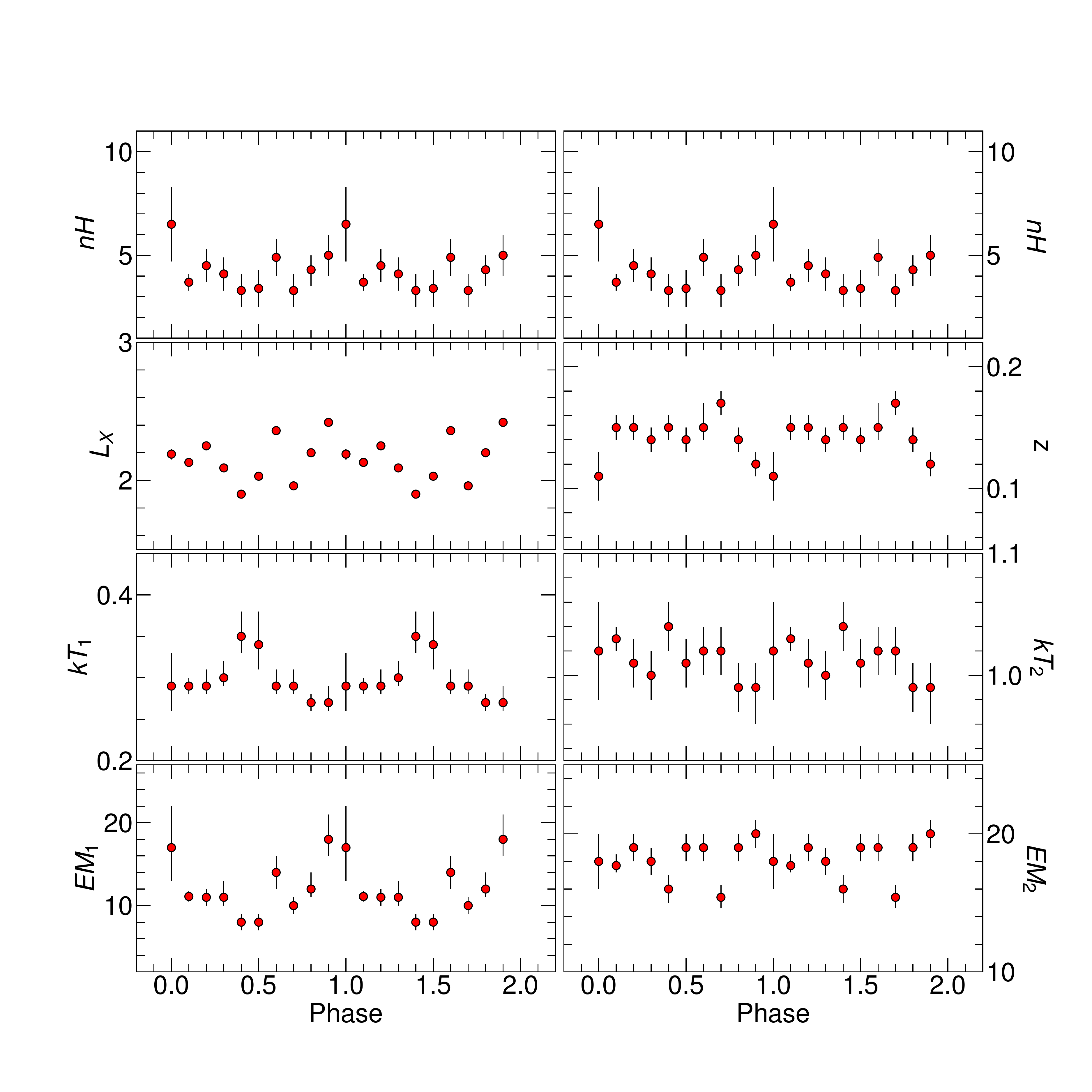}}
\subfigure[TX Cnc]{\includegraphics[height=6.5cm,width=0.45\linewidth]{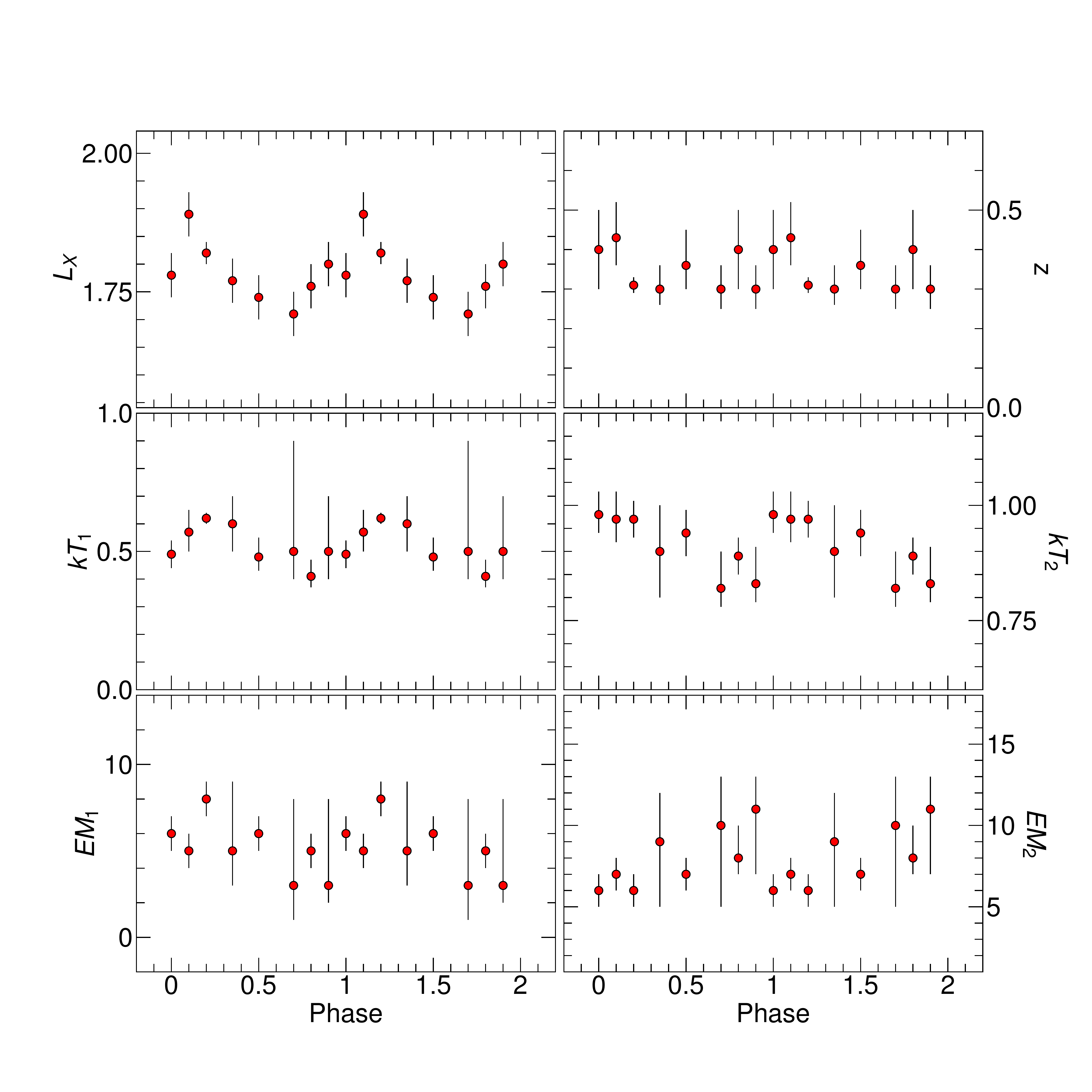}}
    \caption{ Spectral parameters vs phase. $L_X$ is un-absorbed luminosity in units of $10^{30} erg/s$, nH in units of $10^{20}$, Emission measures are in units $10^{52}cm^{-3}$, Temperatures are in keV and abundance (z) is relative to the solar photospheric abundances \citep[][]{1989GeCoA..53..197A}. The errors are within $1\sigma$ confidence limits.}
    \label{fig:4}
\end{figure*}

In order to study the orbital evolution of X-ray spectral parameters, we extracted spectra from each EPIC instrument with a phase interval of 0.1 using the \textsc{phasecalc} task of \textsc{SAS}. The spectra from the MOS1, MOS2, and PN detectors were fitted simultaneously for each phase interval. Similar to the average spectral fitting, we fitted the two-temperature plasma model to each phase bin X-ray spectra. The  $N_H$ was obtained to be constant during the entire orbital phase of each binary except XY UMa. So, we froze the $N_H$ value to the average value and refitted the spectra with the two-temperature \textsc{APEC} models. The best-fit spectral parameters for all phase bins are given in Table \ref{tab:phaseresolvedparams} and shown in Figure \ref{fig:4}.  Here,  $kT_1$  and  $EM_1$ are temperature and corresponding emission measure of the cool component,  $kT_2$ and $EM_2$ are temperature and  emission measure corresponding to the hot component, Z is abundances in terms of solar photospheric abundances (Z$_\odot$), and L$_{XQ}$ is X-ray luminosity  in the 0.3$-$10.0 keV energy band. 

The variability in each spectral parameter was tested by the $\chi^2$ test, where  $\chi^2=\Sigma_i\big(\frac{P_i-P_{av}}{\sigma_{P_{i}}}\big)^2$,  $P_i$ and $\sigma_{P_{i}}$ are the value and corresponding error of the parameter in a particular phase bin, and $P_{av}$ is the average value of the parameter. This $\chi^2$ was compared with the critical value of $\chi^2$ for a 99\% of confidence level for a given degree of freedom. When observed $\chi^2$ is greater than the critical value of $\chi^2$, the parameter was considered to be a variable with a confidence level of 99\%. The results of this variability analysis are the given in the Table \ref{tab:variability_in_spectral_params}, where "V" stands for variable parameters and "C" stands for constant parameter. 
We found the parameter L$_{XQ}$  is variable with 99\% confidence level for all stars.  All the spectral parameters are found to be variable only for the star ER Vul.  The EM$_2$ is found to be varying with the orbital phase for  all stars, excluding TX Cnc, whereas EM$_1$ is found to be variable only for DV Psc, ER Vul, and XY UMa.  The coronal abundances also vary significantly for the stars 44 Boo and ER Vul.


\startlongtable
\begin{deluxetable*}{ccccccccc}
\tablecaption{Best-fit spectral parameters for the EBs in the sample. \label{tab:phaseresolvedparams}}
\tablehead{
\colhead{Phase}&\colhead{N$_H$}&\colhead{kT$_1$}&\colhead{kT$_2$}&\colhead{Z(Z$_\odot)$}&\colhead{EM$_1$}&\colhead{EM$_2$}&\colhead{L$_{XQ}$}&\colhead{$\chi_\nu^2(d.o.f)$}\\
}

\startdata
    \multicolumn{9}{c}{44 Boo}\\
Avg.&$3.5_{-0.2}^{+0.2}$&$0.298_{-0.004}^{+0.004}$&$0.954_{-0.004}^{+0.004}$&$0.129_{-0.003}^{+0.003}$&$5.3_{-0.2}^{+0.2}$&$5.9_{-0.1}^{+0.1}$&$0.736_{-0.002}^{+0.002}$& 1.83(918)\\
0.1~ &\nodata&$ 0.28 ^{+ 0.03 }_{- 0.03 }$&$ 1.01 ^{+ 0.05 }_{- 0.04 }$&$ 0.17 ^{+ 0.04 }_{- 0.03 }$&$ 5.4 ^{+ 0.7 }_{- 0.7 }$&$ 5.6 ^{+ 0.7 }_{- 0.7 }$&$ 0.80 ^{+ 0.02 }_{- 0.02 }$& 1.04 ( 66 )\\
0.2~ &\nodata&$ 0.30 ^{+ 0.01 }_{- 0.01 }$&$ 0.98 ^{+ 0.01 }_{- 0.01 }$&$ 0.15 ^{+ 0.01 }_{- 0.01 }$&$ 5.3 ^{+ 0.2 }_{- 0.2 }$&$ 5.6 ^{+ 0.2}_{- 0.2 }$&$ 0.764 ^{+ 0.005 }_{- 0.005 }$& 1.1 ( 525 )\\
0.3~&\nodata&$ 0.30 ^{+ 0.01 }_{- 0.01 }$&$ 0.96 ^{+ 0.02 }_{- 0.01 }$&$ 0.13 ^{+ 0.01 }_{- 0.01 }$&$ 5.1 ^{+ 0.2 }_{- 0.2 }$&$ 6.2 ^{+ 0.2 }_{- 0.2 }$&$ 0.75 ^{+ 0.01 }_{- 0.01 }$& 1.13 ( 497 )\\
0.4~&\nodata&$ 0.31 ^{+ 0.01 }_{- 0.01 }$&$ 0.92 ^{+ 0.01 }_{- 0.02 }$&$ 0.13 ^{+ 0.01 }_{- 0.01 }$&$ 5.2 ^{+ 0.2 }_{- 0.2 }$&$ 5.6 ^{+ 0.3 }_{- 0.2 }$&$ 0.728 ^{+ 0.005 }_{- 0.005 }$& 1.19 ( 510 )\\
0.5~&\nodata&$ 0.28 ^{+ 0.01 }_{- 0.01 }$&$ 0.90 ^{+ 0.02 }_{- 0.01 }$&$ 0.14 ^{+ 0.01 }_{- 0.01 }$&$ 5.1 ^{+ 0.2 }_{- 0.2 }$&$ 5.3 ^{+ 0.2 }_{- 0.2 }$&$ 0.696 ^{+ 0.005 }_{- 0.005 }$& 1.22 ( 493 )\\
0.6~&\nodata&$ 0.32 ^{+ 0.02 }_{- 0.02 }$&$ 1.00 ^{+ 0.04 }_{- 0.04 }$&$ 0.16 ^{+ 0.03 }_{- 0.02 }$&$ 5.2 ^{+ 0.6 }_{- 0.5 }$&$ 4.6 ^{+ 0.5 }_{- 0.5 }$&$ 0.69 ^{+ 0.01 }_{- 0.01 }$& 1.06 ( 100 )\\
0.7~&\nodata&$ 0.33 ^{+ 0.01 }_{- 0.01 }$&$ 0.99 ^{+ 0.01 }_{- 0.01 }$&$ 0.12 ^{+ 0.01 }_{- 0.01 }$&$ 5.3 ^{+ 0.2 }_{- 0.2 }$&$ 5.6 ^{+ 0.2 }_{- 0.2 }$&$ 0.713 ^{+ 0.006 }_{- 0.006 }$& 1.15 ( 457 )\\
0.8~&\nodata&$ 0.29 ^{+ 0.01 }_{- 0.01 }$&$ 0.98 ^{+ 0.01 }_{- 0.01 }$&$ 0.12 ^{+ 0.01 }_{- 0.01 }$&$ 4.8 ^{+ 0.2 }_{- 0.2 }$&$ 5.9 ^{+ 0.2 }_{- 0.2 }$&$ 0.695 ^{+ 0.005 }_{- 0.005 }$& 1.01 ( 509 )\\
0.9~&\nodata&$ 0.31 ^{+ 0.01 }_{- 0.01 }$&$ 0.98 ^{+ 0.01 }_{- 0.01 }$&$ 0.122 ^{+ 0.004 }_{- 0.004 }$&$ 5.2 ^{+ 0.1 }_{- 0.1 }$&$ 5.8 ^{+ 0.1 }_{- 0.1 }$&$ 0.709 ^{+ 0.003 }_{- 0.003 }$& 1.33 ( 723 )\\
    \multicolumn{9}{c}{DV Psc}\\
Avg.& $1.9_{-0.3}^{+0.3}$&$0.257_{-0.003}^{+0.003}$&$1.004_{-0.008}^{+0.008}$&$0.14_{-0.01}^{+0.01}$&$3.5_{-0.2}^{+0.2}$&$2.98_{-0.08}^{+0.08}$&$0.416_{-0.002}^{+0.002}$&1.26(814) \\
0.0 &\nodata&$ 0.26 ^{+ 0.01 }_{- 0.01 }$&$ 1.0 ^{+ 0.02 }_{- 0.02 }$&$ 0.14 ^{+ 0.02 }_{- 0.02 }$&$ 2.9 ^{+ 0.2 }_{- 0.2 }$&$ 2.8 ^{+ 0.2 }_{- 0.2 }$&$ 0.368 ^{+ 0.005 }_{- 0.005 }$& 1.15 ( 226 )\\
0.1~&\nodata&$ 0.25 ^{+ 0.01 }_{- 0.01 }$&$ 1.06 ^{+ 0.03 }_{- 0.03 }$&$ 0.15 ^{+ 0.02 }_{- 0.02 }$&$ 4.1 ^{+ 0.3 }_{- 0.3 }$&$ 3.7 ^{+ 0.3 }_{- 0.3 }$&$ 0.51 ^{+ 0.01 }_{- 0.01 }$& 1.07 ( 183 )\\
0.2~&\nodata&$ 0.25 ^{+ 0.01 }_{- 0.01 }$&$ 1.0 ^{+ 0.02 }_{- 0.02 }$&$ 0.12 ^{+ 0.01 }_{- 0.01 }$&$ 4.1 ^{+ 0.2 }_{- 0.2 }$&$ 4.0 ^{+ 0.2 }_{- 0.2 }$&$ 0.49 ^{+ 0.01 }_{- 0.01 }$& 1.1 ( 306 )\\
0.3~&\nodata&$ 0.26 ^{+ 0.02 }_{- 0.01 }$&$ 1.01 ^{+ 0.01 }_{- 0.01 }$&$ 0.14 ^{+ 0.01 }_{- 0.01 }$&$ 3.56 ^{+ 0.08 }_{- 0.07 }$&$ 2.98 ^{+ 0.07 }_{- 0.07 }$&$ 0.417 ^{+ 0.002 }_{- 0.002 }$& 1.24 ( 816 )\\
0.4~&\nodata&$ 0.28 ^{+ 0.01 }_{- 0.01 }$&$ 1.04 ^{+ 0.02 }_{- 0.02 }$&$ 0.16 ^{+ 0.02 }_{- 0.02 }$&$ 3.29 ^{+ 0.2 }_{- 0.2 }$&$ 2.57 ^{+ 0.17 }_{- 0.16 }$&$ 0.394 ^{+ 0.005 }_{- 0.005 }$& 0.95 ( 280 )\\
0.5~&\nodata&$ 0.25 ^{+ 0.02 }_{- 0.02 }$&$ 1.01 ^{+ 0.05 }_{- 0.05 }$&$ 0.14 ^{+ 0.04 }_{- 0.03 }$&$ 3.08 ^{+ 0.38 }_{- 0.37 }$&$ 2.4 ^{+ 0.33 }_{- 0.32 }$&$ 0.34 ^{+ 0.01 }_{- 0.01 }$& 0.87 ( 83 )\\
0.7~&\nodata&$ 0.26 ^{+ 0.01 }_{- 0.01 }$&$ 1.03 ^{+ 0.04 }_{- 0.03 }$&$ 0.18 ^{+ 0.04 }_{- 0.03 }$&$ 2.94 ^{+ 0.36 }_{- 0.35 }$&$ 2.29 ^{+ 0.26 }_{- 0.27 }$&$ 0.37 ^{+ 0.01 }_{- 0.01 }$& 1.07 ( 105 )\\
0.8~&\nodata&$ 0.26 ^{+ 0.01 }_{- 0.01 }$&$ 1.02 ^{+ 0.02 }_{- 0.02 }$&$ 0.17 ^{+ 0.02 }_{- 0.02 }$&$ 3.1 ^{+ 0.2 }_{- 0.2 }$&$ 2.3 ^{+ 0.2 }_{- 0.1 }$&$ 0.370 ^{+ 0.005 }_{- 0.005 }$& 0.98 ( 268 )\\
0.9~&\nodata&$ 0.25 ^{+ 0.01 }_{- 0.01 }$&$ 1.02 ^{+ 0.02 }_{- 0.02 }$&$ 0.15 ^{+ 0.02 }_{- 0.02 }$&$ 3.2 ^{+ 0.2 }_{- 0.2 }$&$ 2.6 ^{+ 0.2 }_{- 0.2 }$&$ 0.374 ^{+ 0.005 }_{- 0.005 }$& 0.98 ( 265 )\\
\multicolumn{9}{c}{ER Vul (observation ID: 0785140601)}\\
Avg.&$1.8_{-0.3}^{+0.3}$ &$0.63_{-0.01}^{+0.01}$&$1.10_{-0.04}^{+0.01}$&$0.16_{-0.01}^{+0.01}$&$21_{-1}^{+1}$&$30_{-1}^{+3}$&$4.43_{-0.01}^{+0.01}$&1.28(787) \\
0.4~&\nodata&$ 0.64 ^{+ 0.03 }_{- 0.03 }$&$ 1.12 ^{+ 0.03 }_{- 0.03 }$&$ 0.16 ^{+ 0.01 }_{- 0.01 }$&$ 22 ^{+ 2 }_{- 1 }$&$ 27 ^{+ 2 }_{- 2 }$&$ 4.32 ^{+ 0.04 }_{- 0.04 }$& 1.06 ( 384 )\\
0.5~&\nodata&$ 0.60 ^{+ 0.02 }_{- 0.02 }$&$ 1.06 ^{+ 0.02 }_{- 0.01 }$&$ 0.16 ^{+ 0.01 }_{- 0.01 }$&$ 18 ^{+ 1 }_{- 1 }$&$ 28 ^{+ 1 }_{- 1 }$&$ 4.06 ^{+ 0.02 }_{- 0.02 }$& 1.04 ( 561 )\\
0.6~&\nodata&$ 0.63 ^{+ 0.02 }_{- 0.02 }$&$ 1.1 ^{+ 0.01 }_{- 0.01 }$&$ 0.15 ^{+ 0.01 }_{- 0.01 }$&$ 21 ^{+ 1 }_{- 1 }$&$ 34 ^{+ 1 }_{- 1 }$&$ 4.78 ^{+ 0.02 }_{- 0.02 }$& 1.15 ( 644 )\\
   \multicolumn{9}{c}{ER Vul (Observation ID: 0781500101)}\\
Avg.&$1.4_{-0.1}^{+0.1}$&$0.640_{-0.005}^{+0.01}$&$1.110_{-0.005}^{+0.01}$&$0.162_{-0.002}^{+0.002}$&$20.1_{-0.3}^{+0.3}$&$33.5_{-0.4}^{+0.4}$&$4.78_{-0.01}^{+0.01}$& 1.76(1722)\\
0.0~&\nodata&$ 0.65 ^{+ 0.01 }_{- 0.02 }$&$ 1.11 ^{+ 0.01 }_{- 0.01 }$&$ 0.16 ^{+ 0.01 }_{- 0.01 }$&$ 19.16 ^{+ 0.89 }_{- 0.99 }$&$ 31 ^{+ 1}_{- 1 }$&$ 4.43 ^{+ 0.02 }_{- 0.02 }$& 1.11 ( 936 )\\
0.1~&\nodata&$ 0.59 ^{+ 0.02 }_{- 0.02 }$&$ 1.06 ^{+ 0.01 }_{- 0.01 }$&$ 0.155 ^{+ 0.004 }_{- 0.004 }$&$ 17.9 ^{+ 0.7 }_{- 0.7 }$&$ 35.47 ^{+ 0.97 }_{- 0.95 }$&$ 4.64 ^{+ 0.02 }_{- 0.02 }$& 1.23 ( 940 )\\
0.2~&\nodata&$ 0.65 ^{+ 0.02 }_{- 0.02 }$&$ 1.13 ^{+ 0.01 }_{- 0.01 }$&$ 0.154 ^{+ 0.004 }_{- 0.004 }$&$ 22 ^{+ 1 }_{- 1 }$&$ 39 ^{+ 1 }_{- 1 }$&$ 5.33 ^{+ 0.02 }_{- 0.02 }$& 1.15 ( 1019 )\\
0.3~&\nodata&$ 0.75 ^{+ 0.02 }_{- 0.02 }$&$ 1.27 ^{+ 0.04 }_{- 0.03 }$&$ 0.18 ^{+ 0.01 }_{- 0.01 }$&$ 27 ^{+ 2 }_{- 2 }$&$ 37 ^{+ 2 }_{- 2 }$&$ 5.95 ^{+ 0.04 }_{- 0.04 }$& 1.11 ( 692 )\\
0.4~&\nodata&$ 0.67 ^{+ 0.01 }_{- 0.01 }$&$ 1.14 ^{+ 0.02 }_{- 0.01 }$&$ 0.161 ^{+ 0.004 }_{- 0.004 }$&$ 23 ^{+ 1 }_{- 1 }$&$ 36 ^{+ 1 }_{- 1 }$&$ 5.17 ^{+ 0.02 }_{- 0.02 }$& 1.09 ( 994 )\\
0.5~&\nodata&$ 0.62 ^{+ 0.01 }_{- 0.01 }$&$ 1.1 ^{+ 0.01 }_{- 0.01 }$&$ 0.176 ^{+ 0.005 }_{- 0.005 }$&$ 17.4 ^{+ 0.6 }_{- 0.6 }$&$ 29.7 ^{+ 0.8 }_{- 0.8 }$&$ 4.33 ^{+ 0.02 }_{- 0.02 }$& 1.03 ( 916 )\\
0.6~&\nodata&$ 0.64 ^{+ 0.02 }_{- 0.01 }$&$ 1.11 ^{+ 0.01 }_{- 0.01 }$&$ 0.180 ^{+ 0.005 }_{- 0.005 }$&$ 18.4 ^{+ 0.7 }_{- 0.6}$&$ 30.3 ^{+ 0.8}_{- 0.8 }$&$ 4.53 ^{+ 0.02 }_{- 0.02 }$& 1.11 ( 929 )\\
0.7~&\nodata&$ 0.63 ^{+ 0.01 }_{- 0.01 }$&$ 1.07 ^{+ 0.01 }_{- 0.01 }$&$ 0.170 ^{+ 0.004 }_{- 0.004}$&$ 18.1 ^{+ 0.7 }_{- 0.7 }$&$ 30.8 ^{+ 0.9 }_{- 0.9 }$&$ 4.45 ^{+ 0.01 }_{- 0.01 }$& 1.2 ( 1040 )\\
0.8~&\nodata&$ 0.64 ^{+ 0.01 }_{- 0.01 }$&$ 1.11 ^{+ 0.01 }_{- 0.01 }$&$ 0.173 ^{+ 0.003 }_{- 0.003 }$&$ 19.4 ^{+ 0.5 }_{- 0.4 }$&$ 32.0 ^{+ 0.5 }_{- 0.6 }$&$ 4.71 ^{+ 0.01 }_{- 0.01 }$& 1.4 ( 1305 )\\
0.9~&\nodata&$ 0.62 ^{+ 0.03 }_{- 0.03 }$&$ 1.07 ^{+ 0.03 }_{- 0.02 }$&$ 0.16 ^{+ 0.01 }_{- 0.01 }$&$ 19 ^{+ 1 }_{- 1 }$&$ 31 ^{+ 2 }_{- 2 }$&$ 4.37 ^{+ 0.03 }_{- 0.03 }$& 1.03 ( 651 )\\
    \multicolumn{9}{c}{XY UMa}\\
Avg. &$4.0_{-0.3}^{+0.3}$&$0.29_{-0.01}^{+0.01}$&$1.01_{-0.01}^{+0.01}$&$0.140_{-0.005}^{+0.005}$&$11.8_{-0.6}^{+0.6}$&$18.8_{-0.4}^{+0.4}$&$2.18_{-0.005}^{+0.005}$&1.57(439)\\
0.0~&$ 6.5 ^{+1.8}_{-1.8}$&$ 0.29 ^{+0.04 }_{-0.03 }$&$ 1.02 ^{+0.04 }_{-0.04 }$&$ 0.11 ^{+0.02 }_{-0.02 }$&$ 17 ^{+5 }_{-4 }$&$ 18 ^{+2 }_{-2 }$&$ 2.19 ^{+0.04 }_{-0.04 }$& 1.49 ( 907 )\\
0.1~&$ 3.7 ^{+0.4 }_{-0.4 }$&$ 0.29 ^{+ 0.01 }_{- 0.01 }$&$ 1.03 ^{+ 0.01 }_{- 0.01 }$&$ 0.15 ^{+ 0.01 }_{- 0.01 }$&$ 11.1 ^{+ 0.7 }_{- 0.6 }$&$ 17.7 ^{+ 0.8 }_{- 0.5 }$&$ 2.13 ^{+ 0.01 }_{- 0.01 }$& 1.13 ( 386 )\\
0.2~&$ 4.5 ^{+ 0.8 }_{- 0.8 }$&$ 0.29 ^{+ 0.02 }_{- 0.01 }$&$ 1.01 ^{+ 0.02 }_{- 0.02 }$&$ 0.15 ^{+ 0.01 }_{- 0.01 }$&$ 11 ^{+ 1 }_{- 1 }$&$ 19 ^{+ 1 }_{- 1 }$&$ 2.25 ^{+ 0.02 }_{- 0.02}$& 1.12 ( 376 )\\
0.3~&$ 4.1 ^{+ 0.8 }_{- 0.8 }$&$ 0.30 ^{+ 0.02 }_{- 0.01 }$&$ 1.00 ^{+ 0.02 }_{- 0.02 }$&$ 0.14 ^{+ 0.01 }_{- 0.01 }$&$ 11 ^{+ 2 }_{- 1}$&$ 18 ^{+ 1 }_{- 1 }$&$ 2.09 ^{+ 0.02 }_{- 0.02 }$& 1.05 ( 365 )\\
0.4~&$ 3.3 ^{+ 0.8 }_{- 0.8 }$&$ 0.35 ^{+ 0.03 }_{- 0.02 }$&$ 1.04 ^{+ 0.02 }_{- 0.02 }$&$ 0.15 ^{+ 0.01 }_{- 0.01 }$&$ 8 ^{+ 1 }_{- 1}$&$ 16 ^{+ 1 }_{- 1 }$&$ 1.90 ^{+ 0.02 }_{- 0.02 }$& 1.17 ( 380 )\\
0.5~&$ 3.4 ^{+ 0.9 }_{- 0.9 }$&$ 0.34 ^{+ 0.04 }_{- 0.03 }$&$ 1.01 ^{+ 0.02 }_{- 0.02 }$&$ 0.14 ^{+ 0.01 }_{- 0.01 }$&$ 8 ^{+ 1 }_{- 1 }$&$ 19 ^{+ 1 }_{- 1 }$&$ 2.03 ^{+ 0.02 }_{- 0.02 }$& 0.99 ( 392 )\\
0.6~&$ 4.9 ^{+ 0.9 }_{- 0.9 }$&$ 0.29 ^{+ 0.02 }_{- 0.01 }$&$ 1.02 ^{+ 0.02 }_{- 0.02 }$&$ 0.15 ^{+ 0.02 }_{- 0.01 }$&$ 14 ^{+ 2 }_{- 2 }$&$ 19 ^{+ 1 }_{- 1 }$&$ 2.36 ^{+ 0.02 }_{- 0.02 }$& 1.17 ( 365 )\\
0.7~&$ 3.3 ^{+ 0.8 }_{- 0.8 }$&$ 0.29 ^{+ 0.02 }_{- 0.01 }$&$ 1.02 ^{+ 0.02 }_{- 0.02 }$&$ 0.17 ^{+ 0.01 }_{- 0.01 }$&$ 10 ^{+ 1 }_{- 1 }$&$ 15.4 ^{+ 0.9 }_{- 0.8 }$&$ 1.96 ^{+ 0.02 }_{- 0.02 }$& 1.15 ( 338 )\\
0.8~&$ 4.3 ^{+ 0.7 }_{- 0.8 }$&$ 0.27 ^{+ 0.01 }_{- 0.01 }$&$ 0.99 ^{+ 0.02 }_{- 0.02 }$&$ 0.14 ^{+ 0.01 }_{- 0.01 }$&$ 12 ^{+ 2 }_{- 1 }$&$ 19 ^{+ 1 }_{- 1 }$&$ 2.20 ^{+ 0.02 }_{- 0.02 }$& 1.16 ( 275 )\\
0.9~&$ 5 ^{+ 1 }_{- 1 }$&$ 0.27 ^{+ 0.02 }_{- 0.01 }$&$ 0.99 ^{+ 0.02 }_{- 0.03 }$&$ 0.12 ^{+ 0.01 }_{- 0.01 }$&$ 18 ^{+ 3 }_{- 2 }$&$ 20 ^{+ 1}_{- 1 }$&$ 2.42 ^{+ 0.03}_{- 0.03 }$& 1.2 ( 147 )\\
   \multicolumn{9}{c}{TX Cnc}\\
    Avg. & $1.6_{-0.6}^{+0.6}$ &$0.56_{-0.04}^{+0.04}$&$0.91_{-0.02}^{+0.03}$&$0.35_{-0.03}^{+0.03}$&$5_{-1}^{+1}$ &$7_{-1}^{+1}$&$1.75_{-0.01}^{+0.01}$&1.07(452)\\
0.0~&\nodata&$ 0.49 ^{+ 0.05 }_{- 0.05 }$&$ 0.98 ^{+ 0.05 }_{- 0.04 }$&$ 0.4 ^{+ 0.1 }_{- 0.1 }$&$ 6 ^{+ 1 }_{- 1 }$&$ 6^{+ 1 }_{- 1 }$&$ 1.78 ^{+ 0.04 }_{- 0.04 }$& 0.82 ( 80 )\\\
0.1~&\nodata&$ 0.57 ^{+ 0.08 }_{- 0.07 }$&$ 0.97 ^{+ 0.06 }_{- 0.05 }$&$ 0.43 ^{+ 0.09 }_{- 0.07 }$&$ 5 ^{+ 1 }_{- 1 }$&$ 7 ^{+ 1 }_{- 1 }$&$ 1.89 ^{+ 0.04 }_{- 0.04 }$& 1.16(87)\\
0.2~&\nodata&$ 0.62 ^{+ 0.02 }_{- 0.02 }$&$ 0.97 ^{+ 0.04 }_{- 0.04 }$&$ 0.31 ^{+ 0.02 }_{- 0.02 }$&$8 ^{+ 1 }_{- 1 }$&$ 6 ^{+ 1}_{- 1 }$&$ 1.82 ^{+ 0.02 }_{- 0.02 }$& 1.04(319)\\
0.35 &\nodata&$ 0.6 ^{+ 0.1 }_{- 0.1 }$&$ 0.9 ^{+ 0.1 }_{- 0.1 }$&$ 0.3 ^{+ 0.06 }_{- 0.04 }$&$ 5^{+ 4 }_{- 2 }$&$ 9 ^{+ 3 }_{- 4 }$&$ 1.77 ^{+ 0.04 }_{- 0.04 }$& 1.00 ( 67 )\\
0.5~&\nodata&$ 0.48 ^{+ 0.07 }_{- 0.05 }$&$ 0.94 ^{+ 0.05 }_{- 0.05 }$&$ 0.36 ^{+ 0.09 }_{- 0.06 }$&$ 6 ^{+ 1 }_{- 1 }$&$ 7 ^{+ 1 }_{- 1 }$&$ 1.74 ^{+ 0.04 }_{- 0.04 }$& 0.77(81)\\
0.6~&\nodata&$>$0.45 & $0.9^{+0.3}_{-0.1}$&$ 0.23^{+0.04}_{-0.03 }$&$ 10^{+4 }_{-5}$ &$ 5^{+6}_{-4}$& $1.69 ^{+ 0.04}_{- 0.04 }$& 0.89(78)\\
0.7~&\nodata&$ 0.5 ^{+ 0.4 }_{- 0.1 }$&$ 0.82 ^{+ 0.08 }_{- 0.04 }$&$ 0.3 ^{+ 0.06 }_{- 0.05 }$&$ 3 ^{+ 5 }_{- 2 }$&$ 10 ^{+ 3 }_{- 5 }$&$ 1.71 ^{+ 0.04 }_{- 0.04 }$& 0.97(79)\\
0.8~&\nodata&$ 0.41 ^{+ 0.06 }_{- 0.04 }$&$ 0.89 ^{+ 0.04 }_{- 0.04 }$&$ 0.4 ^{+ 0.1 }_{- 0.1 }$&$ 5 ^{+ 1 }_{- 1 }$&$ 8 ^{+ 2 }_{- 1 }$&$ 1.76 ^{+ 0.04}_{- 0.04 }$& 1.24(81)\\
0.9~&\nodata&$ 0.5 ^{+ 0.2 }_{- 0.1 }$&$ 0.83 ^{+ 0.08 }_{- 0.04 }$&$ 0.30 ^{+ 0.06 }_{- 0.05 }$&$ 3 ^{+ 5 }_{- 1 }$&$ 11 ^{+ 2 }_{- 4 }$&$ 1.80 ^{+ 0.04 }_{- 0.04 }$& 0.94(80)\\
\enddata
\tablecomments{  All the parameters are with 68\% confidence interval for a single parameter value. kT$_{1}$ and kT$_{2}$ are in units of kiloelectronvolts, EM$_{1}$ and EM$_2$ are in units of 10$^{52}$ cm$^{_3}$, the quiescent state X-ray luminosity, $L_{XQ}$ is in units of 10$^{30}$ erg s$^{-1}$ in 0.3-10.0 keV band, and N$_H$ is in units of 10$^{20}$ cm$^{-2}$ }
\end{deluxetable*}

\begin{table}
    \caption{Variability matrix for parameters obtained from Phase-resolved spectral fitting with a 99\% of confidence level.}
    \label{tab:variability_in_spectral_params}
    \centering
    \begin{tabular}{ccccccc}
    \hline
    \hline
         system&$kT_1$&$kT_2$&$EM_1$&$EM_2$&z&$L_{XQ}$\\
    \hline
    44 Boo&C&V&C&V&V&V\\
    DV Psc&C&C&V&V&C&V\\
    ER Vul (1)& V&V&V&V&V&V\\
    XY UMa&C&C&V&V&C&V\\
    TX Cnc&C&C&C&C&C&V\\
    \hline
    \end{tabular}
~\\
\textsc{Note-- } Here V stands for variable C stands for constant. 
\end{table}

\subsubsection{Quiescent coronal length scales}
The EM ($EM=n_en_HdV$;$n_e$ and $n_H$ is the number density of electrons and hydrogen, respectively) provided in Table. \ref{tab:phaseresolvedparams} can be used to constrain the X-ray emitting plasma. 
As EM is a composite of particle density and emitting volume, thus,  the density measurement is required to constrain the emitting volume. For late-type stars, coronal densities were determined by \cite{2004A&A...427..667N} based on O VII and the Ne IX triplet. The O VII triplet corresponds to the low-temperature component of 1-6 MK and electron densities  $n_e$ of $10^{9.5-11} cm^{-3}$.
 As the density diagnosis based on He-like triplets has large errors, so we used the average value of density scatter (i.e. $log(n_e)\approx10$) for the estimation of the quiescent coronal length scale. For estimation of $n_H$ (number density of hydrogen), one can assume fully ionized plasma with a 10\% helium abundance. In this case,  \footnote{As n(He)=0.1n(H) and hence n(H)/n(e)=n(H)/[n(H)+2n(He)]=1/1.2=0.833}$n_H\sim0.8\times n_e$. 
It is reasonable to assume that the emitting volume is uniformly distributed over the surface of the star as the surface filling factor  for very active stars is  close to unity \citep[see ][]{2004ApJ...617..508T}. In this case,  the height of the coronal shell can be estimated as  $V_{coronal} = 4\pi R_{*}^{2}h_{coronal}$. 
 We computed quiescent coronal length scales for both the primary and secondary components of each binary. Next, to check for coronal connection between the components of the system, we computed the parameter $P_c=a-(R_1+h_1+R_2+h_2)$, where ``$a$" is the binary separation, $h_1$ and $h_2$ are average coronal lengths of the primary and secondary, respectively. 
The parameter $P_c$ should be less than zero for a coronally connected system.  The value of $P_C$ for the systems, ER Vul, DV Psc, 44 Boo, XY UMa, and TX Cnc  were estimated to be 1.2, 0.3, 0.06, -0.13, and -0.62 $R_\odot$, respectively.  The value of $P_c$  close to zero  indicates that the quiescent coronae of both the components are in contact with each other. 
The lower values of P$_c$ for the  contact binaries TX Cnc and 44 Boo indicate their coronae are  in contact.
Despite being a semidetached system, XY UMa shows a stronger coronal connection than 44 Boo. On the other hand, DV Psc shows a moderate coronal connection. The large value of $P_c$ for ER Vul indicates the least  possibility of a coronal connection in this system.  However, when the lower limit of electron density (i.e. $n_e=10^{9.5}$ cm$^{-3}$) is taken into account, all EBs show the significant coronal connection. 

\subsubsection{Flares}
As described in section  \ref{temporal_analysis}, a total of seven flaring events were detected from the current sample of CCBs. The spectrum for each flaring region was also extracted for a detailed analysis.  The EPIC spectra for each of the flaring events were fitted simultaneously. To account for the flare emission only the X-ray spectra during the flaring states were fitted with an additional \textsc{apec} model on the top of the quiescent state. The quiescent phase emission close to the flaring phase was taken into account by including its best-fitting two-temperature plasma model as a frozen background contribution.  This approach is equivalent to subtracting the quiescent state coronal component and provides the spectral parameters of flaring plasma. 

 The results of the best-fit flare spectra are given in Table  \ref{tab:flare_table}. The temperature of the flaring plasma is found to be in the range of 14 - 47 MK, whereas abundances during the flares are found to be similar to that from the quiescent state coronae. The X-ray luminosity during the flaring events is found to be 11 - 44 \% more than that from the quiescent state coronae. The energy released during the flaring events is in the range of $1.95\times10^{33}-2.7\times10^{34}$ erg, with flare durations ranging from 42--145 minutes.

\begin{table*}
    \caption{Best-fit spectral  and  loop parameters for flaring events. }
    \label{tab:flare_table}
    \centering
    \renewcommand{\arraystretch}{1.7}
    \begin{tabular}{cccccccccccc}
    \hline
               
         \multirow{2}{*}{System}&\multirow{2}{*}{Flare}&\multirow{2}{*}{Orbital phase}& \multicolumn{6}{c}{Spectral Parameters}&&\multicolumn{2}{c}{Loop Parameters} \\
              \cline{4-9} \cline{11-12}
              &&& $kT_3$&$EM_3$&$Z(Z_\odot)$&$L_{XF}$&E$_L$ (\%)&$\chi_\nu^2$(d.o.f.)&& L & B\\
    \hline
    44 Boo  & F1 &0.56 - 0.67 & $3.76_{-0.62}^{+0.91}$ & $ 0.43_{-0.05}^{+0.06}$&$0.164_{-0.003}^{+0.003}$ &$0.766_{-0.005}^{+0.005}$&11&1.15(544)  &&2-9&162-457\\
    DV Psc & F2 &0.51 - 0.71 & $3.05_{-0.85}^{+2.91}$ & $ 0.58_{-0.09}^{+0.10}$&$0.16_{-0.01}^{+0.01}$    &$0.442_{-0.005}^{+0.005}$&28&0.90(309)  &&4-15&107-301\\
           & F3 &0.46 - 0.70 & $2.86_{-0.58}^{+1.09}$ & $ 0.90_{-0.09}^{+0.10}$&$0.18_{-0.01}^{+0.01}$    &$0.495_{-0.005}^{+0.005}$&44&$1.09(378)$&&5-22&88-247\\
    ER Vul & F4 &0.26 - 0.32 & $3.15_{-0.14}^{+0.16}$ & $18.95_{-0.52}^{+0.51}$&$0.213_{-0.003}^{+0.003}$ &$7.45_{-0.03}^{+0.03}$  &25&1.08(1035) &&29-116&56-158\\
    XY UMa & F5 &0.85 - 1.00 & $6.37_{-1.57}^{+2.58}$ & $ 2.26_{-0.28}^{+0.32}$&$0.126_{-0.004}^{+0.004}$ &$2.67_{-0.03}^{+0.03}$  &22&1.10(249)  &&3-10&285-803\\
           & F6 &0.82 - 0.03 & $4.61_{-1.87}^{+5.00}$ & $ 1.37_{-0.21}^{+0.29}$&$0.12*$                   &$2.42_{-0.02}^{+0.02}$  &11&1.21(279)  &&3-13&182-512\\
           & F7 &0.10 - 0.21 & $4.58_{-0.84}^{+1.28}$ & $ 3.28_{-0.37}^{+0.40}$&$0.141_{-0.004}^{+0.004}$ &$2.90_{-0.03}^{+0.03}$  &33&1.18(269)  &&6-22&151-425\\
    
    \hline
    \end{tabular}
~~\\
\textsc{Notes--} All the parameters are reported at 68\% confidence. kT$_3$ is in keV,  EM$_3$ is in 10$^{52}$ cm$^{-3}$, luminosity during flare L$_{XF}$ is in units of 10$^{30}$ erg s$^{-1}$, the parameter E$_L$ is the percentage of quiescent state  luminosity enhanced during the flaring phase, L is the loop length in $10^9 cm$, and B is the magnetic field in Gauss. \\
*Fixed to the quiescent state value.\\

    \renewcommand{\arraystretch}{1.}

\end{table*}

In order to estimate the loop lengths of flaring events, we used the magneto-hydrodynamic loop model of \cite{2002ApJ...577..422S}. Using this model, the loop length (L) and magnetic field (B) are given as
$$
\begin{aligned}
&L=10^{-5}~{\mathrm EM }^{0.6} ~n_{e}^{-0.4} ~{\mathrm T}^{-1.6} \mathrm{~cm}\\
&B=5\times10^{-4} ~{\mathrm EM}^{-0.2} ~n_{e}^{0.3} ~{\mathrm T}^{1.7} \mathrm{~G} 
\end{aligned}
$$

This model allows deducing flaring loop lengths based on  three observed parameters, namely, pre-flare coronal density ($n_e$ in cm$^{-3}$), flare temperature (T = T$_3$ in K),  and  flare emission measure (EM = EM$_3$ in cm$^{-3}$).  For coronal densities, we used the density scatter of 10$^{9.5-11.0}$ as obtained by \cite{2004A&A...427..667N} to get the bounds for the loop length and magnetic field strength. The estimated values of the loop length and magnetic fields for all flaring events are given in Table  \ref{tab:flare_table}. The loop lengths of the flaring events are estimated to be of the order of $10^{9-11}$cm, whereas the magnetic field strength is estimated to be in the range of a few tens to a few hundreds of Gauss.

\section{Discussion}
\label{discussion}
We have carried out a detailed study of five CCBs. These CCBs are detached, semidetached, and contact EBs with solar-type dwarf companions. 
The X-ray light curves of all CCBs show a very dynamic nature,  with at least the presence of one flaring event in the majority of CCBs. The phase-folded light curves for all CCBs revealed the variability, which we interpret as the orbital modulation as phase-folded X-ray light curves near the primary and secondary eclipses show minimum X-ray flux. The out-of-eclipse variability can be due to the modulation of active regions that usually survive for several orbital cycles. Nonetheless,  the variability seen in the X-ray cannot be unambiguously attributed to orbital/rotational modulation and there might be a considerable degree of stochastic variability.

There are very few examples where orbital modulation in X-rays was evident among the eclipsing binaries. For example, AR Lac, HR 1099, and XY UMa. \citep[e.g.][]{1988MNRAS.235..239A,1990MNRAS.243..557B,1992MNRAS.259..453S,1996ApJ...473..470S,2014ApJ...783....2D} are binaries that had shown X-ray orbital modulation. 
X-ray eclipses are observed in these three systems as well. The stars DV Psc and ER Vul showed a similar phase-folded X-ray light curve. The X-ray light curves of these two systems mimic the O'Connell-like effect with a ratio of brightness at Max I to Max II  of $\sim$1.5. In the earlier X-ray observations, no orbital modulation for ER Vul was found  \citep[][]{1987MNRAS.227..545W,2002ASPC..277..223B,2004IAUS..215..334B}. The X-ray light-curve modeling of ER Vul and DV Psc shows that one side of each binary component is occupied by more active regions than the other side. A coronal connection in these two systems is also visible in the coronal images, where DV Psc shows a pole-to-pole intra-binary connection of active regions and  ER Vul  shows a near-the-equator coronal connection.

 The X-ray light curve of XY UMa shows three minima, with the deep primary eclipse at phase 0.0, a secondary eclipse at phase 0.5, and a dip near phase 0.75. The presence of a third minima near the 0.75 phase could be due to the following reasons: (i) an active region crosses the limb of the primary star, (ii) there exist an intra-binary plasma component that is being eclipsed, or (iii) both of the above reasons are happening at the same time. The X-ray light-curve modeling favors the second scenario to be most likely for the observed third dip. The intra-binary active region is in fact relatively twice as bright as the secondary star.
In the past, XY UMa was observed in X-rays by \cite{1998MNRAS.295..825J} and \cite{2016RAA....16..131G}, and they did not find any X-ray eclipses. However, \cite{1990MNRAS.243..557B} found clear signatures of the primary eclipse in X-rays using EXOSAT observations.

The contact binary systems,  44 Boo and TX Cnc, show an almost similar phase orbital modulation in X-rays. Both  systems are brighter during the primary eclipse. This kind of counterintuitive behavior could be due to the fact that coronae of  both stars in these systems are significantly overlapped. It appears that when the system is in an over-contact state, primary and/or secondary eclipses show a brightening feature rather than a dip. The absence of an X-ray eclipse in these two systems can also be attributed to the orbital inclination, which is 63.$^\circ$5 and 77.$^\circ$5 for TX Cnc and 44 Boo, respectively. So for X-ray light curves to show no eclipses, X-ray emitting regions need to exist anywhere from $0^\circ$ to colatitude of $26.^\circ5$ for TX Cnc and from $0^\circ$ to a colatitude of $12.^\circ5$ for 44 Boo.  In the case of 44 Boo, the X-ray light-curve modeling is also attributed to the presence of polar active regions; hence, the absence of eclipses in 44 Boo is more probably due to the geometry of the system. Whereas in the case of TX Cnc, the eclipse modeling and quiescent coronal length parameter obtained from X-ray spectral analysis support a common envelope scenario for the absence of the X-ray eclipse.   It is also noted that no eclipsing light curve for 44 Boo has been observed in the past \citep[see][]{1998ASPC..154.1093K}. The X-ray light-curve modeling and quiescent state coronal length scales suggest that coronal connection is independent of whether the binary is detached or in contact. The phase-folded X-ray light curves of the sample stars indicate that X-ray eclipses exist as long as the system is not evolved to the over-contact phase of binary evolution.


The UV light curves of DV Psc and XY UMa show the O'Connell effect with Max I/Max II being 1.17$\pm$0.01 and 1.05$\pm$0.01, respectively.  The O'Connell effect is found to be absent in the case of ER Vul. The minimum value of the estimated PC indicates that chromospheric emission dominates in the case of  ER Vul and is comparable to photospheric emission in the cases of DV Psc and XY UMa.  In either case, the positive correlation between UV and X-ray emission indicates that the chromospheric emission is positively correlated to the coronal emission. However, considering the upper bounds of PC, it appears that UV emission is being dominated by photospheric emission. If this is the case, the positive correlation between UV and X-ray flux can be attributed to a correlated photospheric and coronal emissions. The resemblance of the UV light curve of DV Psc to its optical light curves presented by \cite{2014AJ....147...50P} indicates the photospheric domination of the UV emission. If it is true, then the photosphere may be dominated by faculae rather than cool star spots.  The origin of faculae-dominated photospheric UV emission from DV Psc at the age of 1.8$\pm$0.5 Gyr \citep[][]{1974HiA.....3..395W} can be supported by the fact that stars become faculae dominated at the age of $\sim$2.55 Gyr \citep[see][]{2019A&A...621A..21R}.
 Interestingly, we found that the  UV  light curve shows the O'Connell effect when the higher photospheric contribution (even if at the lower limit) is estimated.   The positive correlation between the two emissions reinforces the belief that the photospheres of these systems are dominated by faculae.

The average hot temperature component in our current sample is $\sim$ 1 keV, while the temperature of the cool component is derived to be around two values of  $\sim$ 0.28 and $\sim$ 0.6 keV. These temperatures are similar to those of other active stars like the BY Dra type and RS CVn type \citep{dempsey1993rosat,dempsey1997rosat}. However, the quiescent state RS CVn binaries show the existence of three temperature plasma using XMM-Newton \citep[][]{2012ASInC...6..239P}. The EM corresponding to cool and hot components for these stars are in the range of $10^{52.5-53.3}$ and $10^{52.5-53.5}$ cm$^{-3}$, which is also similar to that for other active dwarf binaries \citep[][]{dempsey1997rosat}. The abundances of the majority of EBs in the sample are $\sim$ 0.15 Z$_\odot$ which is similar to those of other active systems \citep[][]{2008MNRAS.387.1627P,2012ASInC...6..239P,2012MNRAS.419.1219P}. 

For 44 Boo, based on the {\sc EINSTEIN} observations in 1979,  \cite{1984ApJ...277..263C} showed that the coronal temperature is about 1.63 keV, whereas, EUVE observations of 1994 showed a coronal temperature of $\sim$ 0.65 keV \citep[][]{1998ASPC..154..993B}, which is close to the average temperature of 0.64 keV found in this study. 
The XMM-Newton observation of 44 Boo in the present work was also analyzed by \cite{2004A&A...426.1035G} using the RGS and PN data. We found a discrepancy in the fitted spectral parameters.  The main reasons for obtaining the discrepancies in the modeled parameters are (i) they have taken around the first 25 ks of observations from PN for the spectral fitting; however, after 17 ks of observations the data were heavily affected by background flaring events, and (ii) the PN observations suffered from a severe pile-up effect, which was unavoidable.  

In the case of ER Vul, the two-temperature model was fitted by \cite{1987MNRAS.227..545W} using EXOSAT observations of 1984 and 1985. They showed that the corona of ER Vul consisted of 0.52 and 3 keV plasma in 1984 and found that the hotter component had increased by 15\% in 1985. The present observations of 2016, show cool and hot components with temperatures of 0.64 and 1.11 keV, respectively, indicating that the cool temperature has remained almost constant since past observations, whereas in the last 32 yr, the hotter component has decreased by 63\%. The hot temperature is sensitive to a higher level of activity.  Thus, it may be possible that the EXOSAT observations had suffered from the higher level of magnetic activity.  The magnetic activity cycle of 31 yr \citep{1997Ap&SS.257....1Q} supports the decreased X-ray activity level after 32 yr. The cool and hot temperature components for XY UMa are found to be consistent with that of the ROSAT observation in 1992 \citep[][]{1998MNRAS.295..825J}.

The detailed phase-resolved analysis showed luminosity is variable at a 99\% confidence level for each of the CCBs in the current sample, indicating evidence of orbital modulation. 
For the system ER Vul, all the spectral parameters are found to be variable being maximum during phase 0.25 compared with those during phase 0.75, indicating the highly active region near phase 0.25. This result is also supported by the coronal imaging obtained from the X-ray light-curve modeling.  
 Both kT$_1$ and kT$_2$ for the stars DV Psc and XY UMa are found to be constant, but their respective EMs are significantly variable. This behavior indicates that both binary components are equilibrated to a common temperature distribution, but EMs change to account for the change in emission volume seen as the phase evolves.  The coronal abundance for the majority of the stars in the sample does not show any significant variability.  Such a non-variable nature of coronal abundance shows plasma abundances are almost similar in both components. The non-variable spectral parameters in TX Cnc indicate that both components consist of a common coronal envelope as shown in their coronal images.

The derived X-ray luminosity of all stars in the sample lies in the range of $(0.4-4.9)\times10^{30}erg^{-1}$. The  $log_{10}(L_x/L_{bol})$  of  -3.3 -- -3.9 put these stars  in the supersaturation regime  of $L_X/L_{bol}$ versus the  $R_o$ relation. Inverting this result, we suggest that all the binary stars with both components being X-ray active in the supersaturation regime must have a coronal connection to some extent.  Thus, we suggest that the coronal connection between binary components must be taken into consideration while discussing the supersaturation regime. As most of the W UMa type binaries reside in the supersaturation regime, and thus could be a part of the CCB class.

At least one flare-like event is observed in four out of five  CCBs. The total energy released during these flares is in the range of $10^{33.2-34.4}$ erg.   Coronal abundances during the flare are increased by 1.1-1.3 times from the quiescent state abundances, whereas the  X-ray luminosity is increased  11-44\% to that of the quiescent state value. The loop length and loop magnetic field derived from the magnetohydrodynamic loop model are found to be in the range of $10^{9.3-11.1}$ cm and 56-803 G, which are close to the values obtained for RS CVn and other active late-type dwarfs \citep[][]{2012MNRAS.419.1219P,2008MNRAS.387.1627P}. 
Because both components of CCBs are solar-type stars, it is very plausible that the coronal connection is causing complex magnetic topologies in the system, allowing energetic flares to occur. However, energetic flares on CCBs might occur for a variety of reasons, including faster rotation periods, supersaturated coronae, and so on. Both flares that occurred in DV Psc and XY UMa are observed near the secondary and primary eclipses, respectively, indicating that the flares are driven by the same active region in each CCB.

\section{Summary and Conclusions}
\label{conclusion}
 A  detailed analysis of five CCBs has been carried out.   We found that the CCBs are very soft X-ray sources, with a majority of the X-rays coming below 2 keV energy, as found for many binaries in the past. The X-ray light curves of three CCBs, DV Psc, ER Vul, and XY UMa are found to show the eclipse behavior, whereas the other two CCBs, 44 Boo and TX Cnc, do not show eclipsed-like light curves.   Modeling of the X-ray light curves indicates that both components of the CCBs are active where the primary component is more X-ray bright than that of the secondary. Results from the coronal imaging along with the X-ray spectral analysis show that coronae of both components are connected for the detached and semidetached binaries DV Psc, ER Vul, and XY UMa, whereas in the contact binaries 44 Boo and TX Cnc a significant part of the coronae of both components are overlapped.  We suggest that the X-ray eclipse occurs as long as the photospheric surfaces of both components of active binaries do not come into physical contact. In the case of XY UMa, we found a highly active region near the coronal connection region. The X-ray spectral parameters, particularly EMs and X-ray luminosity, for the majority of the CCBs are found to be variable. Excluding TX Cnc, all other CCBs are found to be flaring during their observations with flare luminosity in the range of 10$^{29.6 - 30.9}$, which is 11 - 44 \% more than their quiescent state luminosity.

\section*{Acknowledgements}
This work is based on observations obtained with XMM–Newton, an ESA science mission with instruments and contributions directly funded by ESA Member States and NASA. We thank the referee of this paper for his/her comments and suggestions. 

\section*{Data Availability}
XMM-Newton archive (http://nxsa.esac.esa.int/nxsa-web/\#search),



\bibliographystyle{aasjournal}
\bibliography{./main.bib} 

\label{lastpage}
\end{document}